\documentclass[fleqn,usenatbib]{mnras}
\usepackage{amsmath}
\usepackage[utf8]{inputenc}
\usepackage{graphicx}
\usepackage{txfonts}
\usepackage[para,online,flushleft]{threeparttable}
\usepackage{xcolor}
\usepackage{xfrac}
\usepackage{subcaption}
\definecolor{redak}{rgb}{0.9,0.15,0.05}

\usepackage{svg}
\newcommand{\orcid}[1]{\href{https://orcid.org/#1}{\includesvg[width=10pt]{orcid}}}

\title[Exploding WR stars and CCSN progenitors]{The landscape of binary core-collapse supernova progenitors and the late emergence of Wolf-Rayet winds}
\author[Gilkis et al.]{Avishai~Gilkis$^{\orcid{0000-0001-8949-5131}}$,$^{1}$\thanks{\href{agilkis@ast.cam.ac.uk}{agilkis@ast.cam.ac.uk}} Eva~Laplace$^{\orcid{0000-0003-1009-5691}}$,$^{2,3,4}$\thanks{\href{eva.laplace@kuleuven.be}{eva.laplace@kuleuven.be}} Iair~Arcavi$^{\orcid{0000-0001-7090-4898}}$,$^{5}$ Tomer~Shenar$^{\orcid{0000-0003-0642-8107}}$ $^{5}$ and Fabian~R.~N.~Schneider$^{\orcid{0000-0002-5965-1022}}$ $^{2,6}$ \\
$^{1}$ Institute of Astronomy, University of Cambridge, Madingley Road, Cambridge CB3 0HA, United Kingdom\\
$^{2}$ Heidelberger Institut f{\"u}r Theoretische Studien, Schloss-Wolfsbrunnenweg 35, D-69118 Heidelberg, Germany\\
$^{3}$ Institute of Astronomy, KU Leuven, Celestijnenlaan 200D, 3001
Leuven, Belgium\\
$^{4}$ Anton Pannekoek Institute for Astronomy, University of Amsterdam, Science Park 904, 1098 XH Amsterdam, The Netherlands\\
$^{5}$ The School of Physics and Astronomy, Tel Aviv University, Tel Aviv 6997801, Israel \\
$^{6}$ Astronomisches Rechen-Institut, Zentrum f{\"u}r Astronomie der Universit{\"a}t Heidelberg, M{\"o}nchhofstr. 12-14, 69120 Heidelberg, Germany}

\date{Accepted 2025 May 28. Received 2025 May 23; in original form 2025 March 3}

\pubyear{2025}

\begin{document}
\label{firstpage}
\pagerange{\pageref{firstpage}--\pageref{lastpage}}

\maketitle

% ==========================================================
\begin{abstract}
The majority of core-collapse supernova (CCSN) progenitors are massive stars in multiple systems, and their evolution and final fate are affected by interactions with their companions. These interactions can explain the presence of circumstellar material in many CCSNe, and the inferred low mass in stripped-envelope supernova progenitors. Through binary interactions, stars can gain mass, lose mass, or merge, impacting their final properties. Specific sub-types of binary interaction products have been investigated but few detailed full population models exist. Using thousands of detailed simulations with updated prescriptions for binary interactions and winds at Milky Way and Magellanic Clouds metallicities, we follow the evolution of single massive stars, primaries in interacting binaries and coalescence products following common envelope evolution. We also follow the evolution of the surviving secondary star, with a compact companion formed from the evolutionary end of the primary star or alone if the system was disrupted in the first supernova. The endpoints of our simulations map the rich landscape of CCSN progenitors, and provide detailed mass-loss history and progenitor structures. We identify an important evolutionary phase for stripped-envelope supernova progenitors, in which the wind mass-loss rate of stars stripped by binary interaction rapidly increases in their final evolutionary stages, after core helium burning. These strong winds would give rise to a Wolf-Rayet (WR) spectral appearance, though only for a few millennia, in contrast to hundreds of millennia for their more massive WR counterparts. Such lightweight WR stars in binaries can account for observed properties of type Ib/c supernovae.
\end{abstract}

% ==========================================================

% ==========================================================
\section{Introduction}
\label{sec:intro}
% ==========================================================

Core-collapse (CC) supernovae (SNe) are energetic transient events arising from the explosive deaths of massive stars. A diversity of CCSN types are classified by their light curves and by spectral features like the identification (or not) of hydrogen or helium, or narrow emission lines usually associated with interaction of the ejecta with circumstellar matter (CSM). Some of the main CCSN types are Type~Ib SNe which have helium but not hydrogen \citep{Harkness1987,Porter1987}, Type~Ic SNe which lack both hydrogen and helium \citep*{Filippenko1990,Nomoto1990}, Type~II-P SNe, in which hydrogen is observed and the light curve has a long plateau \citep*[][]{Barbon1979}, and Type~IIn \citep{Schlegel1990}, where `n' indicates narrow emission-line features thought to be a result of interaction between the SN ejecta and CSM. The most common among these are Type~II-P, which are readily connected to single red supergiant (RSG) stars that are the most common evolutionary endpoint of massive stars, and numerous RSGs have been confirmed as Type~II-P progenitors in pre-explosion images \citep{Smartt2015,VanDyk2017}.

Directly identifying the progenitors of CCSNe in pre-explosion observations provides invaluable information and further constraints on the evolutionary channels leading to specific SN types \citep{VanDyk2017,VanDyk2023}. Already in the first SN for which the progenitor was clearly identified\footnote{Earlier SNe for which a progenitor might have been detected, SN~1961V \citep{Zwicky1964} and SN~1978K \citep{Ryder1993,Ryder2016}, are peculiar and it is debated whether they were real SNe or SN impostors that did not signal the end of a star \citep{VanDyk2012,Chiba2020}.}, SN~1987A in the nearby Large Magellanic Cloud (LMC) galaxy \citep*{Sonneborn1987}, the importance of binary interaction was evident. Instead of a RSG, the progenitor was surprisingly blue, and one theory accounting for the blue appearance is that the progenitor merged with a low-mass companion before exploding \citep*{ChevalierSoker1989,Podsiadlowskietal1990}. The second SN with an identified progenitor was SN~1993J in M81, where hydrogen was detected at early times but later disappeared, making this a Type~IIb SN \citep{Nomoto1993}. The associated progenitor identification was inconsistent with a single star \citep{Aldering1994}, suggesting that binary interaction removed most of the hydrogen in the progenitor envelope \citep{Podsiadlowski1993}, and indeed late-time observations reveal an ultraviolet excess consistent with a surviving companion \citep{Maund2004,Fox2014}.

In the decades since the first identification of a CCSN progenitor, numerous progenitors of Type~II-P SNe have been identified in farther galaxies, corresponding to RSGs and now outnumbering all other detected SN progenitors \citep{Smartt2015}. For Type~Ib and Type~Ic SNe, it has been suggested that the progenitors might be Wolf-Rayet (WR) stars \citep*[e.g][]{Ensman1988,Woosley1995}, which are characterised by optically-thick winds and emission line features in their spectra arising from high temperatures and high luminosity to mass ratios \citep{Crowther2007}. However, the observed properties of WR stars are in tension with the light curves and inferred ejecta masses of Type~Ib/c SNe \citep[e.g.][]{Shigeyama1990}, as the fast decline of Type~Ib/c SNe is inconsistent with the estimated masses of WR stars \citep[$\ga 8\,\mathrm{M}_\odot$ in Solar-like metallicity environments, e.g.][]{Shenar2020}, and an alternative scenario postulates that their progenitors are lower-mass helium stars formed through binary interactions \citep*[e.g.][]{Podsiadlowskietal1992,Yoon2015}. The first Type~Ib SN to have a confirmed progenitor identified was iPTF13bvn, for which initial analyses pointed to a WR progenitor (\citealt{Cao2013}; \citealt*{Groh2013}), but later studies favoured a helium star formed by binary interaction with a final mass lower than that expected for WR stars \citep{Bersten2014,Fremling2014,Eldridge2015,Folatelli2016}. The second Type Ib SN with a progenitor identified was SN~2019yvr, where the pre-explosion photometry is consistent with a yellow supergiant (YSG) with a a radius of $\ga 200\,\mathrm{R}_\odot$ \citep{Kilpatrick2021}. The disappearance of the YSG and its association with a Type Ib SN awaits confirmation with follow-up observations. In both the case of iPTF13bvn and of SN~2019yvr it is probable that the progenitor was not a WR star, but an evolved helium star resulting from envelope stripping in binary interaction. Such non-WR stripped helium stars have recently been discovered in the LMC and the Small Magellanic Cloud (SMC) galaxies by \cite{Drout2023} and \cite{Gotberg2023}, uncovering a crucial missing link in the evolution towards hydrogen-deficient CCSNe.

A clear identification of a WR progenitor for a CCSN in pre-explosion images is still absent. The Type Icn SN~2019hgp has been claimed to be the explosion of a WR star \cite{GalYam2022}, though its inferred ejecta mass of $\approx 1.2\,\mathrm{M}_\odot$ challenges the classical WR progenitor scenario, as it is much lower than the typical masses expected in WR stars. \cite{Pellegrino2022} also found low ejecta masses, of $\approx 1$--$2\,\mathrm{M}_\odot$, for a sample of Type~Icn SNe, and a SN associated with neutron star (NS) formation will be accompanied by an ejecta mass of the order of WR masses, i.e. $\ga 8\,\mathrm{M}_\odot$.

Some of the challenges in connecting SNe with their progenitors might be related to stellar multiplicity. Binary interactions are expected to play an important role for SN progenitors, as the majority of massive stars are born in binary systems with initial orbital periods that are short enough so that binary interaction can significantly affect their evolution (\citealt*{Vanbeveren1998}; \citealt{Sana2012}). Binary interactions not only affect the properties and appearance of the progenitor stars (as discussed above, see also \citealt*{Yoon2010}; \citealt{Eldridge2013,Eldridge2015}; \citealt*{Yoon2017}; \citealt{Laplace2020,Sravan2020}), but can also significantly affect the pre-SN core structure through mass loss, mass gain, or merging, all of which lead to a new configuration of the stellar structure (\citealt{Laplace2021}; \citealt*{Schneider2021,Schneider2024}). Another crucial aspect of massive star evolution is the mass loss by stellar winds \citep*{Smith2014,Vink2022}. The combined effects of binary interaction and stellar winds can be decisive for the final evolution of massive stars and the resulting SN type (\citealt{GilkisVinkEldridgeTout2019}; \citealt*{Jung2022}). The mass-loss history of CCSN progenitors, from winds or binary interactions, can in turn considerably affect the light curve and spectral properties of SNe through interaction with CSM \citep*{Moriya2016,Morozova2018,Dessart2022}. The evolution of massive stars, the SN types associated with them and the physics of their final collapse and explosion therefore require that the theoretical study of CCSN progenitors combine detailed binary stellar evolution simulations with the state-of-the-art in stellar wind mass loss. So far, detailed stellar evolution simulations have been used for a relatively small number of CCSN progenitor models, while population synthesis studies use approximations that do not account for stellar structure effects at late evolutionary stages or out of thermal equilibrium. Here, we combine both worlds with detailed simulations on a large scale, allowing us both to uncover intricate properties of CCSN progenitors at late evolutionary stages and to quantitatively estimate rates of evolutionary channels and endpoints.

In the present study we investigate the evolution of CCSN progenitors and generate a comprehensive grid of massive star evolution simulations, for both single stars and binary stars. A principal advantage of our study is the combination of large numbers and detailed simulations. We apply updated hot-star wind mass-loss rate recipes and prescriptions for mass-transfer stability and common envelope evolution (CEE). These models are therefore a representation of the current state-of-the-art in stellar and binary physics. In Section\,\ref{sec:method} we describe the numerical method and underlying physical assumptions of our stellar evolution models and in Section\,\ref{sec:results} we present the properties of the CCSN progenitors in our stellar evolution grid. In Section\,\ref{sec:IMFetal} we present probability-weighted outcomes from our simulations.  In Section\,\ref{sec:WR} we focus on the short-lived WR phase in low-mass stripped CCSNe progenitors. We discuss the implications of our results for observations of CCSNe and their progenitors in Section\,\ref{sec:discussion} and present our conclusions in Section\,\ref{sec:summary}.

%==========================================================
\section{Numerical method}
\label{sec:method}
% ==========================================================

We use version 15140 of the Modules for Experiments in Stellar Astrophysics (\textsc{mesa}) code \citep{Paxton2011,Paxton2013,Paxton2015,Paxton2018,Paxton2019} to evolve non-rotating single and binary stellar models with three metallicities, $Z=0.014$, $Z=0.0056$ and $Z=0.00224$. These metallicities approximately represent the average metal enrichment in the Milky Way, the LMC and the SMC galaxies. The initial masses of the primary stars in binaries are $M_\mathrm{ZAMS,1} = 4$, $5$, $6$, $7$, $8$, $9$, $10$, $11$, $12$, $13$, $14$, $15$, $16$, $17$, $19$, $22$, $25$, $29$, $33$, $39$, $45$, $52$, $60$, $69$, $80$, $92$ and $99\,\mathrm{M}_\odot$. The $92$ initial masses of single stars are those for all the primaries and secondaries in binaries with $M_\mathrm{ZAMS}\ge 4\,\mathrm{M}_\odot$ (for a total of $82$ values after discounting repetitions) and another $10$ values in the higher mass range where the mass spacing is sparser. For binary systems (with orbits assumed to be circular) we simulate three mass ratios, $q\equiv M_\mathrm{ZAMS,2} / M_\mathrm{ZAMS,1} = 0.25$, $0.55$ and $0.85$, and $14$ initial orbital periods, $P_\mathrm{i} = 2$, $3$, $4$, $5$, $10$, $18$, $33$, $60$, $110$, $201$, $367$, $669$, $1219$ and $2223\,\mathrm{d}$. The chosen initial orbital periods reflect the range relevant for most binary interactions \citep[e.g.][]{Sana2012}. For each metallicity, a total of $1134$ combinations of initial conditions are simulated for binary systems, in addition to the $92$ single-star simulations.

% ==========================================================
\subsection{Mixing}
\label{subsec:mix}
% ==========================================================

The Ledoux stability criterion is used to define convective regions, where mixing is treated according to mixing-length theory \citep*[MLT;][]{BohmVitense1958,Henyey1965} with a mixing-length parameter of $\alpha_\mathrm{MLT}=2$. In regions which are stable according to the Ledoux criterion but unstable according to the Schwarzschild criterion, semiconvective mixing is according to \cite*{Langer1983}, with an enhanced efficiency factor of $\alpha_\mathrm{sc}=100$ \citep[e.g.][]{Schootemeijer2019}. We employ the \texttt{use\_superad\_reduction} option for treatment of superadiabatic convection.

Overshooting above convective cores burning hydrogen or helium is computed according to the prescription presented by \cite{Jermyn2022}, which was calibrated from three-dimensional numerical hydrodynamics simulations \citep{Anders2022}, and the resulting overshooting extent for single-star evolution is presented in Appendix\,\ref{sec:appendixa}. Overshooting below convective regions of advanced burning stages (carbon burning and beyond) follows the exponentially decaying prescription of \cite{Herwig2000}, with a decay scale of $f_\mathrm{ov} H_P$, where $f_\mathrm{ov}=0.01$ and $H_P$ is the pressure scale height, and the decay starts at a point $f_\mathrm{ov,0} H_P$ into the convective region with $f_\mathrm{ov,0}=0.005$.

% ==========================================================
\subsection{Wind mass loss}
\label{subsec:winds}
% ==========================================================

Wind mass loss rates of massive stars are notoriously uncertain \citep{Smith2014,Vink2022}. In this study we combine latest theoretical expectations and observational constraints to model the wind mass loss rates. We use several different wind mass-loss recipes for the wide range in luminosity and temperature that stellar models cover throughout their evolution, with the motivation of using empirical rates where they are applicable and theoretical rates within the regime where they were computed. We start by computing an `optically thin' wind mass-loss rate $\dot{M}_\mathrm{thin}$ following \cite{VS21} for surface hydrogen mass fractions of $X_\mathrm{H,s}\ge 0.7$, \cite{Vink2017} for $X_\mathrm{H,s}\le 0.4$, and interpolating in between. Then, we consider three temperature regimes, as follows.

In the high temperature regime, $T_\mathrm{eff} > 70\,\mathrm{kK}$, we take the maximum between $\dot{M}_\mathrm{thin}$ and $\dot{M}_\mathrm{SV20}$, where $\dot{M}_\mathrm{SV20}$ follows the theoretical wind prescription of \cite{Sander2020}. The $\dot{M}_\mathrm{SV20}$ prescription depends on the number of free electrons per atomic mass unit $q_\mathrm{ion}$. \cite{Sander2020} assumed $q_\mathrm{ion}=0.5$, which is appropriate for a helium-dominated composition. To allow for a gradual composition change from hydrogen-rich to helium-rich, and eventually to carbon-rich, we take the approximation $q_\mathrm{ion}=\left(4-3 X_\mathrm{H,s}-2 X_\mathrm{He,s}\right)/\left(12-11 X_\mathrm{H,s}-8 X_\mathrm{He,s}\right)$, calculated assuming full ionisation of hydrogen and helium (which is appropriate in this high temperature regime) and a contribution of two free electrons per carbon ion. We use the $\dot{M}_\mathrm{SV20}$ prescription only for luminosities of $L>10^{4.5}\,\mathrm{L}_\odot$, and simply use $\dot{M}_\mathrm{thin}$ for lower luminosities in the high temperature regime. Few models of stellar winds in the high temperature regime exist, and here we employ recent theoretical models which reproduce the metallicity-dependent luminosity threshold of the WR phenomenon \citep{Shenar2020}.

An intermediate temperature regime is defined as $T_\mathrm{jump} \le T_\mathrm{eff} \le 70\,\mathrm{kK}$, where the bi-stability jump temperature $T_\mathrm{jump}$ is computed following \cite*{Vink2000,Vink2001}, and is $\approx 22\,\mathrm{kK}$. In this temperature range, an Eddington parameter $\Gamma_\mathrm{e}$ and optically-thick wind mass-loss rate $\dot{M}_\mathrm{thick}$ are computed following \cite{Grafener2008}. For $\Gamma_\mathrm{e}\ge 0.446$ the mass-loss rate is taken to be $\dot{M}_\mathrm{thick}$, for $\Gamma_\mathrm{e}\le 0.406$ it is $\dot{M}_\mathrm{thin}$, and for $0.406 < \Gamma_\mathrm{e} < 0.446$ an interpolated value between the two is adopted.

Below $T_\mathrm{jump}$, a `cool' wind mass-loss rate $\dot{M}_\mathrm{cool}$ is computed as the higher between $\dot{M}_\mathrm{thin}$ and the empirical formula of \cite*{dJ88}. Our choice to define the `cool' regime below $T_\mathrm{jump}$ is similar to \cite{Brott2011}, who took a maximum between \cite{Vink2001} and \cite{Nieuwenhuijzen1990} in this regime (the parametrisation derived by \citealt{Nieuwenhuijzen1990} is close to \citealt{dJ88}). We note that this can result in usage of the \cite{dJ88} formula at temperatures considerably higher than those of RSGs, but this affects only short evolutionary phases in low-metallicity models in which $\dot{M}_\mathrm{thin}$ is very low. An asymptotic giant branch wind mass-loss rate $\dot{M}_\mathrm{AGB}$ is computed as done by \cite{MillerBertolami2016} if hydrogen and helium are depleted in the core and the luminosity is below an adopted maximum AGB luminosity of $L_\mathrm{AGB,max}=55000\,\mathrm{L}_\odot$, while otherwise $\dot{M}_\mathrm{AGB}=0$. A red giant wind $\dot{M}_\mathrm{RGB}$ is computed with the mass-loss relation of \cite{Schroder2005} for $T_\mathrm{eff} \le 4800\,\mathrm{K}$, and $\dot{M}_\mathrm{RGB}=0$ for higher temperatures. We then take the mass-loss rate in the cool regime as the maximum of the three rates, i.e. $\dot{M}=\max\left(\dot{M}_\mathrm{cool} , \dot{M}_\mathrm{AGB} , \dot{M}_\mathrm{RGB}\right)$.

We do not include mass loss via eruptions as theorised for late evolutionary stages or observed in luminous blue variable (LBV) stars (\citealt*{Conti1984,Smith2004}; \citealt{Jiang2018,Grassitelli2021,Cheng2024}).

% ==========================================================
\subsection{Mass transfer and common envelope evolution}
\label{subsec:cee}
% ==========================================================

The mass-transfer rate is computed following the extension of the \cite{Kolb1990} prescription as described by \cite{Marchant2021} for Roche-lobe overflow (RLOF), where the Roche-lobe radius follows \cite{Eggleton1983}. The mass transfer efficiency (i.e. how much of the mass lost from the Roche-lobe filling donor is accreted by the second star) follows the prescription described by \cite{GilkisVinkEldridgeTout2019}, so that accretion onto the secondary is limited by its thermal timescale or if it is close to filling its Roche lobe\footnote{Mass gain can spin up stars to critical rotation \citep{Packet1981}, so that further accretion is suppressed, and rotation in principle also reduces the mass transfer efficiency. Our models do not include rotation, and therefore there is no mass transfer efficiency reduction because of fast rotation.}. In each time step of the binary evolution a critical mass-transfer rate, $\dot{M}_\mathrm{th,crit}$, is computed following the local thermal readjustment timescale defined by \cite{CopingWithLoss}. If the mass-transfer rate exceeds $\dot{M}_\mathrm{th,crit}$, a CEE phase is simulated as described by \cite{Marchant2021}. During CEE, the mass-loss rate is reduced to the global envelope thermal timescale. The energy required to remove envelope material is taken from the orbit, and the orbital separation is reduced accordingly. The efficiency of using orbital energy to remove envelope material is assumed to be $100\%$, i.e. none of the energy is radiated away.\footnote{This means that the commonly-used efficiency parameter $\alpha_\mathrm{CEE}$ is $1$ in our simulations.} The envelope ejection is considered unsuccessful if the donor core radius exceeds the Roche lobe of the star, where the core boundary is defined at the mass coordinate where the hydrogen and helium mass fraction equal the central values to within $0.01$, and in this case a coalescence product is constructed and evolved (Section \ref{subsec:channels}). If the envelope of the donor star recedes within its Roche lobe before its core overflows then the CEE phase is over, and binary evolution continues as before at the new orbital separation. The stability of mass transfer and the outcome of CEE in cases of unstable mass transfer as a function of the initial binary configuration and metallicity are presented in Appendix\,\ref{sec:appendixMTstability}.

% ==========================================================
\subsection{Evolutionary channels}
\label{subsec:channels}
% ==========================================================

Stars are evolved until core collapse (defined as the point when the maximal inward velocity in the iron core exceeds $50\,\mathrm{km}\,\mathrm{s}^{-1}$) whenever numerically feasible, and at least until the end of core carbon burning. If core collapse is not reached, but the carbon-oxygen core mass at the end of the simulation is above the Chandrasekhar mass, i.e. $M_\mathrm{CO} > 1.38\,\mathrm{M}_\odot$, we consider the stellar model to represent a CCSN progenitor. We enforce a maximum number of time steps of $30000$, which is the stopping condition for most of the lower masses that are not expected to explode as CCSNe, i.e. $M_\mathrm{ZAMS} \la 8$. Binary-star evolution ends if the primary becomes a white dwarf (WD), defined as the point where the surface gravity reaches $g = 10^7\,\mathrm{cm}\,\mathrm{s}^{-2}$, or the two stars merge during CEE (Section\,\ref{subsec:cee}). If the stars merge, a coalescence product is constructed by accretion of material with the composition of the less dense between the core of the donor star and the entire accretor star onto the denser component on the thermal timescale of the denser component, and assuming the entire mass of the less dense component is retained (i.e. no mass is ejected from the system during the merging process). The coalescence product is then further evolved as a single star.

Because of the short timescale between core carbon depletion and iron core collapse, we switch off binary mass transfer, wind mass loss and thermohaline mixing after core carbon depletion. When stellar models locally exceed the Eddington limit numerical difficulties arise due to the assumption of hydrostatic equilibrium. To overcome these numerical difficulties, we found it helpful to switch to the \texttt{MLT++} method for energy transport in superadiabatic regions \citep{Paxton2013} for models with $\log_{10}\left(T_\mathrm{eff} / \mathrm{K} \right)<4.3$, or $\log_{10}\left(T_\mathrm{eff} / \mathrm{K} \right)<5.025$ and $\log_{10}\left(L / \mathrm{L}_\odot\right)<5.65$, at core carbon depletion (while all other models continue with the \texttt{use\_superad\_reduction} option). The $3925$ final profiles of simulations that succeed in reaching iron core collapse are assessed with the semi-analytical neutrino-driven SN explosion code of \cite{Muller2016} to determine whether a successful explosion occurs and forms a NS (we assume a maximum NS mass of $2\,\mathrm{M}_\odot$), or a black hole (BH) forms with no explosion. The explosion code parameters are set to the calibration by \cite{Schneider2021} to achieve an energy in the range between $0.5\times 10^{51}\,\mathrm{erg}$ and $10^{51}\,\mathrm{erg}$ in explosions of RSGs (previous studies using these assumptions include \citealt*{Schneider2023}; \citealt{Schneider2024}; \citealt{Temaj2024} and \citealt*{Laplace2025}). BH formation by fallback (FB) occurs when the explosion is successful according to the SN explosion code \citep{Muller2016} but releases an energy lower than the binding energy of the envelope, and in these cases the remnant mass is the sum of the NS mass and half the ejecta mass\footnote{The fallback mass is highly uncertain and the choice of half the ejecta mass is somewhat arbitrary.}. The masses of the remnants, NSs or BHs, are used to construct additional simulations of the surviving secondary star.

For cases were iron core collapse was not reached, but a CCSN is expected according to the core mass at the end of carbon burning, the successful simulations were used to derive an approximate condition to distinguish between successful CCSNe producing NSs and BH formation with no SN. In these cases, if the CO core mass exceeds $9.03\,\mathrm{M}_\odot$ then a BH is assumed to form, and otherwise a NS forms with a mass following a fit to the successful simulations (Section\,\ref{subsec:remnants_and_SNe}).

NSs are born with natal kicks, and the ejected mass from the SN removes mass and momentum from the system as well as contributing to the kick velocity. To account for these effects we generate for each endpoint $10^5$ random NS velocities following the distribution inferred by \cite{Hobbs2005} from pulsars in the Galaxy, re-scaled by the ratio of the ejecta mass to the remnant mass \citep[e.g.][]{GiacobboMapelli2020}. Three values from the cumulative distribution function of $a (e^2 -1 )$ from the randomly generated post-SN orbits are taken as the separation for a circular orbit in the continuation runs, representing three equally-likely regimes (i.e. the midpoint of each of the three terciles). We assume that if the NS affects the evolution of the secondary in any way then the separation has already become small enough for tidal circularisation to occur quickly. As there is a non-zero probability for either the disruption of the binary or its survival, owing to the large number of randomly sampled kicks, we also simulate the continued evolution of the secondary as a single star. We do not consider any ejecta-NS interactions \citep*[e.g.][]{Hirai2018}.

For the case of BHs, the remnant mass is computed similarly to \cite{Renzo2022}, with mass loss resulting from neutrino radiation in all cases where BHs form \citep{Nadezhin1980,Lovegrove2013}, and additional mass loss caused by pair instability (PI) resulting from electron-positron pair creation in high mass cores \citep{Barkat1967,Rakavy1967}. We adopt a threshold CO core mass for PI of $34.8\,\mathrm{M}_\odot$, following \cite{Renzo2022}. The instability arising from pair creation can lead to pulses of mass ejection in a process termed pulsational pair instability \citep*[PPI;][]{Woosley2007,Woosley2017}, and the total mass lost before a final collapse to a BH is computed according to the fit derived by \cite{Renzo2022}. At high enough CO core masses, PI leads to a single pulse and the complete disruption of the star. Following the mass loss fit by \cite{Renzo2022}, we find complete disruption for $M_\mathrm{CO}\ga 60\,\mathrm{M}_\odot$, and in the small number of cases this occurs PI leaves no remnant behind and the secondary is further evolved as a single star. When a massive star does not produce a NS in a successful SN explosion and the CO core mass is not massive enough to initiate PI, we assume a BH forms by direct collapse, with a mass equal to the total final stellar mass minus the mass lost by neutrino radiation. For all cases in which a BH forms, whether after PPI or by direct collapse, the orbital separation is updated as a result of the sudden mass loss \citep{Blaauw1961}. As in the NS cases, rapid circularisation is assumed, but for BHs the outcome is assumed to be deterministic with a zero-velocity kick and only one simulation is run for the secondary in a bound binary with a BH. An exception is a small number of simulations where a nominally successful very low energy explosion occurs, and we assume that half of the ejecta mass falls back onto the NS, transforming it into a BH. In these fallback cases, the same kick computation scheme as for NSs is employed, but the kick velocity is re-scaled according to the updated BH mass.

We assume that the accretion onto a NS or a BH is limited to the Eddington rate \citep*[e.g.][]{Podsiadlowski2003}. For WDs the accretion is limited following \cite*{Meng2009}, with a low mass-transfer rate regime in which nova eruptions result in no mass gain, a high mass-transfer rate regime where accretion is limited by an optically thick wind \citep*{Hachisu1996}, and a conservative mass-transfer regime for intermediate rates\footnote{The mass-transfer rates defining the boundaries between the three regimes depend on the mass of the accreting WD and the composition of the accreted material as described by \cite{Meng2009}, but the intermediate regime is roughly $10^{-7}\,\mathrm{M}_\odot\,\mathrm{yr}^{-1} \la \dot{M} \la 10^{-6}\,\mathrm{M}_\odot\,\mathrm{yr}^{-1}$.}. Similarly to the primary stars, merged stars and single stars, we attempt to evolve secondary stars up to iron core collapse, or at least the end of core carbon burning.

% ==========================================================
\subsection{Interventions in failed simulations}
\label{subsec:voodoo}
% ==========================================================

Approximately $76\%$ of the binary simulations ($2570$ runs) ended according to the prescribed stopping conditions, i.e. the star is clearly on the way to CC, WD formation or the two stars are merging during CEE. An inspection of the simulations suggested several approaches to rescue the failed runs.

The most common fix was to lower the threshold for CEE initiation by a factor of $40$\footnote{Initiation of CEE when the mass-transfer rate exceeded $2.5\%$ of the critical rate computed following \cite{CopingWithLoss} worked well most of the time, while in a few cases lower threshold rates were needed.}, applied in $520$ simulations, leading to a successful conclusion in $357$ of these. This was applied in simulations that failed during CEE, and also to simulations that had runaway mass transfer but in which the condition for CEE initiation in our code was not fulfilled.

In $41$ cases with initial primary masses $\ge 60\,\mathrm{M}_\odot$, the simulation crashed because of reverse mass transfer onto the primary (after it has become a WR star) from a secondary evolving at a comparable pace. These were re-run with a zero mass-transfer efficiency for the reverse mass transfer (which is reasonable when the mass-transfer rate is comparable to the WR wind mass-loss rate), resulting in a successful conclusion in $40$ simulations. Another $12$ similar runs with two high-mass stars evolving at a comparable pace crashed even with zero mass transfer efficiency, but succeeded when mass loss via RLOF from the secondary was completely disabled (i.e. donor switching was not allowed), resulting in a successful conclusion for $10$ of these.

Another $77$ simulations stopped when both stars were overflowing their Roche lobes. These were assumed to merge (see \citealt*{Henneco2024}), and their evolution was continued in the same way as simulations in which the two stars merged during CEE.

The inspection of $48$ simulations suggested that their future evolution would reach the formation of a WD. These were continued with a new maximum number of time steps of $10^5$, and indeed $32$ reached the conditions for WD formation.

The steps described above increased the success rate of the binary simulations from $\approx 76\%$ to $\approx 90\%$. The final intervention we applied was to extrapolate the endpoint of simulations that did not reach a clear conclusion if they are expected to produce WDs. For this purpose, we used the relation between the CO core mass at the end of helium burning and the final CO core mass in successful simulations, finding that if the former is below $1.1\,\mathrm{M}_\odot$ then WD formation is expected instead of a CCSN. The WD mass was taken as the final CO core mass in these simulations if it had stopped growing, or was extrapolated if its mass was still increasing at the end of the simulation (this was done by using the CO core mass as function of the remaining envelope mass and extrapolating to the point where the envelope mass reaches zero).

Some simulations of the secondary evolution with a WD or NS companion  required the mass-transfer rate threshold for CEE initiation to be lowered as well (by a factor of $40$, like in the simulations with a non-degenerate companion). The aforementioned intervention methods require the inspection of all unsuccessful simulations, and therefore they are not easily scalable to larger simulation grids.

% ==========================================================
\subsection{Evaluating the properties of WR stars}
\label{subsec:calcWR}
% ==========================================================

Spectroscopically, a star is classified as a WR star when diagnostic spectral lines belonging mainly to helium show substantial emission. This happens when the mass-loss rate is large relative to the stellar surface. A convenient parametrisation of the emission strength is given by the `transformed radius' \citep*{Schmutz1989}, 
\begin{equation}
R_\mathrm{t}=R_* \left(\frac{v_\infty / 2500\,\mathrm{km}\,\mathrm{s}^{-1}}{\sqrt{D} \dot{M} / 10^{-4}\,\mathrm{M}_\odot\,\mathrm{yr}^{-1}}\right)^{2/3}\, ,
    \label{eq:Rt}
\end{equation}
where $R_*$ is the stellar radius, $v_\infty$ the terminal wind velocity, $\dot{M}$ the wind mass-loss rate and $D$ the so-called clumping factor which allows for an inhomogeneous wind density. We note that smaller values of $R_\mathrm{t}$ correspond to higher values of $\dot{M}$, and hence stronger emission. Hydrogen-deficient ($X_\mathrm{H,s} < 0.05$) models with more carbon than nitrogen in their surface composition have $v_\infty = 2000\,\mathrm{km}\,\mathrm{s}^{-1}$ and $D=10$ \citep*[e.g.][]{Sander2012}. Models with $X_\mathrm{H,s} \ge 0.05$ have $v_\infty = 1000\,\mathrm{km}\,\mathrm{s}^{-1}$ and $D=4$ (e.g. \citealt*{Hamann2006}; we note that \citealt{Shenar2016,Shenar2019} used $D=10$, though this would make a minor difference), while models with $X_\mathrm{H,s} < 0.05$ and more nitrogen than carbon in their surface composition  have $v_\infty = 1600\,\mathrm{km}\,\mathrm{s}^{-1}$ and $D=4$. If $\log_{10} R_\mathrm{t} < 1.5$ the model is considered to be a WR star \citep{Shenar2020}.

WR sub-types are assigned according to the surface composition as follows: 
\begin{itemize}
    \item WNh: $X_\mathrm{H,s} \ge 0.05$;
    \item WC: $X_\mathrm{H,s} < 0.05$, $X_\mathrm{N,s} < X_\mathrm{C,s}$, $X_\mathrm{C,s} / 12 + X_\mathrm{O,s} / 16 < X_\mathrm{He,s} / 4$;
    \item WO: $X_\mathrm{H,s} < 0.05$, $X_\mathrm{N,s} < X_\mathrm{C,s}$, $X_\mathrm{He,s} / 4 \le X_\mathrm{C,s} / 12 + X_\mathrm{O,s} / 16$;
    \item WN: otherwise;
\end{itemize}
where the distinction between WC and WO is motivated by requiring $N_\mathrm{He,s} \le N_\mathrm{C,s}+N_\mathrm{O,s}$ for WO stars \citep[e.g.][]{Georgy2012}, where $N_\mathrm{He,s}$, $N_\mathrm{C,s}$ and $N_\mathrm{O,s}$ are the surface composition number fractions of helium, carbon and oxygen.

% ==========================================================
\section{Evolutionary endpoints}
\label{sec:results}
% ==========================================================

% ==========================================================
\subsection{Evolutionary channels and stellar types}
\label{subsec:evol_and_stellar}
% ==========================================================

%TTTTTTTTTTTTTTTTTTTTTTTTTTTTTTTTTTTTTTTTTTTTTTTTTTTTTTTTTTTTTT
\begin{table}
\centering
\caption{Outcomes for stellar evolution calculations.}
\begin{threeparttable}
\begin{tabular}{ccccc}
\hline
Channel & SN/BH & WD & CEE & Other \\
\hline
Single & $214$ & $0$ & - & $62$ \\
Primary & $2096$\tnote{a} & $335$\tnote{b} & $750$\tnote{c} & $221$ \\
\hline
Merged & $522$ & $0$ & - & $225$ \\
\hline
Secondary & $770$ & $0$ & - & $277$ \\
Secondary + WD & $40$ & $66$ & $158$ & $71$\tnote{d} \\
Secondary + NS ($0$) & $1$ & $1$ & $5$ & $1$ \\
Secondary + NS ($1$) & $15$ & $93$ & $884$ & $33$ \\
Secondary + NS ($2$) & $196$ & $51$ & $651$ & $127$ \\
Secondary + NS ($3$) & $289$ & $40$ & $536$ & $160$ \\
Secondary + BH ($0$) & $1001$ & $1$ & $8$ & $31$ \\
Secondary + BH ($1$) & $3$ & $0$ & $2$ & $1$ \\
Secondary + BH ($2$) & $4$ & $0$ & $2$ & $0$ \\
Secondary + BH ($3$) & $4$ & $0$ & $2$ & $0$ \\
\hline
\hline
\end{tabular}
\footnotesize
\begin{tablenotes}
\textit{Notes.} The first two rows describe the results of simulations starting from the ZAMS, the third row describes simulations of a coalescence product, and the rest describe simulations of the secondary star that start after the primary reaches the end of its evolution.
The column headings describe the following. Channel: evolutionary channel, as described in the text, with the outcomes in the row labeled `Primary' being an intermediate stage in the system's evolution (as it continues to one of the rows below it) while the rest of the rows describe the final outcome of the stellar system.
SN/BH: models which have reached the end of core carbon burning with a carbon-oxygen core mass above $1.38\,\mathrm{M}_\odot$. WD: models which reached a surface gravity of $g > 10^7\,\mathrm{cm}\,\mathrm{s}^{-2}$. CEE: models in which coalescence occurs during common envelope evolution. Other: simulations which stopped because of other reasons, mostly numerical.\\
\item[a] Including $8$ cases which were assumed to produce ECSNe, whose follow-up simulation outcomes are detailed in the row labeled `Secondary + NS ($0$)'.\\
\item[b] Including $110$ simulations which did not reach the surface gravity stopping condition but were extrapolated to construct follow-up simulations with a WD companion.\\
\item[c] The number of merged star simulations is smaller than merging binaries because in $3$ cases the coalescence product construction failed.\\
\item[d] Including $6$ cases in which the WD accretes mass and exceeds the Chandrasekhar limit, and is therefore assumed to explode in a Type~Ia SN. In all these simulations the companion mass is too low for any future CCSN to occur.
\end{tablenotes}
\end{threeparttable}
\label{tab:outcomes}
\end{table}
%TTTTTTTTTTTTTTTTTTTTTTTTTTTTTTTTTTTTTTTTTTTTTTTTTTTTTTTTTTTTTT
The outcomes of all the simulations are summarised in Table\,\ref{tab:outcomes}. We break down the outcomes into the following evolutionary channels:
\begin{itemize}
    \item `Single' -- evolution of a single star;
    \item `Primary' -- binary evolution from the ZAMS of both stars until the evolutionary end of the primary;
    \item `Merged' -- evolution of a coalescence product of the two stars;
    \item `Secondary' -- evolution of the secondary star after disruption of the binary by a SN;
    \item `Secondary + WD' -- evolution of the secondary with a WD companion formed from the primary;
    \item `Secondary + NS' -- evolution of the secondary with a NS companion formed from the primary in an electron-capture SN (ECSN) with no natal kick ($0$) or by an iron core-collapse SN with a natal kick resulting in a small ($1$), medium ($2$) or large ($3$) post-SN separation;
    \item `Secondary + BH' -- evolution of the secondary with a BH companion formed from the primary by direct collapse or a PPISN ($0$) or by fallback with a natal kick for the BH resulting in a small ($1$), medium ($2$) or large ($3$) post-SN separation.
\end{itemize}

Most of the simulations reached the end of core carbon burning and are therefore expected to be progenitors of CCSNe or BHs. Several of our simulations experienced numerical difficulties, in particular for initial masses in the range between $8\,\mathrm{M}_\odot$ and $10\,\mathrm{M}_\odot$, in which cores are partially degenerate and off-centre neon ignition occurs, which is notoriously challenging to model \citep[e.g.][]{Woosley2015}. These are usually low-mass and might be the progenitors of WDs or ECSNe, and are thought not to contribute to CCSN events \citep[e.g.][]{Nomoto1987}. In $8$ simulations which stopped because of numerical reasons and the final CO core mass was $1.35\,\mathrm{M}_\odot < M_\mathrm{CO} < 1.38\,\mathrm{M}_\odot$ (between WDs and CCSNe, as we show in Appendix\,\ref{sec:appendixb}) we assume an ECSN takes place, with a zero-velocity kick for the newborn NS. The outcome of the follow-up simulations after ECSNe is summarised in the row labeled `Secondary + NS ($0$)' in Table\,\ref{tab:outcomes}. The three rows labeled `Secondary + NS ($1/2/3$)' show the outcomes of simulations for representative small/medium/large post-SN separations (corresponding to the first/second/third tercile in the cumulative distribution function of $a (e^2 -1 )$, respectively) for the binaries including the surviving companion and a NS\footnote{An example of the implementation of the different kick outcomes is presented in Appendix\,\ref{sec:appendixNSexamples}.}. Similarly, the rows labeled `Secondary + BH ($1/2/3$)' show the outcomes of simulations where a BH formed by fallback onto a NS which had already attained a kick velocity, and the row labeled `Secondary + BH ($0$)' shows the outcome for the rest of the simulations involving a binary with a BH. The row labeled just `Secondary' includes all the cases where the NS kick disrupts the binary after the first SN, as well as cases where the binary was disrupted because of PI mass loss. The simulations from the `Primary' channel ending in coalescence during CEE are continued in the `Merged' channel. We do not consider the further evolution for cases where a star merges with a WD, NS or BH companion during CEE, though these might also give rise to energetic transients. Simulations in the `Secondary + X' channels ending with WD formation result in WD + WD, NS + WD and (in one case) BH + WD binaries, where orbital decay by gravitational waves might also lead to energetic transients, and these will be described further in a separate work. In the single-star simulations, the lowest initial mass yielding a CCSN is $8\,\mathrm{M}_\odot$ for $Z=0.00224$, $8.5\,\mathrm{M}_\odot$ for $Z=0.0056$ and $8.8\,\mathrm{M}_\odot$ for $Z=0.014$. This trend of the CCSN mass threshold increasing with metallicity agrees with previous studies \citep[e.g.][]{Eldridge2004,Ibeling2013}.

%FFFFFFFFFFFFFFFFFFFFFFFFFFFFFFFFFFFFFFFFFFFFFFFFFFFFFFFFFFFFFF
\begin{figure*}
   \centering
   \includegraphics[width=1\textwidth]{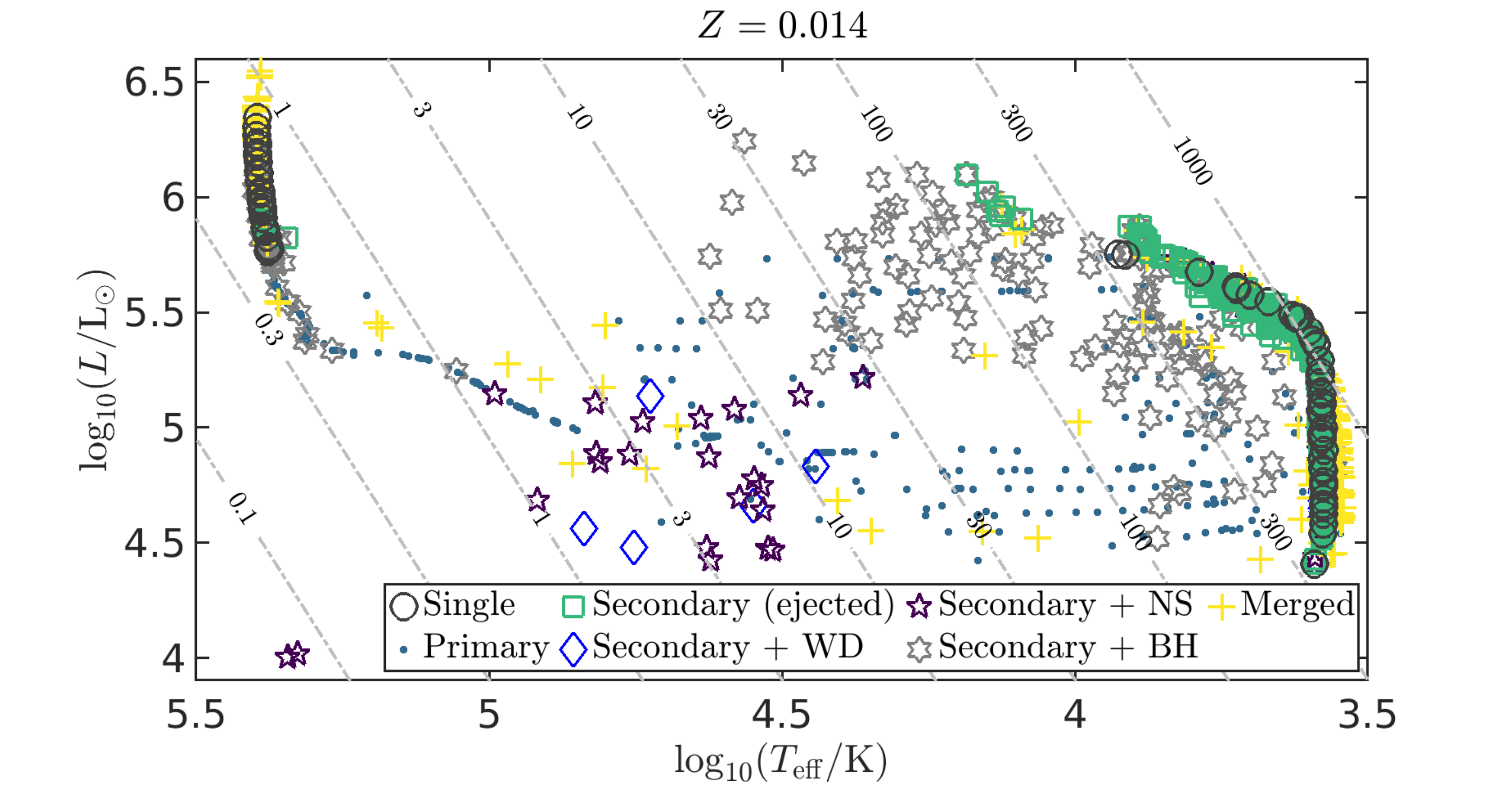}
    \caption{Temperature and luminosity of evolutionary endpoints with metallicity $Z=0.014$ marked by evolutionary channel. Lines of constant radius according to the temperature-luminosity relation are plotted, with the numbers indicating the radius in units of $\mathrm{R}_\odot$.}
    \label{fig:HRDchan}
\end{figure*}
%FFFFFFFFFFFFFFFFFFFFFFFFFFFFFFFFFFFFFFFFFFFFFFFFFFFFFFFFFFFFFF
The effective temperature and luminosity of the models at the evolutionary endpoints of all simulations with a metallicity of $Z=0.014$ that reached the end of core carbon burning are presented in Fig.\,\ref{fig:HRDchan}, marked by evolutionary channel (similar plots for $Z=0.0056$ and $Z=0.00224$ are presented in Appendix\,\ref{sec:endpointsZ}). All endpoints marked as `Secondary' (except for `Secondary + WD') have a corresponding `Primary' point marking the evolutionary end of the faster-evolving and initially more massive primary star in the binary system. There is a clustering of endpoints in low  effective temperatures of $\log_{10}(T_\mathrm{eff}/ \mathrm{K}) \approx 3.6$, corresponding to cool supergiants (CSGs), and a sequence of WR stars in the hot temperatures of $\log_{10}(T_\mathrm{eff}/ \mathrm{K}) \approx 5.4$. We note that the  effective temperature of a WR star in \textsc{mesa} simulations (in which the outer boundary has zero velocity) does not generally correspond to reported base effective temperatures inferred from observations, because the latter depend heavily on the treatment of the optically-thick, expanding atmospheres of WR stars, which is uncertain \citep[e.g.][]{Groh2014,Lefever2023}. Between the sequences of supergiants and WR stars are endpoints with various degrees of envelope stripping, from partially-stripped stars \citep[e.g.][]{Ramachandran2024} with some remaining hydrogen and radii between tens and hundreds of $\mathrm{R}_\odot$ down to a small number of `ultra-stripped' progenitors (\citealt{Tauris2013}; \citealt*{Tauris2015}) with $R\la 0.1\,\mathrm{R}_\odot$ formed via interaction with a NS companion. The detailed evolution of an `ultra-stripped' progenitor is presented in Appendix\,\ref{sec:appendixUSSN}.

%FFFFFFFFFFFFFFFFFFFFFFFFFFFFFFFFFFFFFFFFFFFFFFFFFFFFFFFFFFFFFF
\begin{figure*}
   \centering
\includegraphics[width=1\textwidth]{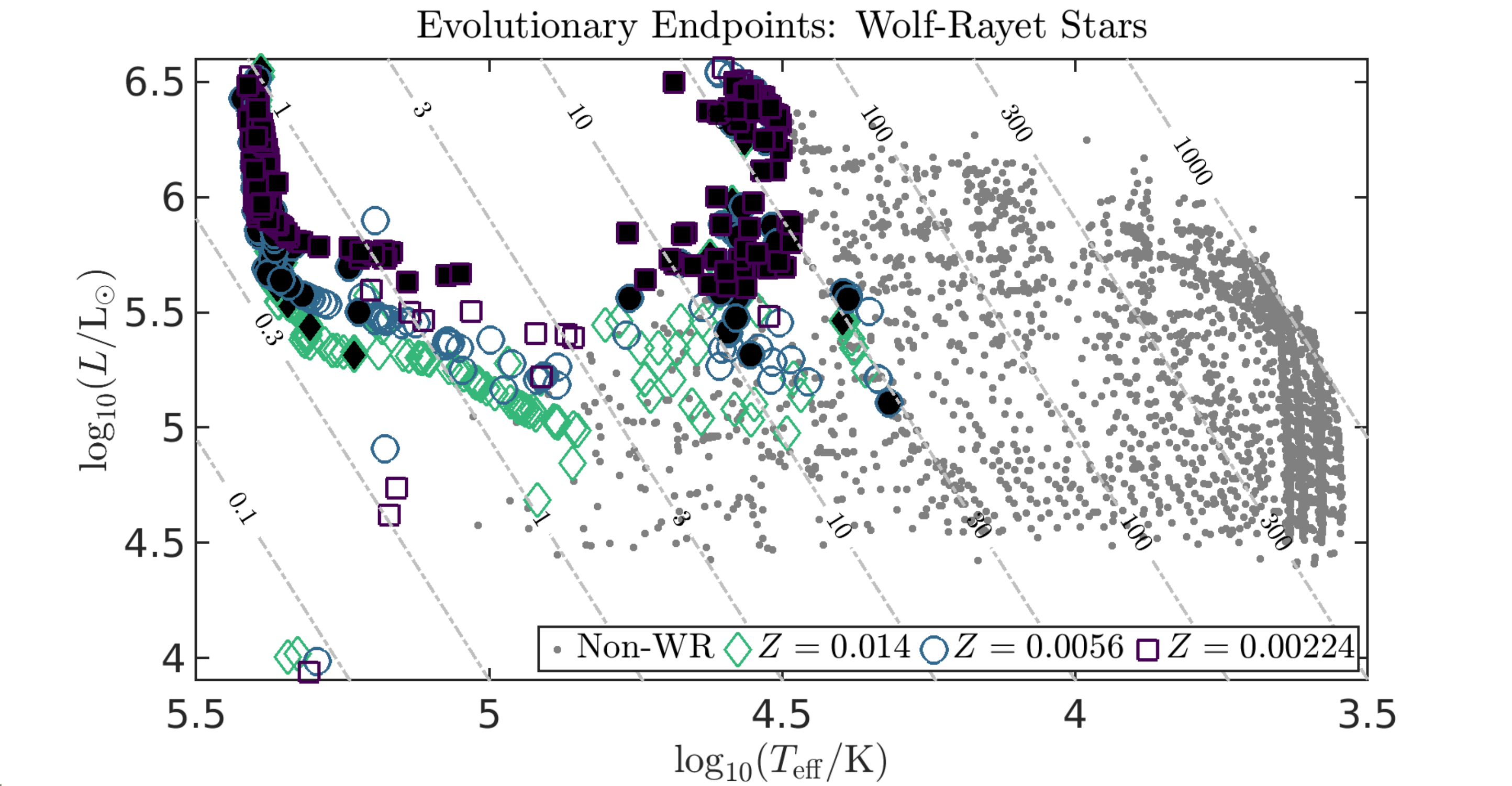}
    \caption{Temperature and luminosity of evolutionary endpoints, with WR stars coloured according to their metallicity. WR endpoints predicted to collapse to BHs are marked by filled symbols.}
    \label{fig:HRDtype}
\end{figure*}
%FFFFFFFFFFFFFFFFFFFFFFFFFFFFFFFFFFFFFFFFFFFFFFFFFFFFFFFFFFFFFF
%FFFFFFFFFFFFFFFFFFFFFFFFFFFFFFFFFFFFFFFFFFFFFFFFFFFFFFFFFFFFFF
\begin{figure*}
   \centering
\includegraphics[width=1\textwidth]{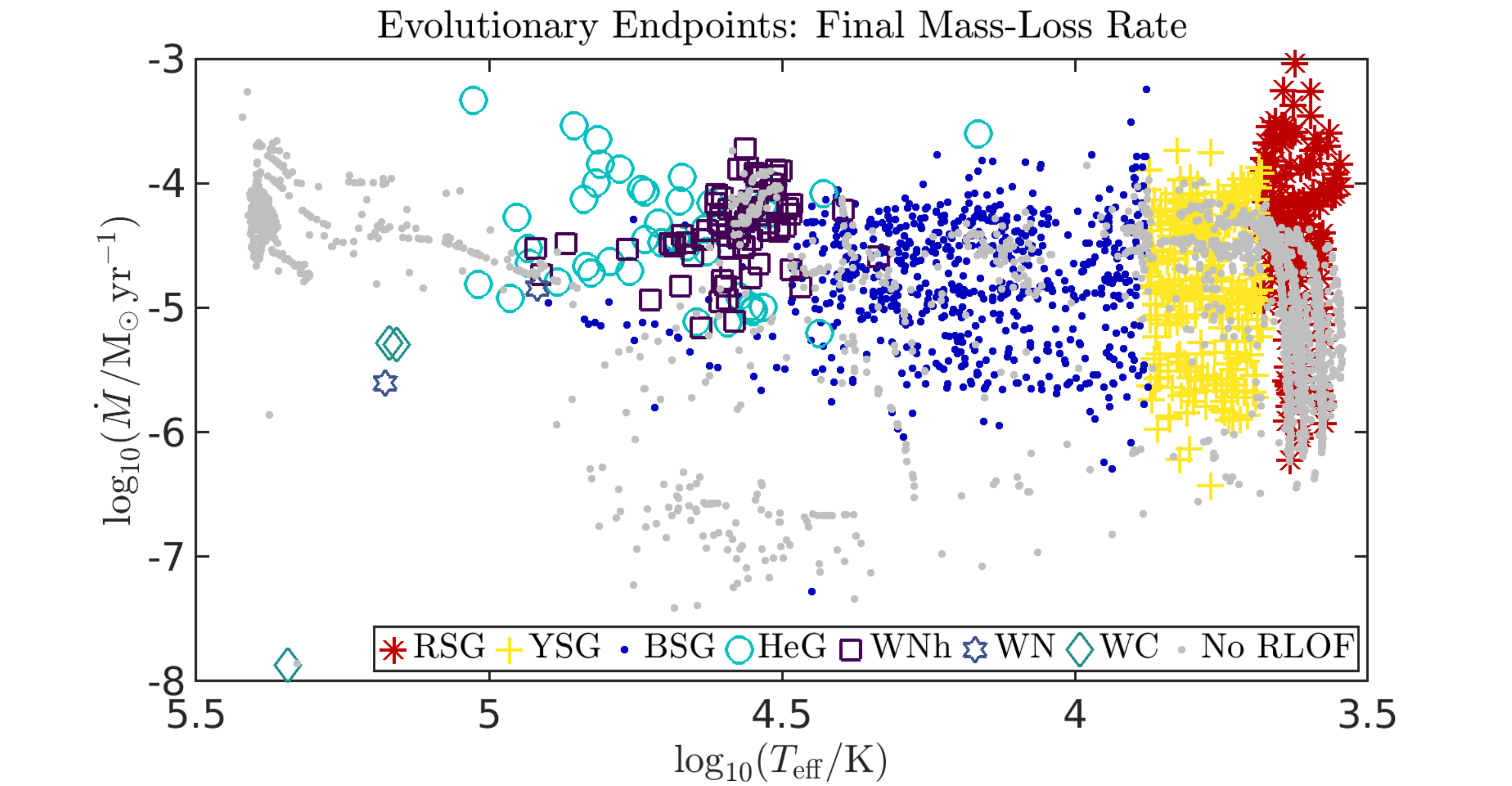}
    \caption{The final mass-loss rate as function of effective surface temperature for all evolutionary endpoints, marked by stellar type for cases where the progenitor fills its Roche lobe at the point of core collapse, while points marked as `No RLOF' have negligible mass-transfer rates compared to the wind mass loss.}
    \label{fig:Mdot}
\end{figure*}
%FFFFFFFFFFFFFFFFFFFFFFFFFFFFFFFFFFFFFFFFFFFFFFFFFFFFFFFFFFFFFF
In Fig.\,\ref{fig:HRDtype} we show the evolutionary endpoints for all metallicities, highlighting those that are WR stars when they reach core-collapse. There is a clear differentiation between metallicities in the sequence of hot WR endpoints. Despite the aforementioned mismatch between computed and observed WR temperatures, there is a correlation between the two and most of the hot ($T_\mathrm{eff} \ga 70\,\mathrm{kK}$) WR stars can be seen to diverge into three sequences according to metallicity, as different opacities at different metallicities influence the efficiency of stripping (\citealt*{Gotberg2017}; \citealt{Laplace2020}). These sequences appear to have a luminosity threshold of $\log_{10} (L / \mathrm{L}_\odot) \approx 5$, $\approx 5.2$ and $\approx 5.5$ for $Z = 0.014$, $Z = 0.0056$ and $Z = 0.00224$, respectively, similar to the WR threshold defined by \cite{Shenar2020}. Most endpoints with $\log_{10} (L / \mathrm{L}_\odot) \ga 5.5$ are predicted to collapse directly to a BH, while the lower luminosity WR stars are mostly expected to explode. Consequently, we find WR stars as primary progenitors of BHs at all metallicites, but their prevalence as CCSN progenitors decreases as the metallicity decreases. A few very hot endpoints, i.e. $T_\mathrm{eff} > 100\,\mathrm{kK}$, are classified as WR stars despite their relatively low wind mass-loss rates (Fig.\,\ref{fig:Mdot}), because they are very compact and would therefore have small transformed radii according to Eq.\,\ref{eq:Rt}. The remaining endpoints classified as WR stars have cooler effective temperatures, clustering around $T_\mathrm{eff}\approx 40\,\mathrm{kK}$, and are all expected to be of sub-type WNh (Section\,\ref{subsec:calcWR}). 

Fig.\,\ref{fig:Mdot} shows the mass-loss rate of all evolutionary endpoints, which is of interest as the mass-loss rate in late evolutionary stages (shortly before CC) is decisive in forming CSM. Stars with a radius larger than $0.9$ of their Roche-lobe radius (corresponding to both those close to filling, and those over-filling, their Roche lobes) are marked by their stellar type, while for the remainder (marked as `No RLOF') the total mass loss is the wind mass loss.  Non-WR star types are classified as follows:
\begin{itemize}
    \item RSG: $T_\mathrm{eff} \le 4.8\,\mathrm{kK}$, $X_\mathrm{H,s}\ge 0.05$;
    \item YSG: $4.8\,\mathrm{kK} < T_\mathrm{eff} < 7.5\,\mathrm{kK}$, $X_\mathrm{H,s}\ge 0.05$;
    \item blue supergiant (BSG): $T_\mathrm{eff}\ge 7.5\,\mathrm{kK}$, $X_\mathrm{H,s}\ge 0.05$;
    \item helium giant (HeG): $X_\mathrm{H,s} < 0.05$.
\end{itemize}
The non-WR endpoints with $X_\mathrm{H,s} < 0.05$ span a wide range of radii, from $\approx 0.5\,\mathrm{R}_\odot$ (similar to the helium stars recently discovered in the SMC and LMC; \citealt{Drout2023,Gotberg2023}) up to $\ga 200\,\mathrm{R}_\odot$ (higher mass counterparts to $\upsilon$~Sagittarii; \citealt{Laplace2020,Gilkis2023}), but we collectively refer to them as helium giants to reflect their post core helium burning evolutionary stage. Most of the hotter WR endpoints are not Roche-lobe filling, as expected for relatively compact stars. A few exceptions are the very hot and highly-stripped endpoints, which went through a CEE phase and end their evolution in a very tight binary where RLOF is possible for compact stars. Many endpoints classified as WNh are engaging in mass transfer at the end of their evolution, as the remaining hydrogen envelopes can be rather puffed up. Among the CSGs, there is a wide range of final mass-loss rates, with some experiencing RLOF and some not. It is worth noting that there are RSGs with very high ($\ga 10^{-4}\,\mathrm{M}_\odot\,\mathrm{yr}^{-1}$) RLOF mass-transfer rates which are stable and do not lead to CEE, and such systems might have a significant amount of CSM when they explode.

% ==========================================================
\subsection{Stellar remnants and SN types}
\label{subsec:remnants_and_SNe}
% ==========================================================

%FFFFFFFFFFFFFFFFFFFFFFFFFFFFFFFFFFFFFFFFFFFFFFFFFFFFFFFFFFFFFF
\begin{figure*}
   \centering
\includegraphics[width=1\textwidth]{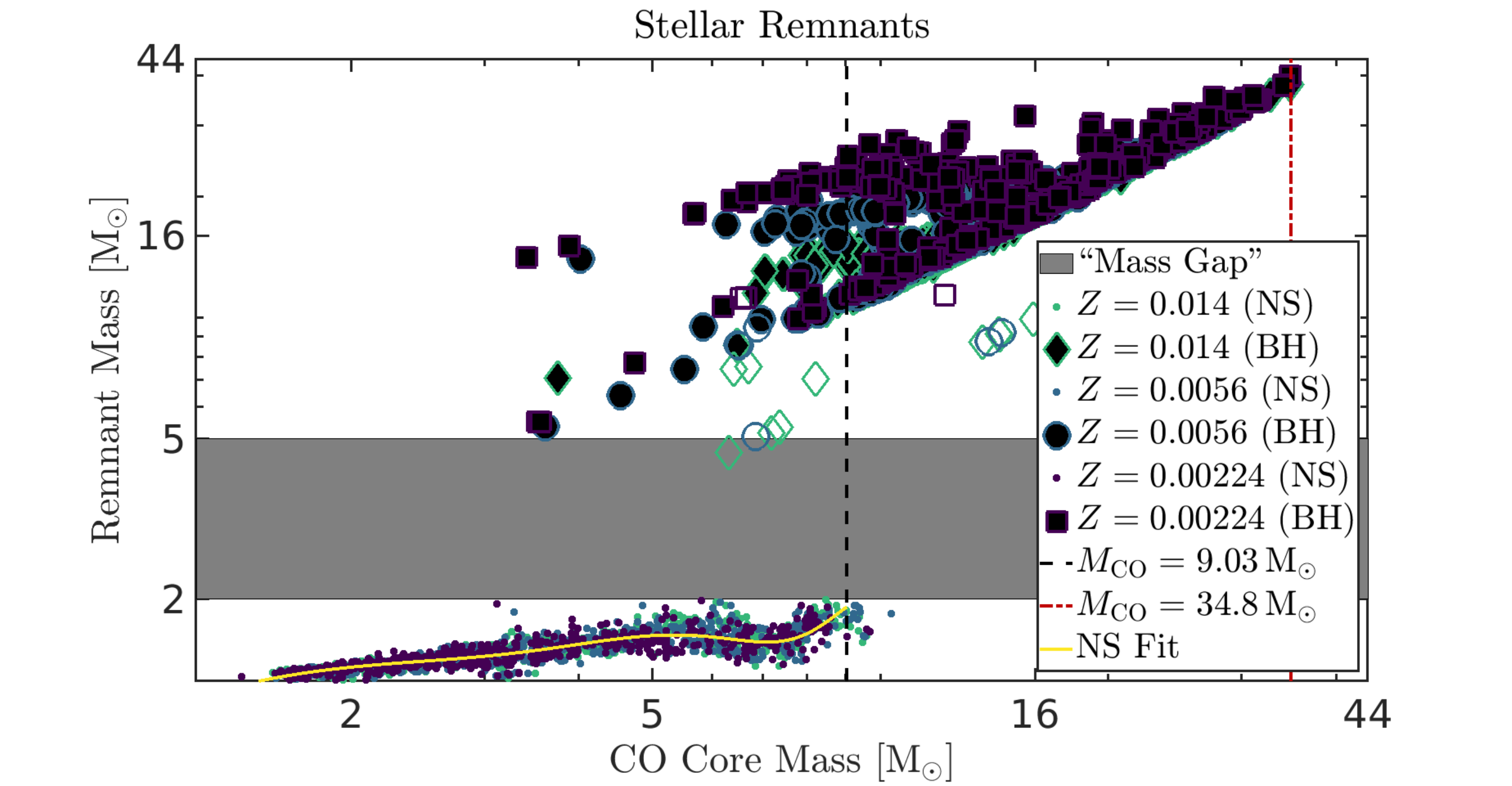}
    \caption{The stellar remnant mass as function of the carbon-oxygen core mass at the end of core carbon burning computed by applying the prescription of \citet{Muller2016} to models that reached iron core collapse. BHs formed by fallback are marked by open symbols. The empirical gap between masses of NSs and BHs in the Galaxy \citep{Ozel2010} is marked by a gray-shaded region. A threshold carbon-oxygen core mass of $M_\mathrm{CO}=9.03\,\mathrm{M}_\odot$ for BH formation (black dashed line) reproduces $96.5\%$ of the results computed using the \citet{Muller2016} prescription. The threshold for PPI at $M_\mathrm{CO}=34.8\,\mathrm{M}_\odot$ (red broken line) marks the transition to the computation of remnant masses following the fit derived by \citet{Renzo2022}. A $6\mathrm{th}$-order polynomial fit of the remnant NS mass as function of the CO core mass (Eq.\,\ref{eq:MNS}) is plotted in yellow.}
    \label{fig:remnants}
\end{figure*}
%FFFFFFFFFFFFFFFFFFFFFFFFFFFFFFFFFFFFFFFFFFFFFFFFFFFFFFFFFFFFFF
Fig.\,\ref{fig:remnants} shows the remnant mass computed with the semi-analytical prescription of \cite{Muller2016} applied to the $3925$ final profiles of simulations that reach iron core collapse. Because a significant fraction\footnote{In addition to the $3925$ simulations that reach iron core collapse, $1239$ simulations reach the end of core carbon burning (but not iron core collapse) with CO cores massive enough to be BH or SN progenitors.} of simulations do not reach iron core collapse, we derive a simple prescription to determine the outcome and remnant mass for those cases. We find that if we assume models with carbon-oxygen cores more massive than $9.03\,\mathrm{M}_\odot$ to collapse to BHs (and the rest to explode in SNe producing NSs) then we reproduce $96.5\%$ of the outcomes computed with the \cite{Muller2016} prescription\footnote{The value of $9.03\,\mathrm{M}_\odot$ was computed by scanning the entire range of $M_\mathrm{CO}$ and selecting the value which maximises the prediction success.}. Our explodability criterion is then a combination of using \cite{Muller2016} when possible and the threshold of $M_\mathrm{CO}=9.03\,\mathrm{M}_\odot$ otherwise, and we apply this criterion to all simulations that reach the end of core carbon burning. For simulations that reach the end of core carbon burning with $M_\mathrm{CO} < 9.03\,\mathrm{M}_\odot$ but do not reach iron core collapse, we assume a successful explosion and a NS mass that follows a $6\mathrm{th}$-order polynomial fit to the successful explosion simulations,
\begin{multline}
M_\mathrm{NS} = -0.0002374\,M_\mathrm{CO}^6 + 0.007645\, M_\mathrm{CO}^5  - 0.09583\, M_\mathrm{CO}^4 \\  + 0.5954\, M_\mathrm{CO}^3 - 1.938\, M_\mathrm{CO}^2 + 3.24\, M_\mathrm{CO} - 0.8327\, ,
    \label{eq:MNS}    
\end{multline}   
with $M_\mathrm{CO}$ in units of $\mathrm{M}_\odot$ for $1.38\,\mathrm{M}_\odot < M_\mathrm{CO} < 9.03\,\mathrm{M}_\odot$.

%FFFFFFFFFFFFFFFFFFFFFFFFFFFFFFFFFFFFFFFFFFFFFFFFFFFFFFFFFFFFFF
\begin{figure*}
   \centering
\includegraphics[width=1\textwidth]{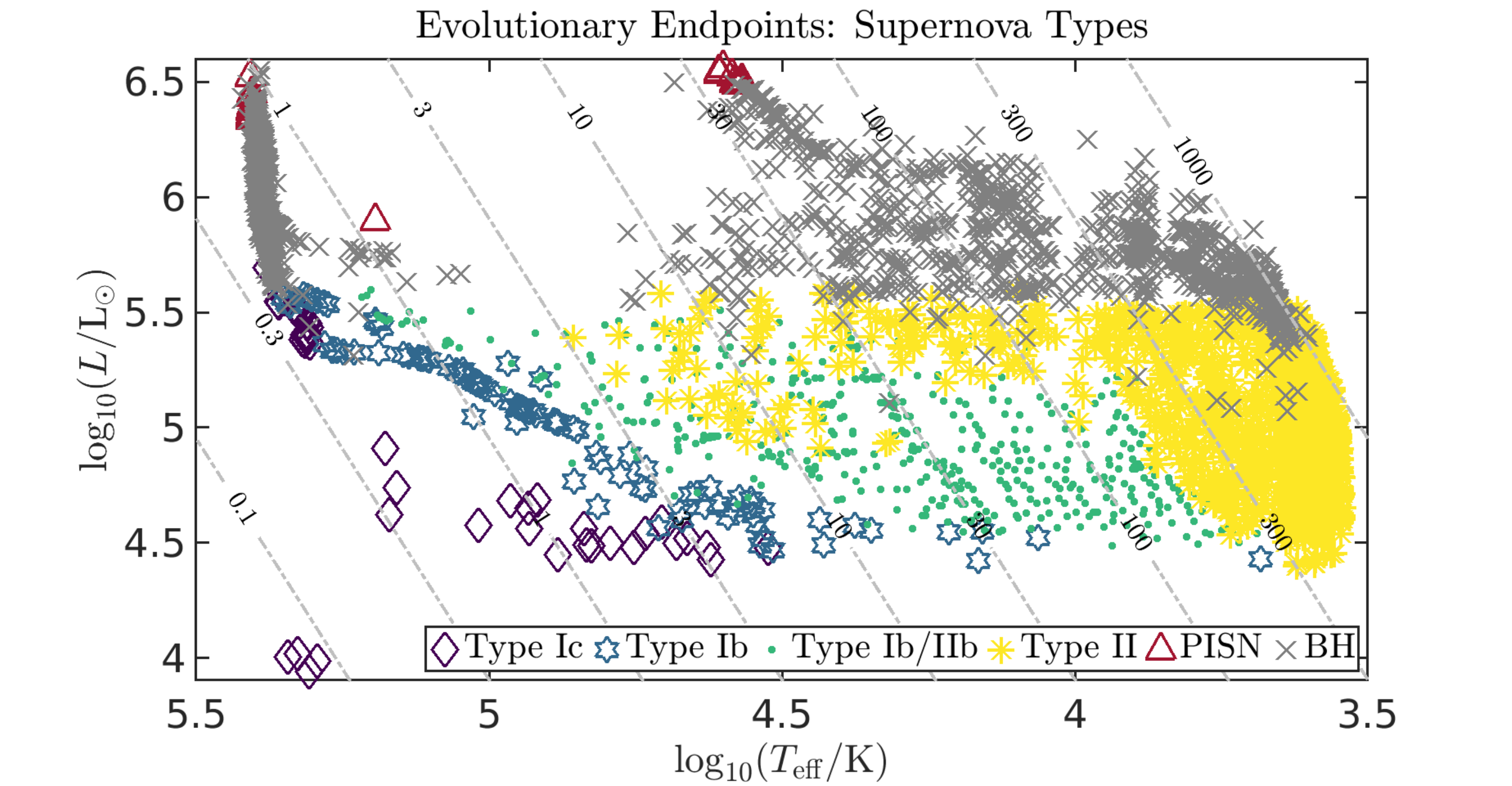}
    \caption{Evolutionary endpoints marked by SN type or BH formation. Models expected to explode are classified into different CCSN sub-types according to the stellar composition, except for models with extremely high CO core masses which are expected to explode in PISNe. Models marked as producing BHs include direct collapse, fallback and PPISNe.}
    \label{fig:HRDsupernovae}
\end{figure*}
%FFFFFFFFFFFFFFFFFFFFFFFFFFFFFFFFFFFFFFFFFFFFFFFFFFFFFFFFFFFFFF
In Fig.\,\ref{fig:HRDsupernovae} we present the pre-SN characteristics of all the models which we assume explode in SNe, assigning SN types according to the total hydrogen mass and the total helium mass at the end of the evolution, marking those we assume collapse to BHs. The hydrogen mass needed to give rise to Type II spectral features is not known, with estimates ranging between $M_\mathrm{H}\approx 0.001\,\mathrm{M}_\odot$ \citep{Dessartetal2011} and $M_\mathrm{H}\approx 0.033\,\mathrm{M}_\odot$ \citep{Hachinger2012}. Pre-SN images of Type~Ib progenitors favour the higher threshold \citep{Gilkis2022}, though only two Type~Ib progenitors are known, and the later case of SN~2019yvr still awaits confirmation with follow-up observations. Therefore, for $0.001\,\mathrm{M}_\odot \le M_\mathrm{H} \le 0.033\, \mathrm{M}_\odot$ we assign a tentative classification of Type~Ib/IIb. The other classifications are Type~II for $M_\mathrm{H} > 0.033\,\mathrm{M}_\odot$, Type~Ic for $M_\mathrm{He} < 0.5\,\mathrm{M}_\odot$ \citep[e.g.][]{Yoon2010}, and Type~Ib otherwise.

Evolutionary endpoints with very high CO core masses are expected to explode as PISNe and leave no remnant of any kind behind, and we mark these separately from the other SNe but do not assign a specific sub-type. These PISNe are expected to be extremely rare (Section\,\ref{sec:IMFetal}). For most evolutionary endpoints, there is an apparent threshold luminosity above which BH formation is expected. This connects to the CO core mass threshold (Fig.\,\ref{fig:remnants}) through the relation between the core mass and luminosity, and taking the fit for this relation from \cite{Temaj2024} we find for $M_\mathrm{CO}=9.03\,\mathrm{M}_\odot$ a luminosity of $\log_{10} (L / \mathrm{L}_\odot) = 5.58$, coinciding with the luminosity threshold in Fig.\,\ref{fig:HRDsupernovae}.

%FFFFFFFFFFFFFFFFFFFFFFFFFFFFFFFFFFFFFFFFFFFFFFFFFFFFFFFFFFFFFF
\begin{figure*}
   \centering
\includegraphics[width=1\textwidth]{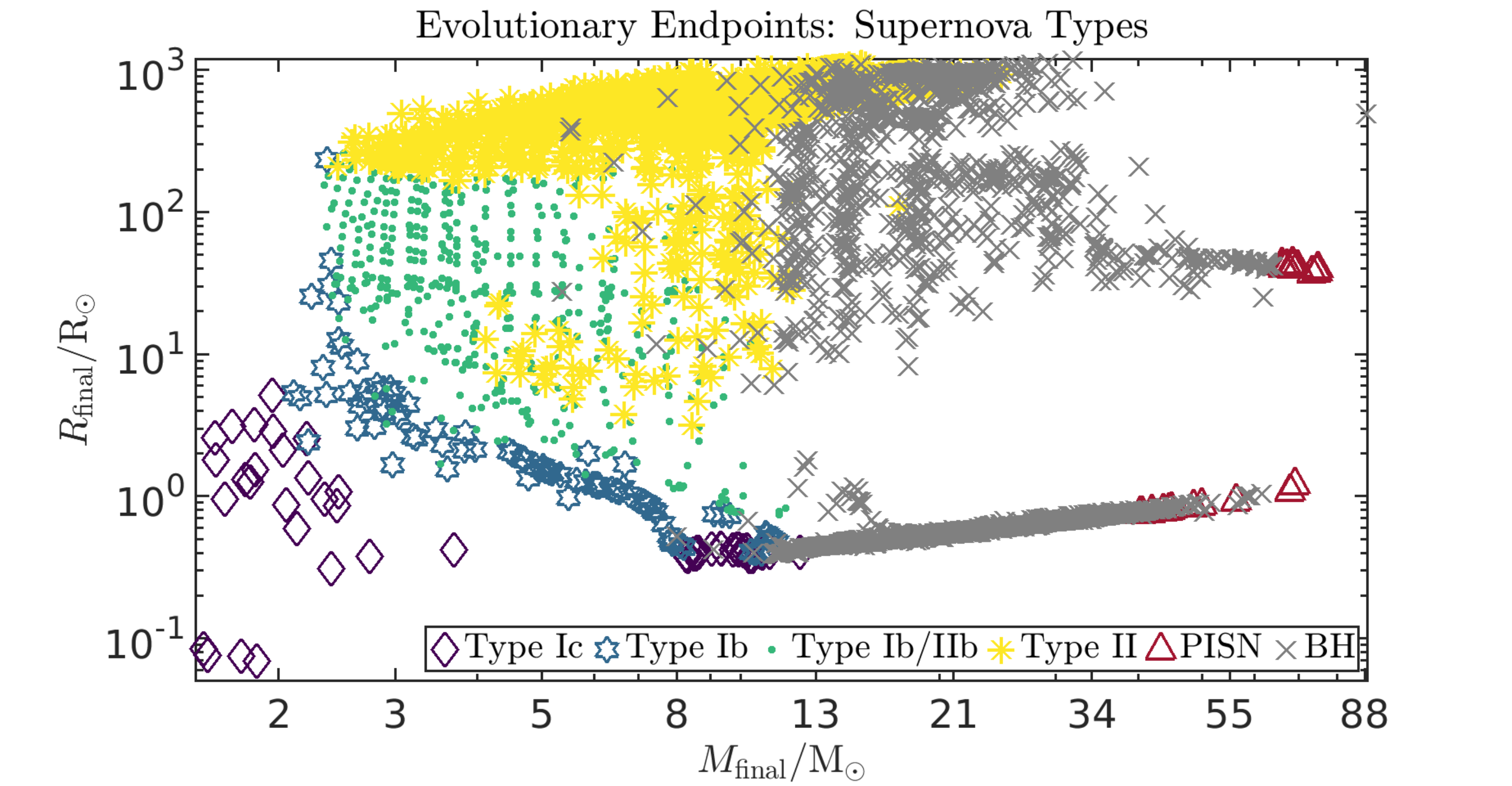}
    \caption{Mass and radius of evolutionary endpoints, with SN types marked in the same manner as in Fig.\,\ref{fig:HRDsupernovae}.}
    \label{fig:MRsupernovae}
\end{figure*}
%FFFFFFFFFFFFFFFFFFFFFFFFFFFFFFFFFFFFFFFFFFFFFFFFFFFFFFFFFFFFFF
In Fig.\,\ref{fig:MRsupernovae} we show the mass and radius of the evolutionary endpoints, marked by our SN type classification. The mass and radius are key properties in shaping the SN light curves \citep[e.g.][]{Arnett1980,Nakar2010}. Unsurprisingly, Type~Ib and Type~Ic are mostly assigned to compact progenitors, though there are a few low-mass helium giants. If, however, the maximum hydrogen that can be `hidden' is closer to the high values derived by \cite{Hachinger2012} then we can expect many Type~Ib SNe from progenitors with extended envelopes. Type~II SNe are expected for a range of radii, from $\approx 3\,\mathrm{R}_\odot$ up to  $\ga 1000\,\mathrm{R}_\odot$, possibly accounting for various Type~II sub-types, e.g. IIb, II-L and II-P. Progenitors with high luminosity-to-mass ratios are known to pulsate at the end of their lives \citep{Heger1997,YoonCantiello2010}, so these radii are just indicative. We note that some of the endpoints with the lowest hydrogen masses above the threshold for our Type~II designation are expected to explode as Type~IIb, and further sub-classifications would require radiation hydrodynamics simulations, but it is expected that there will be a sequence from Type~IIb, to Type~II-L and then Type~II-P according to the hydrogen mass \citep[e.g.][]{Dessart2024}.

% ==========================================================
\section{Probability-weighted evolutionary channels and outcomes}
\label{sec:IMFetal}
% ==========================================================

In the previous section we presented the evolutionary endpoints in our simulations, with no mention of the likelihood of the different evolutionary channels or endpoints. To relate the simulations to expected occurrence rates, we compute a probability, or weight, for each evolutionary sequence and endpoint. We assume the initial mass function (IMF) follows $\mathrm{d} N / \mathrm{d} M \propto M^{-2.3}$, which is appropriate for the range of masses we consider in this study \citep{Kroupa2001}. We assume a flat distribution in the mass ratio \citep{Shenar2022}, a distribution of initial periods according to $\mathrm{d} N / \mathrm{d} (\log P) \propto (\log P)^{-0.55}$ and a binary fraction of $70\%$ \citep{Sana2012}. These assumed probability distributions provide the weight of each combination of initial conditions and the corresponding single or binary stellar evolution simulation. For the follow-up simulations after the evolutionary end of the first star, the same probability is assigned as for the first simulation for cases involving WD formation, NS formation by an ECSN, disruption by a PISN or BH formation by either a PPISN or direct collapse. Simulations that follow a secondary star ejected after the binary is disrupted by a SN are assigned the probability of the first evolutionary stage multiplied by the probability of disruption computed by the random kick realisations described in Section\,\ref{subsec:channels}. For simulations of systems that survive the first SN and remain bound, the probability is that of the first evolutionary stage multiplied by the probability of the system remaining bound divided by three, according to the choice of three equally-spaced post-SN separations in the computed cumulative distribution function. We compute the probabilities separately for each metallicity.

%FFFFFFFFFFFFFFFFFFFFFFFFFFFFFFFFFFFFFFFFFFFFFFFFFFFFFFFFFFFFFF
\begin{figure*}
\centering
\includegraphics[width=\textwidth]{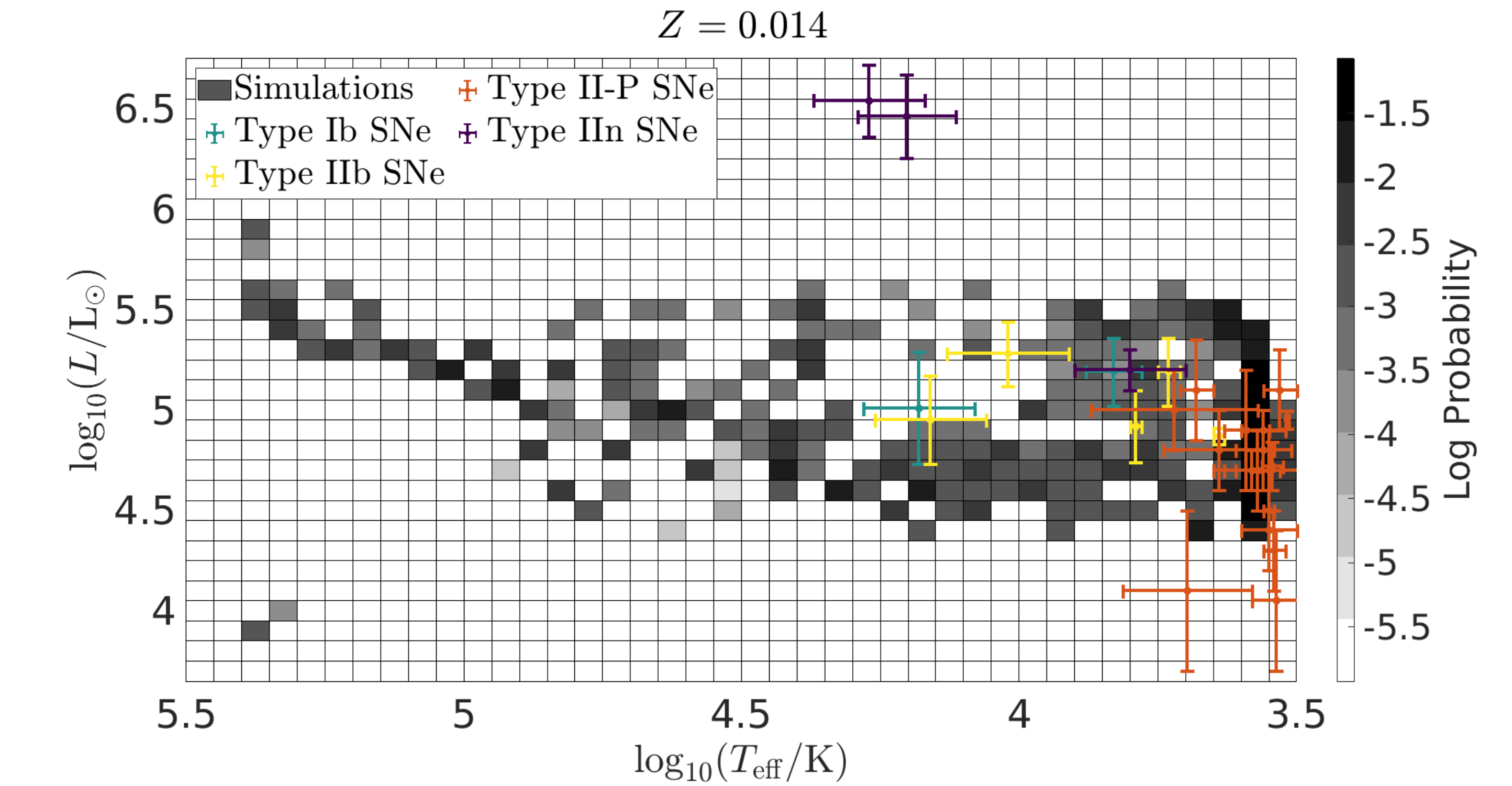} \\
\caption{Weighted probability map of CCSN progenitors for $Z=0.014$. Known progenitors of Type~Ib and Type~IIb SNe (from the analysis of \citealt{Gilkis2022}) are over-plotted. Progenitors of Type~II-P SNe are from \citet{Smartt2015} with the addition of SN~2017eaw \citep{Kilpatrick2018}, SN~2018aoq \citep{ONeill2019,VanDyk2023}, SN~2020jfo \citep{Kilpatrick2023} and SN~2024ggi \citep{Xiang2024}. The progenitors of Type~IIn are those of SN~2010jl \citep*{Niu2024}, SN~2015bh \citep{Boian2018,Jencson2022} and SN~2016jbu \citep{Brennan2022a,Brennan2022b}. Similar plots for $Z=0.0056$ and $Z=0.00224$ are presented in Appendix\,\ref{sec:appendixzmaps}.}
\label{fig:progmap}
\end{figure*}
%FFFFFFFFFFFFFFFFFFFFFFFFFFFFFFFFFFFFFFFFFFFFFFFFFFFFFFFFFFFFFF
In Fig.\,\ref{fig:progmap} we show the probability of endpoints in bins of luminosity and effective temperature, for $Z=0.014$ (Similar plots for $Z=0.0056$ and $Z=0.00224$ are provided in Appendix\,\ref{sec:appendixzmaps}). The darkest squares are the regions of highest probability, like the sequences of CSGs and hot WR stars, with lighter shaded squares representing rarer endpoints, for example the `ultra-stripped' endpoints at the lowest luminosity and highest $T_\mathrm{eff}$. We over-plot known CCSN progenitors observed in nearby galaxies with a metallicity similar to the Milky Way. We note that we did not include SN~2005gl in Fig.\,\ref{fig:progmap} because the effective temperature of its progenitor was not directly constrained. However, the progenitor was detected in pre-explosion Hubble Space Telescope imaging and was identified as a very luminous source ($L \ga 10^6\,\mathrm{L}_\odot$), consistent with LBV stars \citep{GalYam2007,GalYam2009}. Although its exact temperature is uncertain, its high luminosity and apparent disappearance post-explosion make it one of the most compelling direct detections of a Type~IIn SN progenitor.

%TTTTTTTTTTTTTTTTTTTTTTTTTTTTTTTTTTTTTTTTTTTTTTTTTTTTTTTTTTTTTT
\begin{table*}
\centering
\caption{Probability of stellar evolution channels.}
\begin{threeparttable}
\begin{tabular}{cccc}
\hline
Massive-star Channels & $Z=0.00224$ & $Z=0.0056$ & $Z=0.014$ \\
\hline
* $\xrightarrow{\mathrm{CCSN}}$ NS & $23.0716\% $ & $22.7981\% $ & $22.2851\% $ \\
* $\xrightarrow{\mathrm{CC}}$ BH & $6.9862\% $ & $5.6676\% $ & $5.0984\% $ \\
* $\xrightarrow{\mathrm{PPISN}}$ BH & $0.88466\% $ & $0.47083\% $ & $0.3253\% $ \\
* $\xrightarrow{}$ PISN & $0.20445\% $ & $0.20867\% $ & $0\% $ \\
* + * $\xrightarrow{\mathrm{merge}}$ * $\xrightarrow{\mathrm{CCSN}}$ NS & $18.8574\% $ & $24.1326\% $ & $27.8621\% $ \\
* + * $\xrightarrow{\mathrm{merge}}$ * $\xrightarrow{\mathrm{CC}}$ BH & $1.3976\% $ & $2.1017\% $ & $2.8881\% $ \\
* + * $\xrightarrow{\mathrm{merge}}$ * $\xrightarrow{\mathrm{PPISN}}$ BH & $0.24293\% $ & $0.37835\% $ & $0.59481\% $ \\
* + * $\xrightarrow{\mathrm{merge}}$ * $\xrightarrow{}$ PISN & $0.25783\% $ & $0.19727\% $ & $0\% $ \\
* + * $\xrightarrow{}$ WD + * $\xrightarrow{\mathrm{CCSN}}$ WD + NS & $4.5182\% $ & $4.4704\% $ & $4.6587\% $ \\
* + * $\xrightarrow{}$ WD + * $\xrightarrow{\mathrm{AIC}}$ NS + WD & $0.95163\% $ & $0.41507\% $ & $0.34054\% $ \\
* + * $\xrightarrow{\mathrm{CCSN, ~ disrupted}}$ * $\xrightarrow{}$ WD & $7.1345\% $ & $5.643\% $ & $5.3805\% $ \\
* + * $\xrightarrow{\mathrm{CCSN, ~ disrupted}}$ * $\xrightarrow{\mathrm{CCSN}}$ NS & $10.0962\% $ & $9.3773\% $ & $7.4156\% $ \\
* + * $\xrightarrow{\mathrm{CCSN, ~ disrupted}}$ * $\xrightarrow{\mathrm{CC}}$ BH & $2.4569\% $ & $2.4009\% $ & $3.2091\% $ \\
* + * $\xrightarrow{\mathrm{CC,FB, ~ disrupted}}$ * $\xrightarrow{\mathrm{CCSN}}$ NS & $0\% $ & $0\% $ & $0.013617\% $ \\
* + * $\xrightarrow{\mathrm{CC,FB, ~ disrupted}}$ * $\xrightarrow{\mathrm{CC}}$ BH & $0\% $ & $0\% $ & $0.010781\% $ \\
* + * $\xrightarrow{\mathrm{CCSN}}$ NS + * $\xrightarrow{}$ merge/T{\.Z}O & $11.75\% $ & $11.4887\% $ & $10.353\% $ \\
* + * $\xrightarrow{\mathrm{CCSN}}$ NS + * $\xrightarrow{}$ NS + WD & $0.42777\% $ & $0.36893\% $ & $0.56956\% $ \\
* + * $\xrightarrow{\mathrm{CCSN}}$ NS + * $\xrightarrow{\mathrm{CCSN}}$ NS + NS & $0.54753\% $ & $0.6394\% $ & $0.3679\% $ \\
* + * $\xrightarrow{\mathrm{CCSN}}$ NS + * $\xrightarrow{\mathrm{CC}}$ NS + BH & $0.0091839\% $ & $0.012355\% $ & $0.025749\% $ \\
* + * $\xrightarrow{\mathrm{CC}}$ BH + * $\xrightarrow{}$ merge& $0.17964\% $ & $0.41979\% $ & $0.20716\% $ \\
* + * $\xrightarrow{\mathrm{CC}}$ BH + * $\xrightarrow{}$ BH + WD & $0.056843\% $ & $0\% $ & $0\% $ \\
* + * $\xrightarrow{\mathrm{CC}}$ BH + * $\xrightarrow{\mathrm{CCSN}}$ BH + NS & $2.1276\% $ & $2.3477\% $ & $2.6272\% $ \\
* + * $\xrightarrow{\mathrm{CC}}$ BH + * $\xrightarrow{\mathrm{CC}}$ BH + BH & $6.0087\% $ & $5.4472\% $ & $5.629\% $ \\
* + * $\xrightarrow{\mathrm{CC}}$ BH + * $\xrightarrow{\mathrm{PPISN}}$ BH + BH & $0.021803\% $ & $0\% $ & $0\% $ \\
* + * $\xrightarrow{\mathrm{PPISN}}$ BH + * $\xrightarrow{\mathrm{CCSN}}$ BH + NS & $0.22692\% $ & $0.14423\% $ & $0\% $ \\
* + * $\xrightarrow{\mathrm{PPISN}}$ BH + * $\xrightarrow{\mathrm{CC}}$ BH + BH & $1.1924\% $ & $0.56113\% $ & $0.13777\% $ \\
* + * $\xrightarrow{\mathrm{PPISN}}$ BH + * $\xrightarrow{\mathrm{PPISN}}$ BH + BH & $0.31957\% $ & $0.11062\% $ & $0\% $ \\
* + * $\xrightarrow{\mathrm{PPISN}}$ BH + * $\xrightarrow{\mathrm{PISN}}$ BH & $0.058679\% $ & $0.024179\% $ & $0\% $ \\
* + * $\xrightarrow{\mathrm{PISN}}$ * $\xrightarrow{\mathrm{CCSN}}$ NS & $0\% $ & $0.059586\% $ & $0\% $ \\
* + * $\xrightarrow{\mathrm{PISN}}$ * $\xrightarrow{\mathrm{CC}}$ BH & $0\% $ & $0.11432\% $ & $0\% $ \\
* + * $\xrightarrow{\mathrm{PISN}}$ * $\xrightarrow{\mathrm{PPISN}}$ BH & $0.013132\% $ & $0\% $ & $0\% $ \\
\hline
Other Channels & & & \\
\hline
* $\xrightarrow{}$ WD & $44.9614\% $ & $48.5351\% $ & $49.4141\% $ \\
* + * $\xrightarrow{\mathrm{merge}}$ * $\xrightarrow{}$ WD & $33.2885\% $ & $34.2222\% $ & $35.9605\% $ \\
* + * $\xrightarrow{}$ WD + * $\xrightarrow{}$ merge & $27.8686\% $ & $28.9782\% $ & $20.7941\% $ \\
* + * $\xrightarrow{}$ WD + * $\xrightarrow{}$ WD + WD & $7.2608\% $ & $7.7337\% $ & $5.7217\% $ \\
* + * $\xrightarrow{}$ WD + * $\xrightarrow{}$ Type Ia SN & $1.6664\% $ & $0.20009\% $ & $0\% $ \\
\hline
\hline
\end{tabular}
\footnotesize
\begin{tablenotes}
\textit{Notes.} * represents a normal star, i.e. not a WD, NS or BH. Massive-star channels refer to all evolutionary pathways involving the formation of at least one NS or BH, or the occurrence of at least one PISN, and sum up to $100\%$ for each metallicity. The occurrence rate of the other channels is given relative to the massive-star channels, and these are lower bounds because initial masses below $4\,\mathrm{M}_\odot$ can contribute as well.
\end{tablenotes}
\end{threeparttable}
\label{tab:weightedchannels}
\end{table*}
%TTTTTTTTTTTTTTTTTTTTTTTTTTTTTTTTTTTTTTTTTTTTTTTTTTTTTTTTTTTTTT
In Table\,\ref{tab:weightedchannels} we present all the evolutionary pathways with non-zero probability for at least one metallicity in our simulations. The probabilities are computed by summing the weights of all the simulations for each channel and dividing by the sum of the weights of all massive-star channels, defined as any evolutionary channel where either a NS or BH forms, or a PISN occurs. The reasoning for this is that we consider the massive-star channels to be completely covered by our range of initial conditions, as evident by the lowest initial primary mass givin rise to a CCSN being $5\,\mathrm{M}_\odot$ (through merging), and no CCSN occurring for any simulation with an initial primary mass of $4\,\mathrm{M}_\odot$. The massive-star channels include the formation of a NS by accretion-induced collapse (AIC), even though this channel does not include any SN, as this requires the formation of an ONe WD for which the minimal initial primary mass is $7\,\mathrm{M}_\odot$.

The two most common evolutionary channels in the massive-star regime as we define it are the straightforward evolution of a single star to CC and explosion in a CCSN and the merging of two stars leading to a CCSN progenitor, the latter resulting from many intermediate-mass stars favoured by the IMF merging to produce a star massive enough to reach CC \citep[e.g.][]{deMink2014,Zapartas2019}. The third most common channel is a CCSN followed by the evolution of a bound post-SN binary ending in a fatal CEE phase, where the NS merges with the core of the star engulfing it. We do not follow the evolution beyond the point where the core of the engulfing star fills its Roche lobe, which would either result in a SN-like transient \citep{Chevalier2012,Soker2018,Schroder2020} or the formation of a star with a NS core, i.e. a Thorne-{\.Z}ytkow object \citep[T{\.Z}O;][]{ThorneZytkow1975,ThorneZytkow1977,Cannon1993,Tout2014,Farmer2023}.

We also present the relative likelihood of low-mass evolutionary channels, i.e those ending only with WDs. For example, the probability-weighted outcomes from our simulations represent a similar number of systems where a single star or a coalescence product becomes a WD as all the massive-star channels combined. A less likely outcome is that a CO WD gains mass by accretion and eventually exceeds the Chandrasekhar mass and explodes as a Type~Ia SN. The rates of these low-mass channels are lower limits, as these can arise from initial primary masses lower than $4\,\mathrm{M}_\odot$. The channels leading to single WDs should considerably outnumber any massive-star evolutionary channel.

%TTTTTTTTTTTTTTTTTTTTTTTTTTTTTTTTTTTTTTTTTTTTTTTTTTTTTTTTTTTTTT
\begin{table}
\centering
\caption{Probability of CCSNe.}
\begin{threeparttable}
\begin{tabular}{cccc}
\hline
CCSN Types & $Z=0.00224$ & $Z=0.0056$ & $Z=0.014$ \\
\hline
Type Ic & $0.042925\% $ & $0.93162\% $ & $2.8246\% $ \\
Type Ib & $1.2539\% $ & $2.8616\% $ & $11.8205\% $ \\
Type Ib/IIb & $18.8433\% $ & $24.7071\% $ & $18.3254\% $ \\
Type II & $79.8599\% $ & $71.4997\% $ & $67.0296\% $ \\
\hline
Other Transients &  & & \\
\hline
Collapse to BH & $28.6177\% $ & $25.9347\% $ & $27.1265\% $ \\
PPISN & $3.5573\% $ & $1.9043\% $ & $1.1346\% $ \\
PISN & $0.57647\% $ & $0.63906\% $ & $0\% $ \\
CEE coalescence (NS) & $12.6825\% $ & $12.1549\% $ & $11.1037\% $ \\
CEE coalescence (BH) & $0.19389\% $ & $0.44413\% $ & $0.22218\% $ \\
\hline
\hline
\end{tabular}
\footnotesize
\begin{tablenotes}
\textit{Notes.} The probabilities of Type Ib/c and Type II CCNSe sum up to $100\%$ for each metallicity. The occurrence rate of the other transient events is given relative to the CCSNe.
\end{tablenotes}
\end{threeparttable}
\label{tab:weightedCCSNe}
\end{table}
%TTTTTTTTTTTTTTTTTTTTTTTTTTTTTTTTTTTTTTTTTTTTTTTTTTTTTTTTTTTTTT
In Table\,\ref{tab:weightedCCSNe} we present the probability-weighted CCSN types for each metallicity. Type~II SNe are by far the most common for all metallicities. The rate of Type~Ib/c SNe appears to strongly depend on metallicity, with an order of magnitude more Ib/c SNe predicted at $Z=0.014$ compared to $Z=0.00224$. The Type~Ic rate is much lower than that of Type~Ib, as this SN type requires an extreme stripping of the envelope to remove a significant amount of helium. The expected rate of direct collapse to a BH is about one for every four CCSNe.

%TTTTTTTTTTTTTTTTTTTTTTTTTTTTTTTTTTTTTTTTTTTTTTTTTTTTTTTTTTTTTT
\begin{table}
\centering
\caption{Distribution of progenitor types.}
\begin{threeparttable}
\begin{tabular}{cccc}
\hline
CCSN Progenitors & $Z=0.00224$ & $Z=0.0056$ & $Z=0.014$ \\
\hline
Type Ic & & & \\
\hline
RSG & $0\% $ & $0\% $ & $0\% $ \\
YSG & $0\% $ & $0\% $ & $0\% $ \\
BSG & $0\% $ & $0\% $ & $0\% $ \\
HeG & $95.3128\% $ & $89.8243\% $ & $32.5516\% $ \\
WNh & $0\% $ & $0\% $ & $0\% $ \\
WN & $0\% $ & $0.0018145\% $ & $9.749\% $ \\
WC & $4.6872\% $ & $8.2551\% $ & $57.6994\% $ \\
WO & $0\% $ & $1.9189\% $ & $0\% $ \\
\hline
Type Ib & & & \\
\hline
RSG & $0\% $ & $0\% $ & $0\% $ \\
YSG & $0\% $ & $0\% $ & $0\% $ \\
BSG & $2.6371\% $ & $0.00099044\% $ & $1.4718\% $ \\
HeG & $97.3629\% $ & $80.3959\% $ & $45.0455\% $ \\
WNh & $0\% $ & $0\% $ & $0\% $ \\
WN & $0\% $ & $19.1799\% $ & $50.0165\% $ \\
WC & $0\% $ & $0.42326\% $ & $3.4663\% $ \\
WO & $0\% $ & $0\% $ & $0\% $ \\
\hline
Type Ib/IIb & & & \\
\hline
RSG & $0\% $ & $0\% $ & $0\% $ \\
YSG & $9.3624\% $ & $3.66\% $ & $9.6009\% $ \\
BSG & $89.1575\% $ & $82.9636\% $ & $68.9961\% $ \\
HeG & $0\% $ & $0\% $ & $3.8481\% $ \\
WNh & $1.4801\% $ & $13.3764\% $ & $17.5549\% $ \\
WN & $0\% $ & $0\% $ & $0\% $ \\
WC & $0\% $ & $0\% $ & $0\% $ \\
WO & $0\% $ & $0\% $ & $0\% $ \\
\hline
Type II & & & \\
\hline
RSG & $82.8889\% $ & $86.995\% $ & $91.3688\% $ \\
YSG & $7.0269\% $ & $7.4249\% $ & $6.4821\% $ \\
BSG & $9.8746\% $ & $5.3786\% $ & $1.8007\% $ \\
HeG & $0\% $ & $0\% $ & $0\% $ \\
WNh & $0.20958\% $ & $0.20152\% $ & $0.34846\% $ \\
WN & $0\% $ & $0\% $ & $0\% $ \\
WC & $0\% $ & $0\% $ & $0\% $ \\
WO & $0\% $ & $0\% $ & $0\% $ \\
\hline
Collapse to BH & & & \\
\hline
RSG & $23.1226\% $ & $15.5134\% $ & $2.0927\% $ \\
YSG & $16.1497\% $ & $15.3177\% $ & $16.324\% $ \\
BSG & $33.1841\% $ & $24.6606\% $ & $21.3267\% $ \\
HeG & $0\% $ & $0\% $ & $0\% $ \\
WNh & $10.2457\% $ & $3.81\% $ & $0.66518\% $ \\
WN & $4.862\% $ & $4.7628\% $ & $0.22823\% $ \\
WC & $11.5958\% $ & $28.4208\% $ & $32.8768\% $ \\
WO & $0.84001\% $ & $7.5146\% $ & $26.4864\% $ \\
\hline
\hline
\end{tabular}
\footnotesize
\begin{tablenotes}
\textit{Notes.} The probabilities sum up to $100\%$ for each SN type (or BH formation) and metallicity.
\end{tablenotes}
\end{threeparttable}
\label{tab:weightedprogs}
\end{table}
%TTTTTTTTTTTTTTTTTTTTTTTTTTTTTTTTTTTTTTTTTTTTTTTTTTTTTTTTTTTTTT
In Table\,\ref{tab:weightedprogs} we detail the probability-weighted stellar progenitor of each SN type, for each metallicity. Type~Ic SNe originate almost exclusively from helium giants for sub-Solar metallicities, while for $Z=0.014$ they arise from a mix of helium giants and WN or WC types of WR stars. Type~Ib SNe originate from similar progenitors as Type~Ic SNe, with some endpoints classified as BSGs being similar in nature to `helium giants' but with slightly more hydrogen. In Type~Ib SNe, the dominant WR sub-type is WN, and for Type~Ic SNe it is WC, in accordance with the composition definitions for both WR and SN types. The ambiguous SNe which we classify as `Ib/IIb' mostly originate from BSGs, with some contributions also from YSGs and WNh stars, i.e. partially stripped stars. Type~II SNe are dominated by RSG progenitors, as expected. Helium giants are the only stellar type predicted to always explode and avoid direct collapse to a BH, a result of these stars occupying the lower end of the final mass distribution and therefore expected to be easier to explode. With increasing metallicty, the progenitors of direct collapse shift to stellar types with higher degrees of envelope stripping. The fraction of direct collapse events originating from WR stars increases with metallicity, from $\approx 30\%$ at $Z=0.00224$ to $\approx 45\%$ at $Z=0.0056$ and $\approx 60\%$ at $Z=0.014$. The rest of the direct collapse progenitors are CSGs, though notably RSGs are much less favoured to collapse directly to BHs at $Z=0.014$ compared to YSGs and BSGs, with a more even distribution between CSG sub-types at the lower metallicities. Our results suggest that searches for disappearing stars in nearby galaxies (whose metallicities are similar to that of our Galaxy) should focus on WR stars and YSGs/BSGs, rather than RSGs.

% ==========================================================
\section{Short-lived lightweight WR stars}
\label{sec:WR}
% ==========================================================

In Section\,\ref{sec:results} we showed that many massive star evolution endpoints should be WR stars. In many of these cases, the WR appearance only emerges at a very late stage in their evolution, after core helium depletion. In the present section we elaborate on this short-lived WR phase, and quantify its duration.

\begin{figure*}
   \centering
\includegraphics[width=1\textwidth]{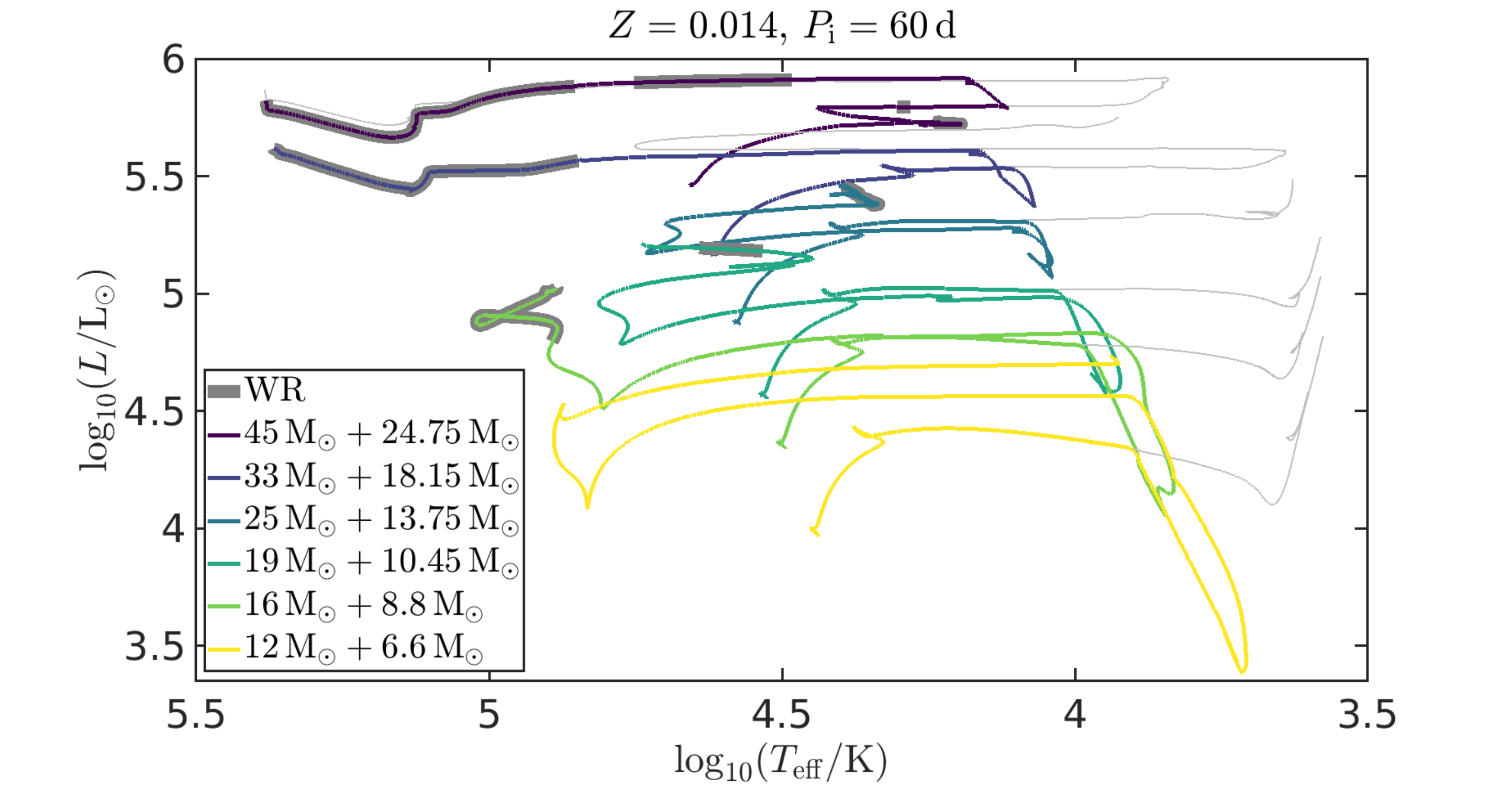}
    \caption{Effective temperature and luminosity evolution for the primary in a binary system with component masses detailed in the inset, compared to single-star evolution tracks (thin gray lines) with the same initial mass as that of the primary in the binaries. The evolutionary phase in which stellar models would have a WR spectral appearance are highlighted with thick dark gray marking.}
    \label{fig:HRDtracks}
\end{figure*}

In Fig.\,\ref{fig:HRDtracks} we present several binary evolutionary tracks with $Z=0.014$, highlighting the WR phase, and compare them to single-star tracks. For the most massive initial mass in Fig.\,\ref{fig:HRDtracks}, $M_\mathrm{ZAMS}=45\,\mathrm{M}_\odot$, single-star evolution results in a WR star, as the stellar wind is strong enough to remove the hydrogen envelope with no need for the aid of a companion. For this mass, the single-star and binary tracks are close to overlapping for most of the evolution, except for a brief expansion in the single-star track to radii that are not allowed by the presence of a companion in the binary simulation. For $M_\mathrm{ZAMS}=33\,\mathrm{M}_\odot$, the binary interaction is crucial in enabling a long-lived WR phase. For $M_\mathrm{ZAMS}=12\,\mathrm{M}_\odot$ there is no WR phase; the stripped star will be an intermediate-mass helium MS star with a weak wind \cite[e.g.][]{Drout2023,Gotberg2023}, and after core helium depletion expand into a helium giant. The three intermediate masses ($M_\mathrm{ZAMS}=16$, $19$ and $25\,\mathrm{M}_\odot$) all have a short WR phase, which they enter shortly prior to core collapse. Notably, for the $M_\mathrm{ZAMS}=16\,\mathrm{M}_\odot$ case this is the immediate evolutionary stage after the helium MS, with the primary star in the binary system transitioning from a an intermediate-mass helium MS star to a WR star.
%FFFFFFFFFFFFFFFFFFFFFFFFFFFFFFFFFFFFFFFFFFFFFFFFFFFFFFFFFFFFFF
%FFFFFFFFFFFFFFFFFFFFFFFFFFFFFFFFFFFFFFFFFFFFFFFFFFFFFFFFFFFFFF

\begin{figure*}
   \centering
\includegraphics[width=1\textwidth]{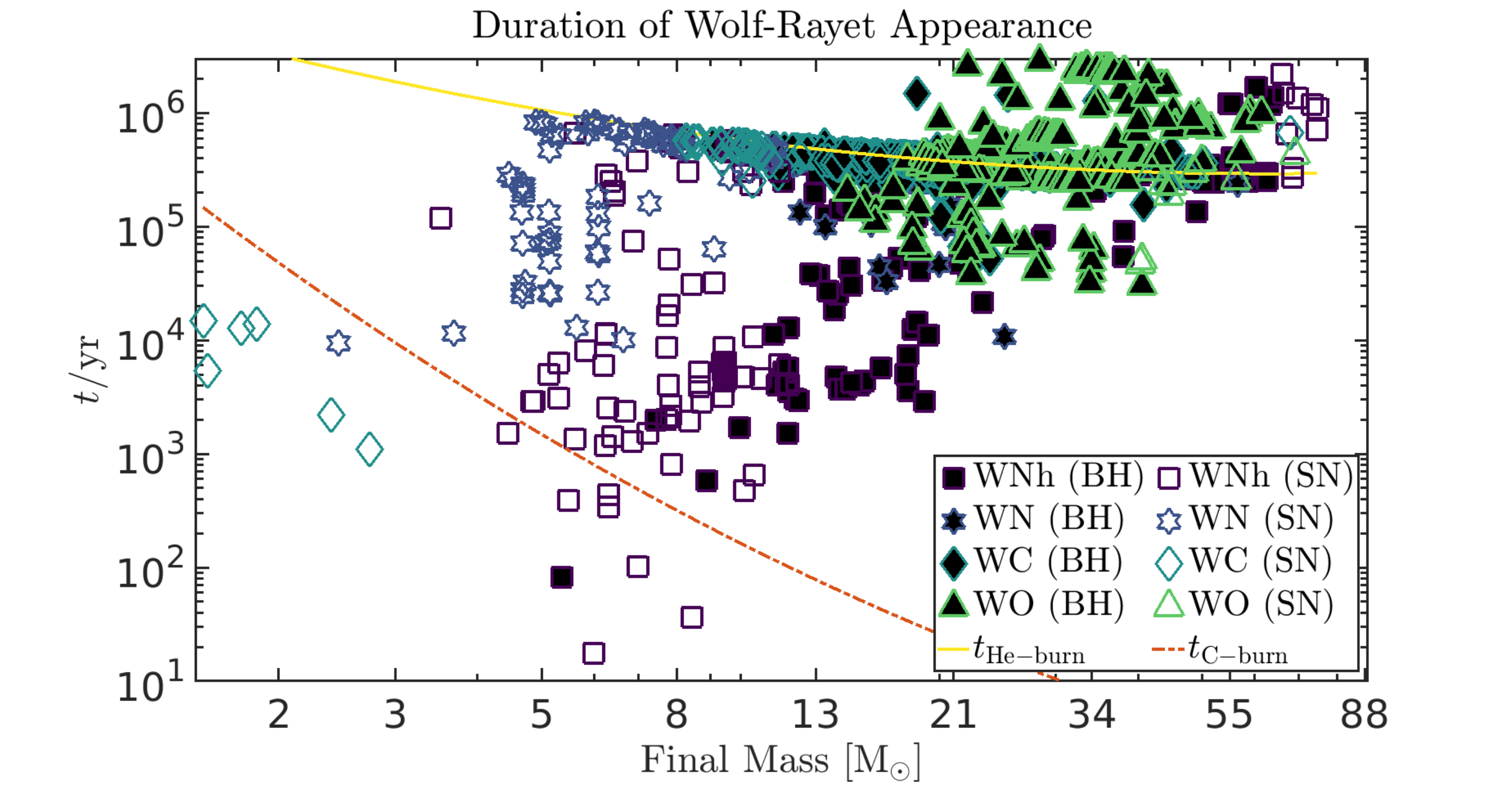}
    \caption{The duration of the WR spectral appearance as a function of the final stellar mass, with models marked by their final WR type and whether they are expected to explode or directly collapse to a BH. Only models expected to explode or implode while appearing as WR stars are shown. Quadratic fits in log-space to the helium and carbon burning timescales as function of final mass are presented for comparison, according to Eq.\,\ref{eq:thelium} and Eq.\,\ref{eq:tcarbon}.}
    \label{fig:WRtime}
\end{figure*}

In Fig.\,\ref{fig:WRtime} we present the duration of the WR appearance in stellar models for those that reach the end of their evolution as WR stars. We mark the WR types of the endpoints as descirbed in Section\,\ref{subsec:calcWR}. Some of the most massive stellar models become WR stars already during core hydrogen burning, and therefore have a WR duration exceeding $1\,\mathrm{Myr}$, eventually all predicted to collapse to BHs (except a few that reach PI). The majority of WR stars have a WR duration of a few hundred millennia, i.e. approximately the core helium burning lifetime. For comparison, we show the core helium-burning timescale using a fit to our results,
\begin{equation}
    \log_{10} (t_\mathrm{He-burn}) =0.46077(\log_{10} M)^2 -1.6613(\log_{10} M) + 6.9627\, ,
    \label{eq:thelium}
\end{equation}
and similarly for the core carbon-burning timescale,
\begin{equation}
    \log_{10} (t_\mathrm{C-burn}) =0.87661(\log_{10} M)^2 -4.6733(\log_{10} M) + 6.0102\, ,
    \label{eq:tcarbon}
\end{equation}
where $M$ is the final mass in units of $\mathrm{M}_\odot$. The WR stars span final masses as low as $\approx 5\,\mathrm{M}_\odot$, though the final mass is naturally lower than the stellar mass during earlier stages which might be those more commonly observed. The WR stars that are expected to explode are generally those with final masses $\la 12\,\mathrm{M}_\odot$, and in a few cases have a WR duration as short as several decades, centuries or millennia.
%FFFFFFFFFFFFFFFFFFFFFFFFFFFFFFFFFFFFFFFFFFFFFFFFFFFFFFFFFFFFFF
%FFFFFFFFFFFFFFFFFFFFFFFFFFFFFFFFFFFFFFFFFFFFFFFFFFFFFFFFFFFFFF

Using the same weighting as described in Section\,\ref{sec:IMFetal}, we present the relative importance of different WR evolutionary cases in Table\,\ref{tab:weightedWRs}. We define three evolutionary cases for WR stars, according to whether they are expected to first show WR features during core hydrogen burning (`Early'), after core helium depletion (`Late'), or otherwise (`Classical'). We compute the expected fractions out of the living populations of WR stars by multiplying the endpoint probabilities with the preceding duration of the WR appearance. As shown in Table\,\ref{tab:weightedWRs}, while `Late' WR stars comprise $\approx 0.8\%$--$3\%$ of living WR stars, their prevalence as CCSN progenitors is an order of magnitude higher. `Early' WR stars arise from the upper end of the IMF, and contribute to BH formation and PISNe, but not to regular CCSNe. We note that their majority fraction among living WR stars at $Z=0.0056$ and $Z=0.014$ does not mean that we predict most WR stars to be core hydrogen burning at those metallicities, as the evolutionary cases are defined by the core properties at the start of the WR phase and not throughout its duration.  
%TTTTTTTTTTTTTTTTTTTTTTTTTTTTTTTTTTTTTTTTTTTTTTTTTTTTTTTTTTTTTT
\begin{table}
\centering
\caption{Weighted fraction of evolutionary stage at first WR appearance.}
\begin{threeparttable}
\begin{tabular}{cccc}
\hline
 & $Z=0.00224$ & $Z=0.0056$ & $Z=0.014$ \\
\hline
Living WR stars & & & \\
\hline
Early & $35.5482\% $ & $59.5389\% $ & $71.2164\% $ \\
Classical & $61.3533\% $ & $39.0407\% $ & $27.9437\% $ \\
Late & $3.0985\% $ & $1.4204\% $ & $0.83994\% $ \\
\hline
WR CCSN progenitors & & & \\
\hline
Early & $0\% $ & $0\% $ & $0\% $ \\
Classical & $64.0591\% $ & $92.0606\% $ & $89.6941\% $ \\
Late & $35.9409\% $ & $7.9394\% $ & $10.3059\% $ \\
\hline
WR BH progenitors & & & \\
\hline
Early & $4.3247\% $ & $41.2661\% $ & $67.2833\% $ \\
Classical & $91.149\% $ & $57.8863\% $ & $32.5257\% $ \\
Late & $4.5263\% $ & $0.84755\% $ & $0.19099\% $ \\
\hline
WR PISN progenitors & & & \\
\hline
Early & $41.8724\% $ & $97.7392\% $ & - \\
Classical & $58.1276\% $ & $2.2608\% $ & - \\
Late & $0\% $ & $0\% $ & - \\
\hline
\hline
\end{tabular}
\footnotesize
\begin{tablenotes}
\textit{Notes.} `Early' cases are those that first exhibit the WR phenomenon during core hydrogen burning, `late' cases are those that become WR stars only after core helium depletion, and `classical' are the rest.
\end{tablenotes}
\end{threeparttable}
\label{tab:weightedWRs}
\end{table}
%TTTTTTTTTTTTTTTTTTTTTTTTTTTTTTTTTTTTTTTTTTTTTTTTTTTTTTTTTTTTTT

% ==========================================================
\section{Discussion}
\label{sec:discussion}
% ==========================================================

% ==========================================================
\subsection{Impact of stellar winds}
\label{subsec:discusswinds}
% ==========================================================

In the present study we use the new theoretical WR wind prescription presented by \cite{Sander2020}, in which the mass-loss rate depends on the stellar mass as well as the luminosity, in contrast to empirical recipes which depend only on luminosity. We find that stripped stars transition from the regime of optically-thin winds to WR winds as their luminosity increases, so that shortly before exploding their appearance would be that of a WR star. The high pre-supernova wind mass loss also creates CSM into which the ejecta will collide, altering the supernova light curve and spectral appearance \citep[e.g.][]{Wu2022}. Our prediction of increased mass loss as stripped stars approach the Eddington limit in late evolutionary phases is conceptually similar to a possible explanation for high mass-loss rates in the most massive RSGs \citep{Vink2023}, where for both WR stars and RSGs the enhanced mass loss can lead to large amounts of CSM.

A caveat in our models is that we do not include a temperature dependence for the WR winds. \cite{Sander2023} predict an inverse relation between temperature and WR mass loss. Therefore, our short-lived lightweight WR models, which are cooler than the massive `classical' WR stars, would have even higher mass-loss rates. Including a temperature dependence in future models is therefore imperative. 

In a somewhat cooler regime compared to WR stars and stripped stars -- MS stars -- there are lingering issues related to the wind mass loss recipes. While we follow \cite{Vink2001} in including a jump in the mass-loss rate around the temperature of recombination of \ion{Fe}{iv} into \ion{Fe}{ii} (the so-called bi-stability jump at $\approx 22\,\mathrm{kK}$), recent observations find no evidence of an increase in the mass-loss rates in this temperature \citep{deBurgos2024}. The usage of \cite{Vink2001} for MS massive stars has been recently challenged, with alternative prescriptions proposed (\citealt{Bjorklund2021,Bjorklund2023}; \citealt*{Krticka2021,Krticka2024}). As most massive stars are in close binaries, mass loss through RLOF is usually much more significant compared to wind mass loss, and the update to wind mass loss in stripped stars (which are more compact than MS stars) is more decisive \citep[e.g.][]{GilkisVinkEldridgeTout2019}.

Winds of very massive stars \citep[VMSs; $M_\mathrm{ZAMS}\ga 100\,\mathrm{M}_\odot$; e.g.][]{Crowther2010,Bestenlehner2020}, which can become WR stars already on the MS, are highly uncertain \citep{Grafener2008,Sabhahit2022}. Although the IMF strongly disfavours such high-mass stars, they are expected to be the main progenitors of PISNe. \cite{Sabhahit2023} use a new framework for low metallicity VMS mass loss, and find that a metallicity as low as $Z_\odot / 20$ is needed for stars to retain enough mass to explode in PISNe. As our models have PISNe already at LMC metallicity ($Z=0.0056$), the findings of \cite{Sabhahit2023} indicate that we might be underestimating the wind mass loss in the highest luminosity regime. Recently, \cite{Angus2024} claimed that SN~2020acct matches models of PPISNe, with its local galactic environment having a metallicity similar to that of the LMC. However, the metallicity threshold for PISNe and PPISNe might be different, as the CO core mass threshold is lower for the latter.

For cool stars -- LBVs and CSGs -- we mostly follow the mass-loss recipe of \cite{dJ88}, which is markedly outdated, and updated empirical and theoretical results should be considered \citep[e.g.][]{Zapartas2025}. For example, \cite{Yang2023} find a `kink' in the mass-loss rate dependence on luminosity in SMC RSGs, where a steep luminosity dependence is apparent at high luminosity. Theoretical models of winds driven by turbulence \citep{Kee2021} or shocks \citep{Fuller2024} predict a runaway increase in mass loss at very late stages of RSG evolution, potentially creating large amounts of CSM. High RSG mass loss, as well as LBV eruptions \citep{Humphreys1984,Humphreys1994,Smith2017,Cheng2024}, can ease the formation of WR stars \citep[even at low metallicity, e.g.][]{Schootemeijer2024} and impact our predictions for the numbers of WR stars.

% ==========================================================
\subsection{Binary interaction}
\label{subsec:discussbin}
% ==========================================================

Our usage of a physically-motivated criterion for mass-transfer stability based on the stellar structure and response to mass loss based on \cite{CopingWithLoss} allows for stable mass transfer in systems with high mass ratios, i.e. mass transfer from a massive star to a companion of much lower mass \citep[see also][]{Ge2010,Ge2015,Ge2020}. As a result, some CCSN progenitors reach mass-loss rates approaching $10^{-3}\,\mathrm{M}_\odot\,\mathrm{yr}^{-1}$, providing significant amounts of CSM. Commonly-used simple stability criteria \citep[e.g.][]{Hjellming1987,Soberman1997,VignaGomez2018} based only on the mass ratio and stellar type might overlook this mechanism for CSM generation.

In cases where mass transfer is unstable, a CEE phase is initiated and the outcome is determined following an energy formalism. We find with an increase in stellar mass a decrease in successful envelope ejection. This ejection failure results in many systems merging during CEE. When the engulfed companion is a normal star (i.e. not a compact object) we construct a coalescence product and follow its further evolution as a single star. The endpoints of this merged-star evolutionary channel are varied, including stripped hydrogen-deficient stars (similar to the `fatal CEE' scenario discussed by \citealt{Lohev2019}) and hydrogen-rich endpoints expected to explode as Type~II SNe \citep[e.g.][]{Zapartas2019}.

We do not follow the further evolution after a CEE phase where the envelope fails to eject, ending in the coalescence of a non-degenerate stellar core with a compact object. We find a relatively high rate (about $10\%$ of the overall CCSN rate) of stars merging with NSs during CEE. If this result is robust, the observational signature of these merging events should be quite commonplace, with one possibility being Type~IIn SNe \citep[e.g.][]{Chevalier2012}. These results, however, are sensitive to the CEE implementation and in particular to our choice of the efficiency of utilising orbital energy to unbind the envelope ($\alpha_\mathrm{CEE}\approx 1$).

Another evolutionary channel where the CEE phase ends in coalescence involves a WD companion. This channel can lead to the WD becoming the stellar core, with further evolution powered by shell burning. For particularly massive CO WDs merging with cores of AGB stars, the final outcome might be a Type~Ia SN \citep[e.g.][]{Bear2018}. For an ONe WD becoming the core of the merged star, an ECSN fate is possible.

Our mass transfer stability prescription results in almost no BHs entering CEE, supporting the stable mass transfer channel as a dominant channel towards merging BHs \citep[e.g.][]{vanSon2023}. However, we do not focus here on sources of gravitational wave radiation, which can be affected by the details of PPISNe \citep[e.g.][]{Hendriks2023} or arise from chemically homogeneous evolution \citep[][]{Mandel2016,duBuisson2020}, the latter requiring to follow stellar rotation which is not included in our models.

% ==========================================================
\subsection{CCSN rate and progenitors}
\label{subsec:discussprogenitors}
% ==========================================================

The rate of Type~II SNe relative to the rate of the other (stripped) CCSN types in our simulations is about $2$--$4$ times higher, depending on metallicity and the minimum hydrogen mass needed for a Type~IIb SN (Table\,\ref{tab:weightedCCSNe}). This ratio is roughly consistent with the volume-limited sample analysed by \cite{Li2011}, who found twice as may Type~II SNe (including Type~IIb) as Type~Ib/c SNe. Observational studies have found the relative rates of different CCSN types to depend on environment, i.e. host galaxy mass and/or metallicity \citep[e.g.][]{Arcavi2010,Graur2017}. We find the rate of Type~Ib/c SNe relative to Type~II SNe increasing with metallicity, consistent with \cite{Graur2017}.

The highest luminosity of a CSG expected to explode in our models is $\log_{10} (L / \mathrm{L}_\odot)\approx 5.5$, while the most luminous RSG progenitor identified is closer to $\log_{10} (L / \mathrm{L}_\odot)\approx 5.1$. This mismatch is apparent in our probability map (Fig.\,\ref{fig:progmap}), and has been termed the `Red Supergiant Problem' by \cite{Smartt2009}. The non-detection of hotter progenitors (e.g. WR or hot helium giants) can be attributed to their faint optical emission \citep{Yoon2012}, but the most luminous RSGs should be readily observable. Yet, the statistical significance of the `Red Supergiant Problem' is debated \citep*{Davies2020a,Davies2020b,Kochanek2020,Rodriguez2022,Beasor2025}. Alternatively, the prediction that RSGs with $\log_{10} (L / \mathrm{L}_\odot)\ga 5.1$ should explode might be wrong \citep[e.g.][]{Horiuchi2014}, the luminosity of the most massive RSGs might be underestimated because of obscuration by dust \citep[e.g.][]{Walmswell2012} or RSGs experience enhanced mass loss that results in much higher final temperatures \citep*[e.g.][]{Gofman2020}.

At face value our results are inconsistent with the identified progenitors of Type~IIn SNe, as we predict stars in the high luminosity and low temperature part of the HRD to collapse to BHs. This tension might be alleviated if a significant fraction of Type~IIn SNe originate from the fatal coalescence of a NS with the core of a RSG as proposed by \cite{Chevalier2012}. While several studies (e.g. \citealt*{Soker2018,Soker2019}; \citealt{Schroder2020}) considered this scenario for rare SNe, the fate of a CEE phase involving a NS inside a RSG is extremely uncertain. For example, \cite{Tanaka2023} showed that small variations in the envelope ejection efficiency ($\alpha_\mathrm{CEE}$) results in orders-of-magnitude differences in the rate of merging NS binaries. Successful envelope ejection during CEE is crucial for producing NS binaries with small separations that will merge in a short enough time to produce detectable gravitational waves. Our results favour a high fraction of CEE events resulting in a `failure' to eject the envelope, which would manifest as an increased number of SN-like events (likely of Type~IIn) at the `expense' of merging NSs.

% ==========================================================
\section{Summary and conclusions}
\label{sec:summary}
% ==========================================================

We have constructed a grid of massive star evolution simulations using updated theoretical mass-loss rates for hot stars. We take into account several evolutionary channels, including single-star evolution, the evolution of coalescence products following a CEE phase that failed to eject the envelope, the evolution of the primary star in an interacting binary system, and the evolution of the surviving secondary star, with a compact companion formed from the evolutionary end of the primary star or alone if the system was disrupted in the first SN. The comprehensive mapping from initial conditions to evolutionary endpoints (using thousands of detailed stellar evolution simulations) allows us to lay out the landscape of SN progenitors and quantitatively predict the rates of massive star outcomes, and in the future to generate synthetic populations of massive stars (including X-ray binaries) to compare with observed populations. Testing such predictions against observations will provide critical constraints on stellar modelling. 

We summarise our findings in the following key points.
\begin{itemize}
    \item WR stars come from several channels, the largest fraction being from post-interaction binaries. Their properties differ depending on metallicity.
    \item Intermediate-mass helium stars stripped by binary interaction can evolve to become WR stars late in their evolution, only a few millennia before exploding.
    \item The final mass of WR stars expected to explode is as low as a few $\mathrm{M}_\odot$.
    \item The majority of Type Ib and Ic SN progenitors are helium giants, with short-lived lightweight WR stars making a significant contribution as well.
    \item Stripped star winds are crucial both for the pre-supernova evolution and for the outer stellar layers and CSM shaping the supernova appearance.
\end{itemize}

Our grid introduces a step forward in modelling progenitors of CCSNe, with its broad range of initial conditions and evolutionary channels and endpoints potentially explaining a wide range of energetic transients associated with massive stars that are being discovered with current and upcoming surveys. Future extensions of the grid will include lower metallicities, rotation, and variation on key uncertain processes in massive star evolution, such as CEE and wind mass loss.

% ==========================================================
\section*{Acknowledgments}
% ==========================================================

We thank an anonymous referee for a constructive review of the paper.
We thank Guy Flint for his careful reading of the manuscript and for his helpful suggestions which improved the paper.
This research was supported by a grant from the Pazy Foundation (grant number 216312).
EL and FRNS acknowledge support from the European Research Council (ERC) under the European Union’s Horizon 2020 research and innovation program, grant agreement number 945806. 
EL acknowledges funding through a start-up grant from the Internal Funds KU Leuven (STG/24/073) and a Veni grant (VI.Veni.232.205) from the Netherlands Organization for  Scientific Research (NWO).
IA acknowledges support from the ERC under the European Union’s Horizon 2020 research and innovation program, grant agreement number 852097.
TS acknowledges support from the ERC under the European Union’s Horizon 2020 research and innovation program, grant agreement number 101164755 and from the Israel Science Foundation (ISF) under grant number 2434/24.
FRNS is supported by the Deutsche Forschungsgemeinschaft (DFG, German Research Foundation) under Germany’s Excellence Strategy EXC 2181/1-390900948 (the Heidelberg STRUCTURES Excellence Cluster).

\section*{Data Availability Statement}

The input files necessary to reproduce our stellar evolution simulations and associated data products are available at:
\begin{itemize}
    \item \href{https://zenodo.org/records/15639618}{https://zenodo.org/records/15639618} ($Z=0.00224$);
    \item \href{https://zenodo.org/records/15644339}{https://zenodo.org/records/15644339} ($Z=0.0056$);
    \item \href{https://zenodo.org/records/15646740}{https://zenodo.org/records/15646740} ($Z=0.014$).
\end{itemize}

\bibliographystyle{mnras}
\input{LightweightWRs.bbl}

% ==========================================================
\appendix
% ==========================================================

\section{Single-star evolutionary models}
\label{sec:appendixa}

%FFFFFFFFFFFFFFFFFFFFFFFFFFFFFFFFFFFFFFFFFFFFFFFFFFFFFFFFFFFFFF
\begin{figure}
\centering
\includegraphics[width=0.48\textwidth]{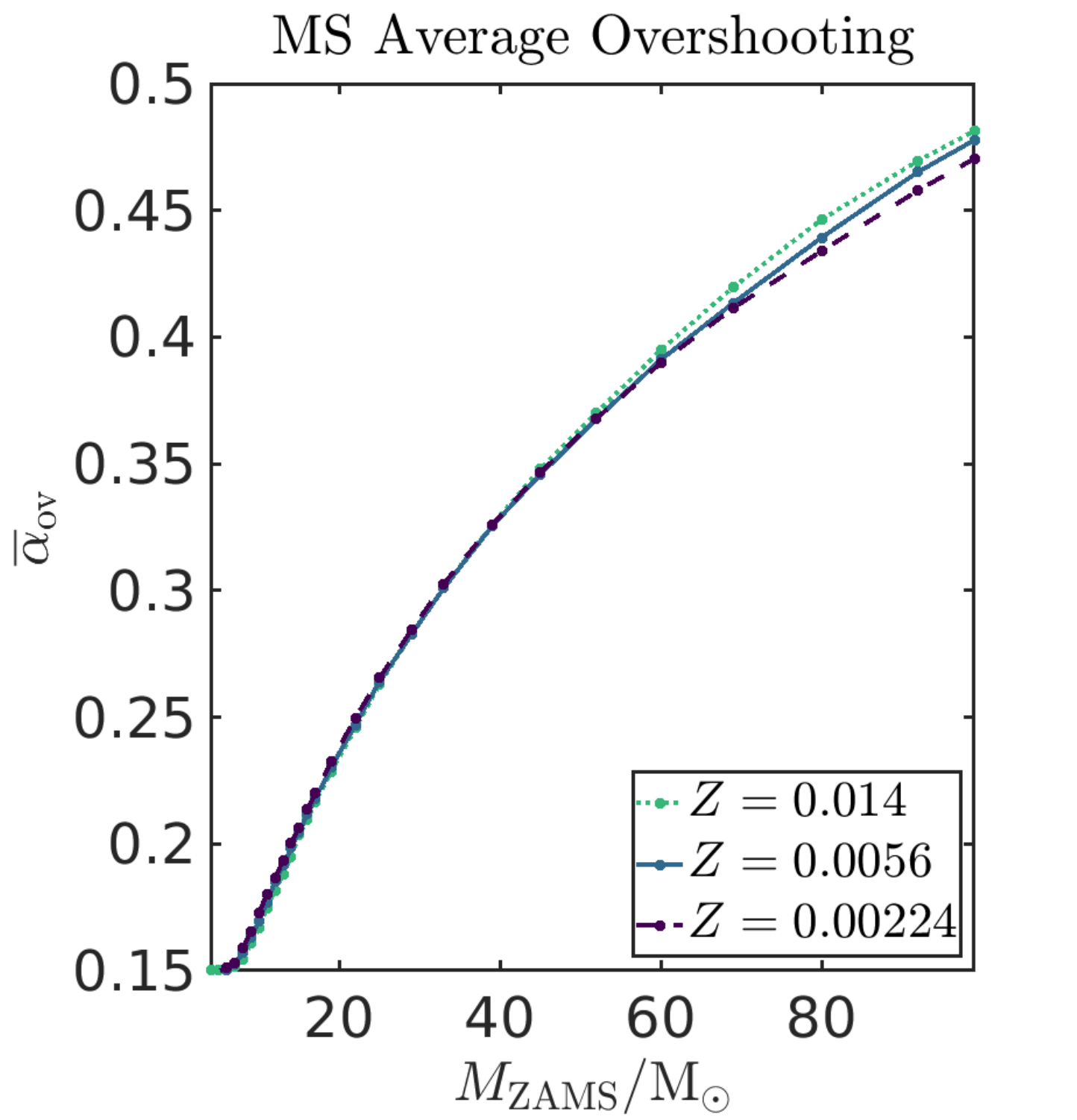} \\
\caption{The time-averaged overshooting extent during the main-sequence evolution in our single-star models, expressed in units of pressure scale height $H_P$.}
\label{fig:ov}
\end{figure}
%FFFFFFFFFFFFFFFFFFFFFFFFFFFFFFFFFFFFFFFFFFFFFFFFFFFFFFFFFFFFFF
%FFFFFFFFFFFFFFFFFFFFFFFFFFFFFFFFFFFFFFFFFFFFFFFFFFFFFFFFFFFFFF
\begin{figure*}
\centering
\includegraphics[width=\textwidth]{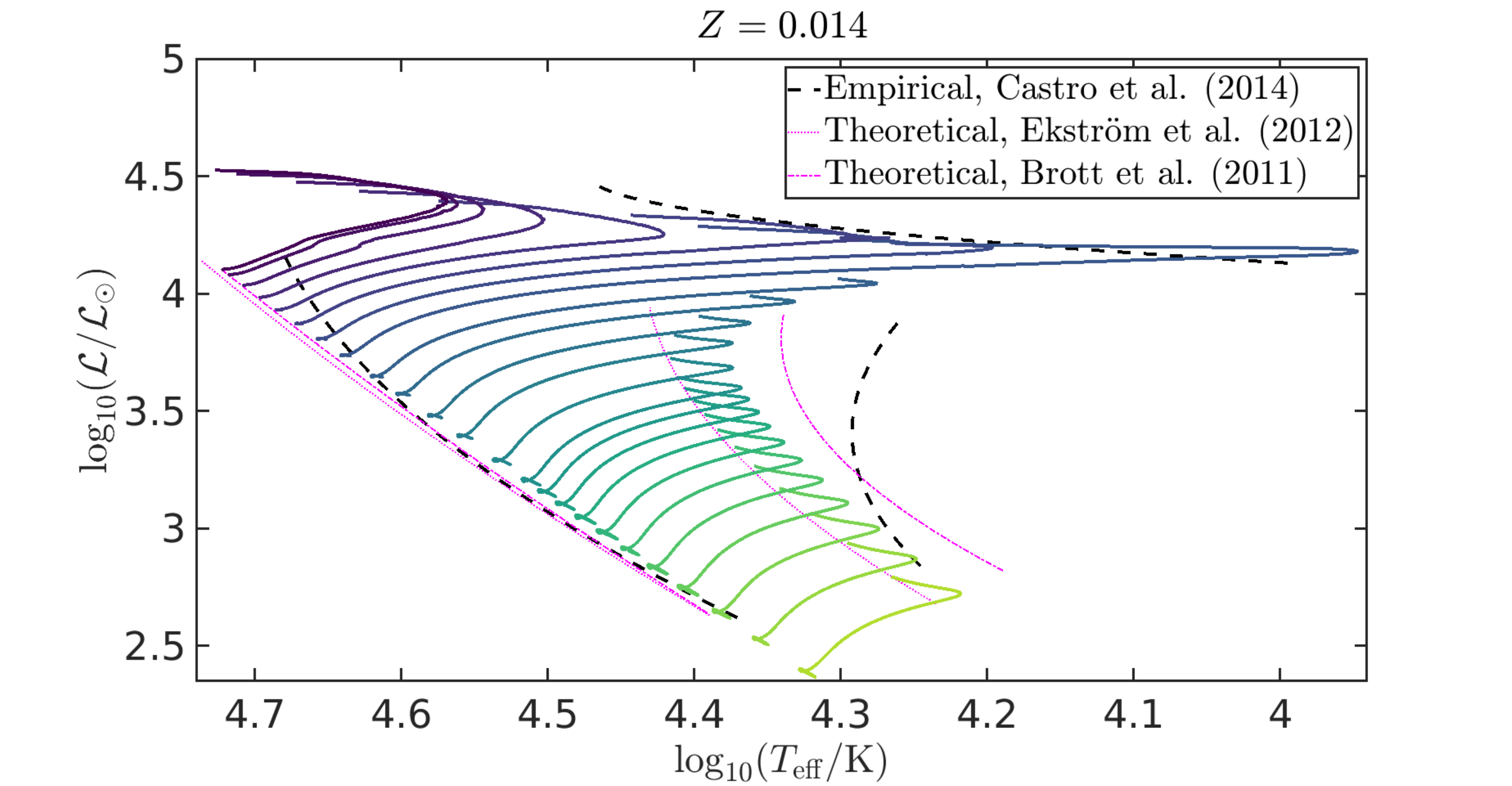} \\
\caption{The spectroscopic Hertzsprung-Russell diagram showing some of our $Z=0.014$ single-star models with $M_\mathrm{ZAMS,1} \ge 7\,\mathrm{M}_\odot$ during the main-sequence part of the evolution in comparison with the models of \citet{Brott2011} and \citet{Ekstrom2012} and with observations \citep{Castro2014}.}
\label{fig:shrd}
\end{figure*}
%FFFFFFFFFFFFFFFFFFFFFFFFFFFFFFFFFFFFFFFFFFFFFFFFFFFFFFFFFFFFFF
%FFFFFFFFFFFFFFFFFFFFFFFFFFFFFFFFFFFFFFFFFFFFFFFFFFFFFFFFFFFFFF
\begin{figure*}
\centering
\includegraphics[width=\textwidth]{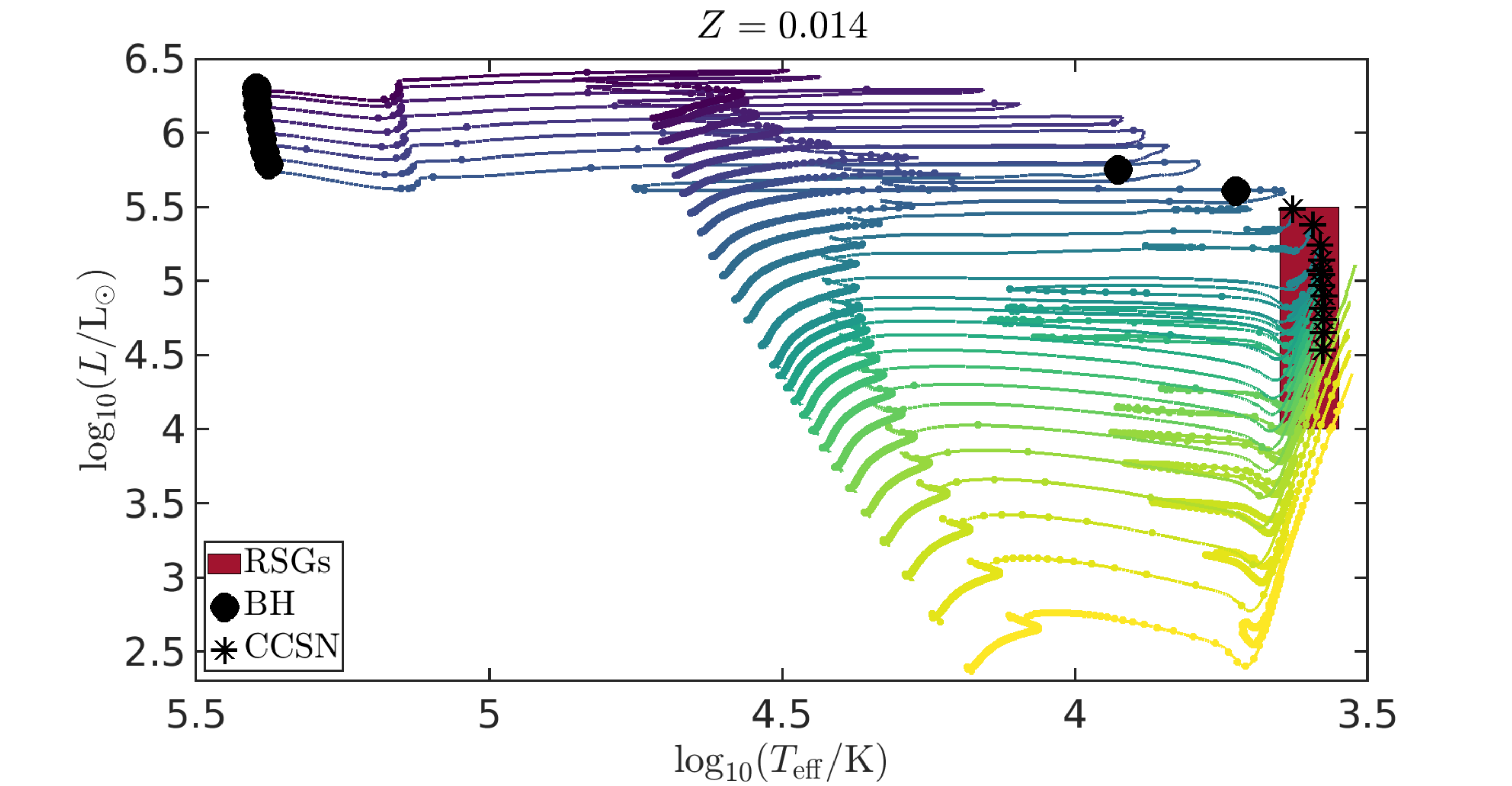} \\
\caption{Single-star evolutionary tracks with initial masses between $4\,\mathrm{M}_\odot$ and $99\,\mathrm{M}_\odot$ and a metallicity $Z=0.014$, and points marked every $50\,\mathrm{kyr}$. The red-shaded box marks the characteristics of Galactic red supergiants \citep{Beasor2021}. Evolutionary endpoints that lead to CCSNe are marked with black asterisks, and endpoints predicted to collapse to BHs are marked with filled black circles. For initial masses $\le 8\,\mathrm{M}_\odot$ the evolution ends in the AGB or super-AGB phase, with cooler temperatures than RSGs; we do not follow the later evolution to WD formation in these cases.}
\label{fig:rsgshrd}
\end{figure*}
%FFFFFFFFFFFFFFFFFFFFFFFFFFFFFFFFFFFFFFFFFFFFFFFFFFFFFFFFFFFFFF
Here we present the key results from our $Z=0.014$ single-star evolutionary tracks. In Fig.\,\ref{fig:ov} we show the time-averaged overshooting extent during the main-sequence evolution as computed using the \cite{Jermyn2022} prescription, showing a gradual increase from $\alpha_\mathrm{ov}\approx 0.15$ for the lowest masses up to $\alpha_\mathrm{ov}\approx 0.5$ for the most massive stars simulated. In Fig.\,\ref{fig:shrd} we present the main-sequence evolution of our models in the so-called spectroscopic Hertzsprung-Russell diagram, where $\mathcal{L}\equiv T^4 / g$, and we show that our models are at least as good as other theoretical models \citep{Brott2011,Ekstrom2012} in reproducing the observationally-derived landscape \citep{Castro2014}. We note that the empirical terminal age main sequence (TAMS) feature described by \cite{Castro2014}, while often used for comparison, is based on a heterogeneous sample of Galactic OB stars and may be affected by selection or systematic biases. More homogeneous studies, such as those using VLT-FLAMES data in the LMC \citep[e.g.][]{Schneider2018}, do not show an equivalent post-TAMS gap, and the location of the TAMS in SMC stars remains uncertain pending final results from the BLOeM survey \citep{Shenar2024}.

In Fig.\,\ref{fig:rsgshrd} we show the entire evolution of $27$ single-star simulations, highlighting the good reproduction of Galactic RSG properties \citep[e.g.][]{Beasor2021}. Fig.\,\ref{fig:rsgshrd} also shows the mass dependence of the final fate of evolved massive stars. The tracks for initial masses $< 28\,\mathrm{M}_\odot$ all end as RSGs and are predicted to explode in CCSNe, while higher initial masses lead to hotter endpoints that are predicted to directly collapse to BHs. Among the more luminous endpoints, those with high enough initial masses become WR stars, as their strong stellar winds are able to remove the hydrogen envelope even without the aid of binary interaction. We find the minimum mass for a single star to become a WR star to be $M_\mathrm{ZAMS} = 36\,\mathrm{M}_\odot$ in our Solar-like metallicity $Z=0.014$ models. The endpoints of the tracks with initial masses $28\,\mathrm{M}_\odot < M_\mathrm{ZAMS} < 36\,\mathrm{M}_\odot$ are either YSGs or BSGs.

As discussed by \cite{Shenar2020}, the minimum initial mass for a single star simulation to reach a WR phase differs between codes and changes according to mixing assumptions (e.g. rotational mixing; \citealt{Limongi2018}) and also depends on metallicity. Our $Z=0.014$ mass threshold is slightly higher than values reported in the literature \citep[e.g.][]{Dray2003,Limongi2018}. For $Z=0.0056$, we find a non-monotonic behaviour, with the the lowest initial mass yielding a WR star being $M_\mathrm{ZAMS}= 42\,\mathrm{M}_\odot$, but several tracks with higher masses not becoming WR stars, and robust WR formation occurs for $M_\mathrm{ZAMS} \ge 51\,\mathrm{M}_\odot$. These masses are higher than some predictions \citep*[e.g. with the \textsc{stars} code;][]{Eldridge2008} but lower than others \citep[e.g. \textsc{genec};][]{Eggenberger2021}. For $Z=0.00224$, none of the single-star tracks lost their entire hydrogen through winds and also do not reach effective surface temperatures higher than $65\,\mathrm{kK}$. However, tracks with $M_\mathrm{ZAMS} > 54\,\mathrm{M}_\odot$ do reach a point where they would be considered to be WR stars according to our transformed radius criterion (Section\,\ref{subsec:calcWR}) of sub-type WNh.

\section{Mass-transfer stability}
\label{sec:appendixMTstability}

\begin{figure*}
\centering
    \begin{subfigure}{0.33\textwidth}   
        \includegraphics[width=1\textwidth]{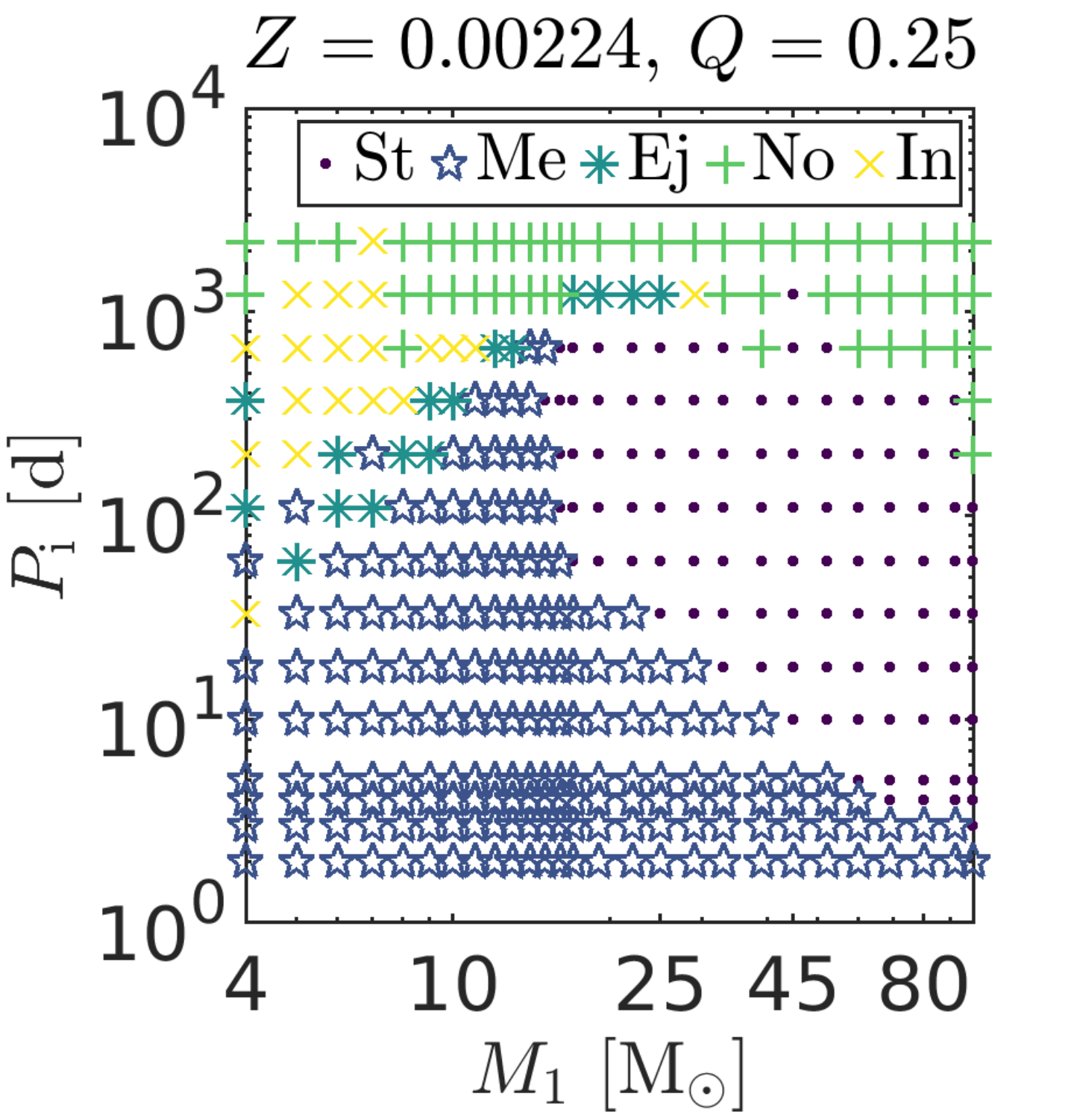}
        \label{fig:stabmapSMC25}
    \end{subfigure}
    \begin{subfigure}{0.33\textwidth}   
        \includegraphics[width=1\textwidth]{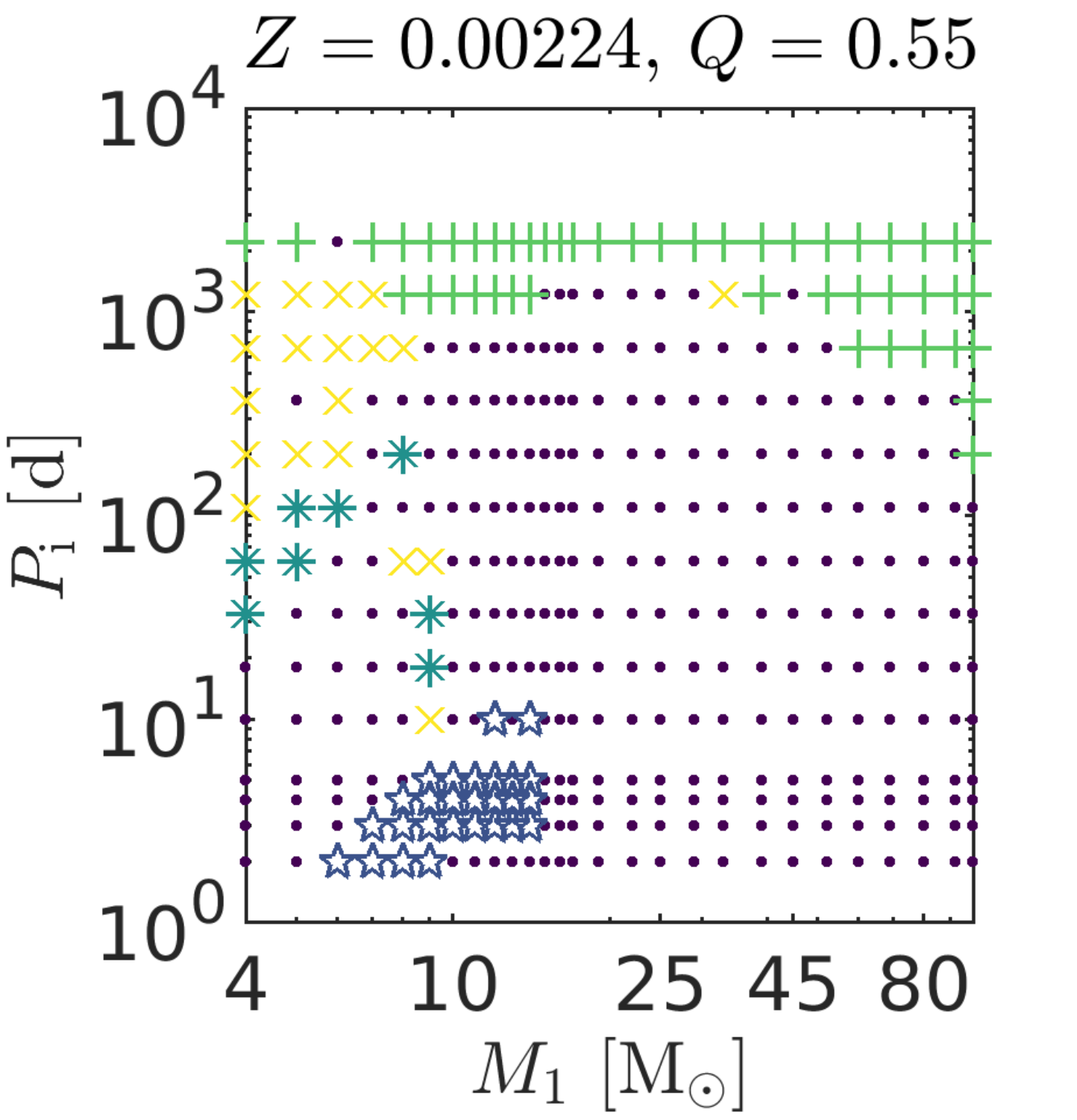}
        \label{fig:stabmapSMC55}
    \end{subfigure}
    \begin{subfigure}{0.33\textwidth}   
        \includegraphics[width=1\textwidth]{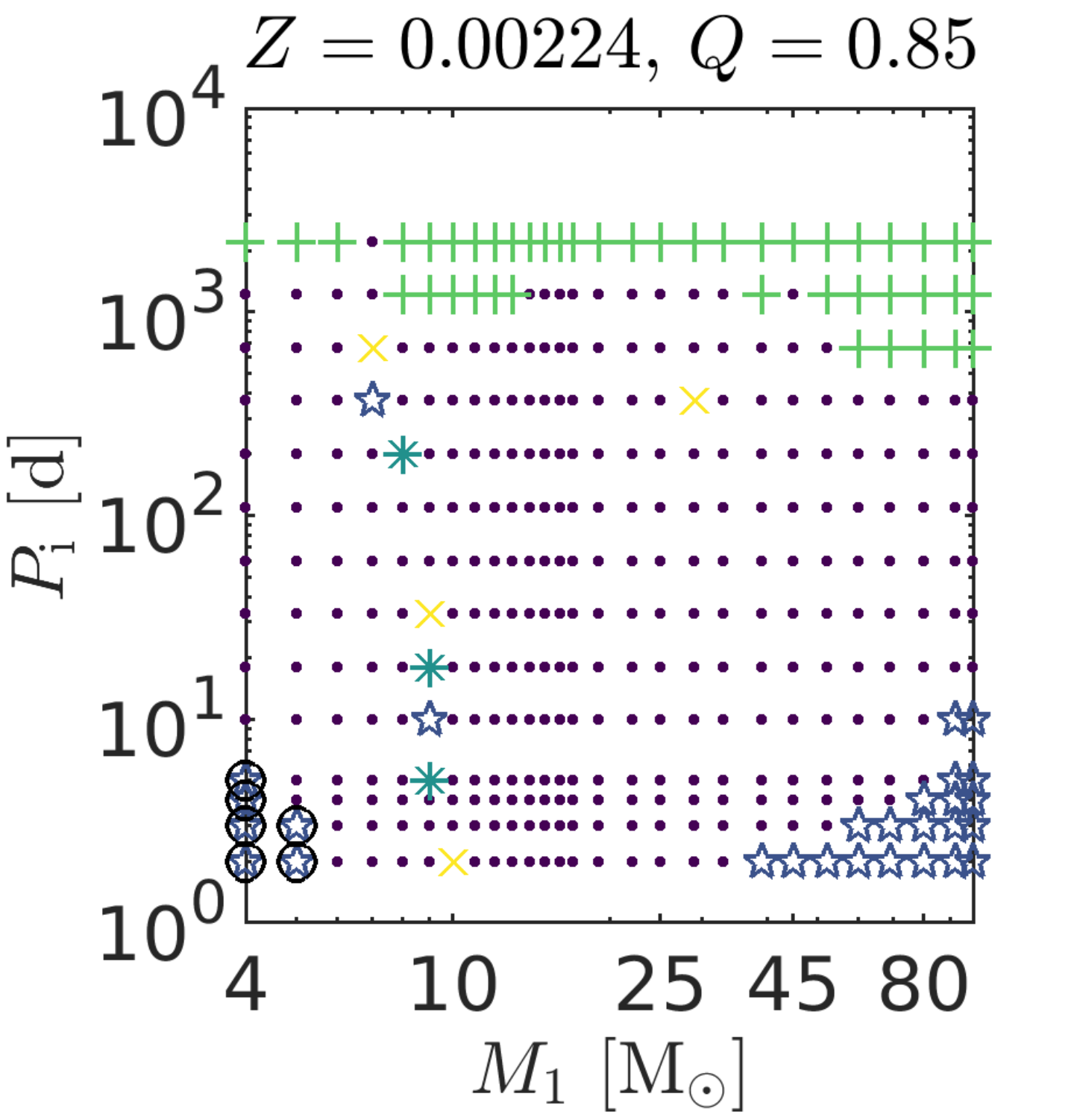}
        \label{fig:stabmapSMC85}
    \end{subfigure}\\
    \begin{subfigure}{0.33\textwidth}   
        \includegraphics[width=1\textwidth]{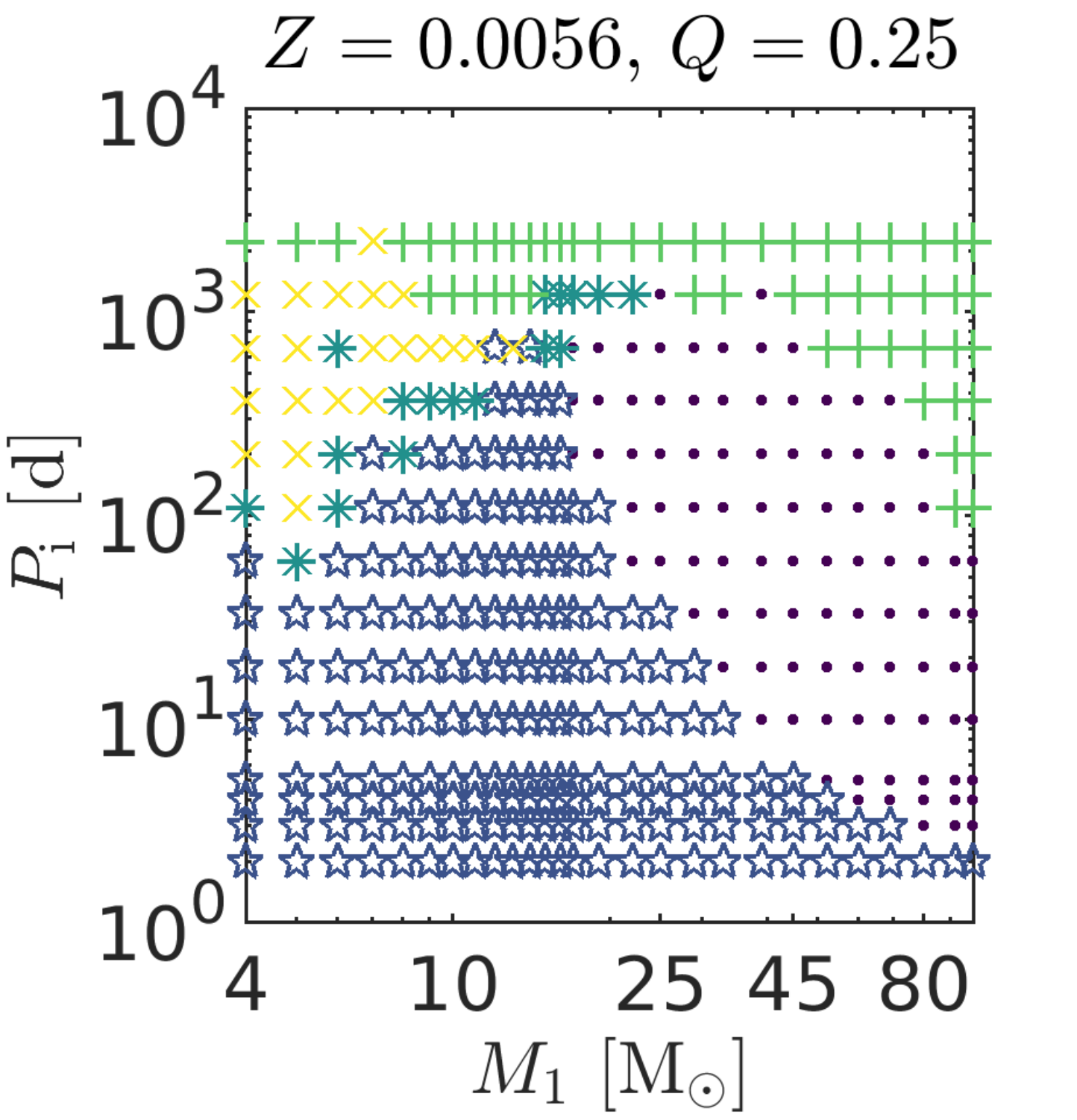}
        \label{fig:stabmapLMC25}
    \end{subfigure}
    \begin{subfigure}{0.33\textwidth}   
        \includegraphics[width=1\textwidth]{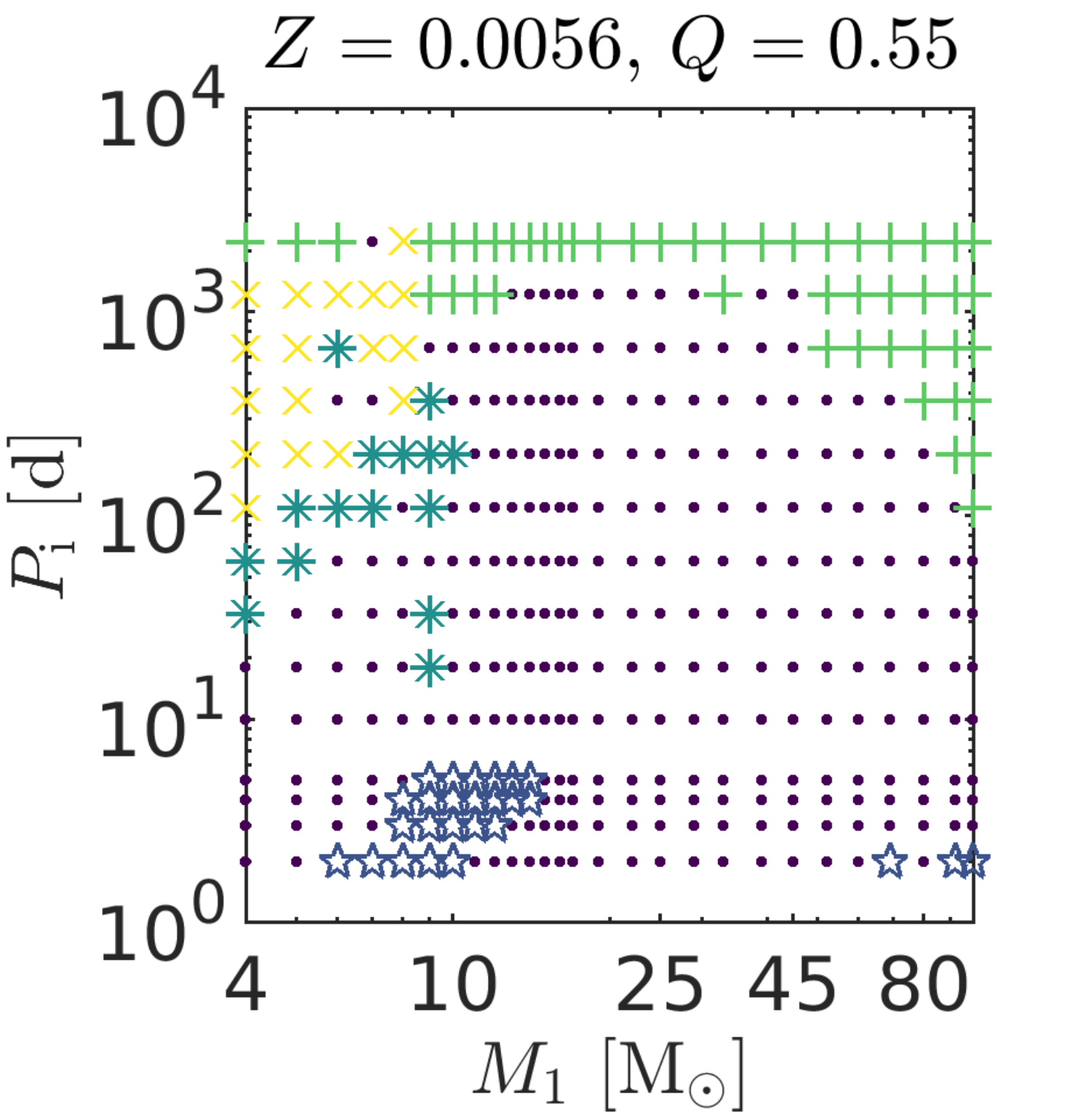}
        \label{fig:stabmapLMC55}
    \end{subfigure}
    \begin{subfigure}{0.33\textwidth}   
        \includegraphics[width=1\textwidth]{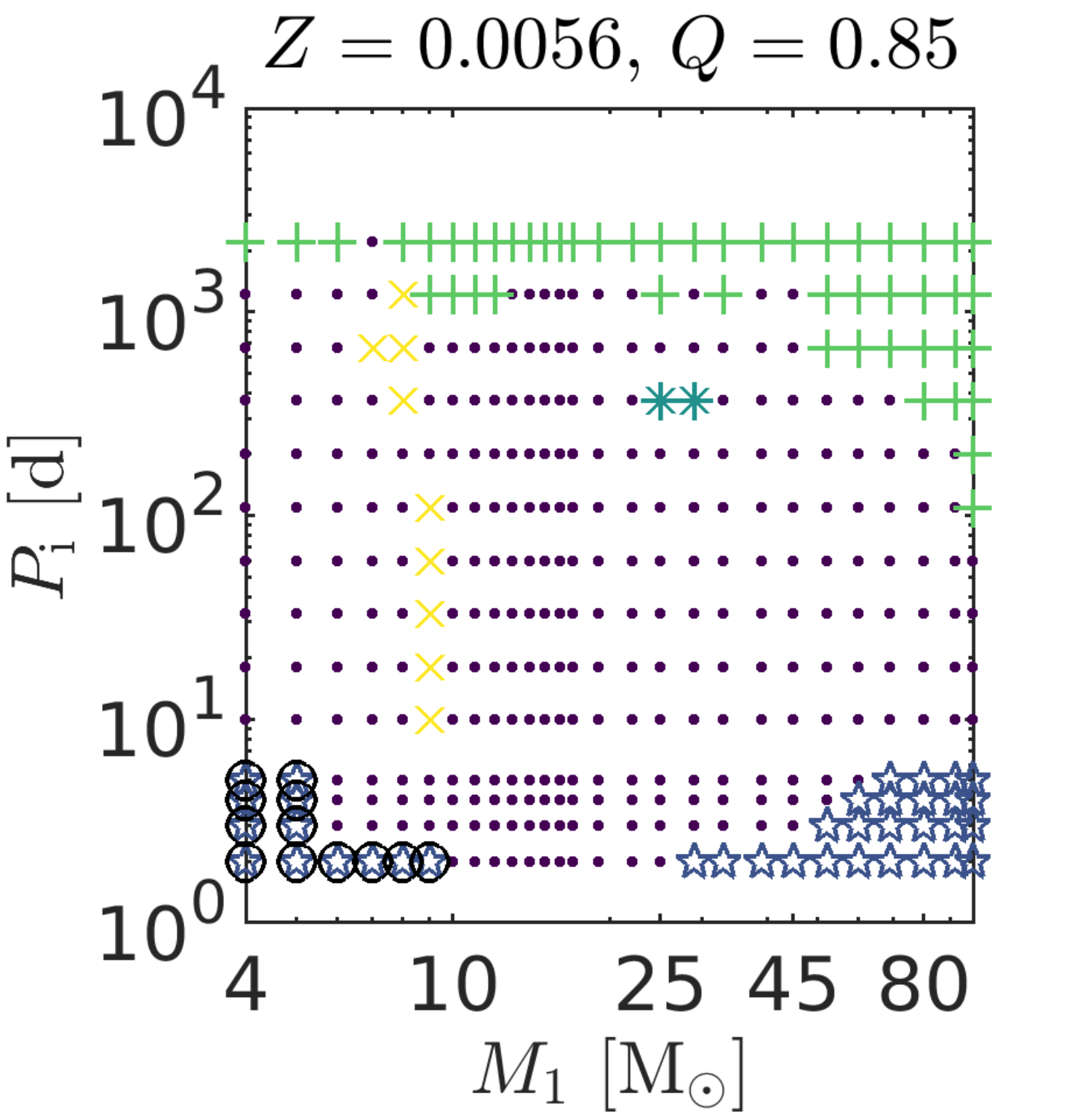}
        \label{fig:stabmapLMC85}
    \end{subfigure}\\
    \begin{subfigure}{0.33\textwidth}   
        \includegraphics[width=1\textwidth]{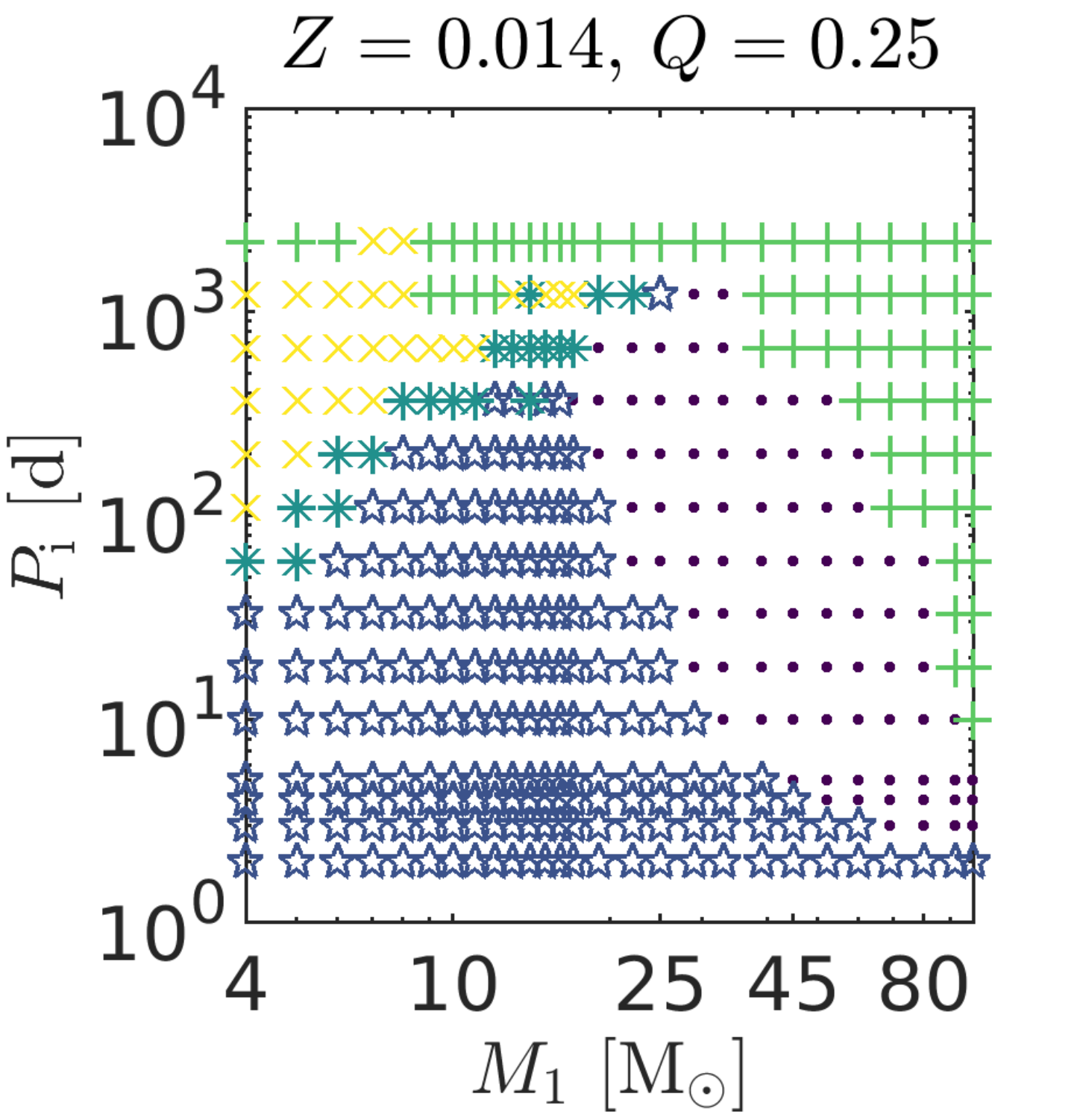}
        \label{fig:stabmapMW25}
    \end{subfigure}
    \begin{subfigure}{0.33\textwidth}   
        \includegraphics[width=1\textwidth]{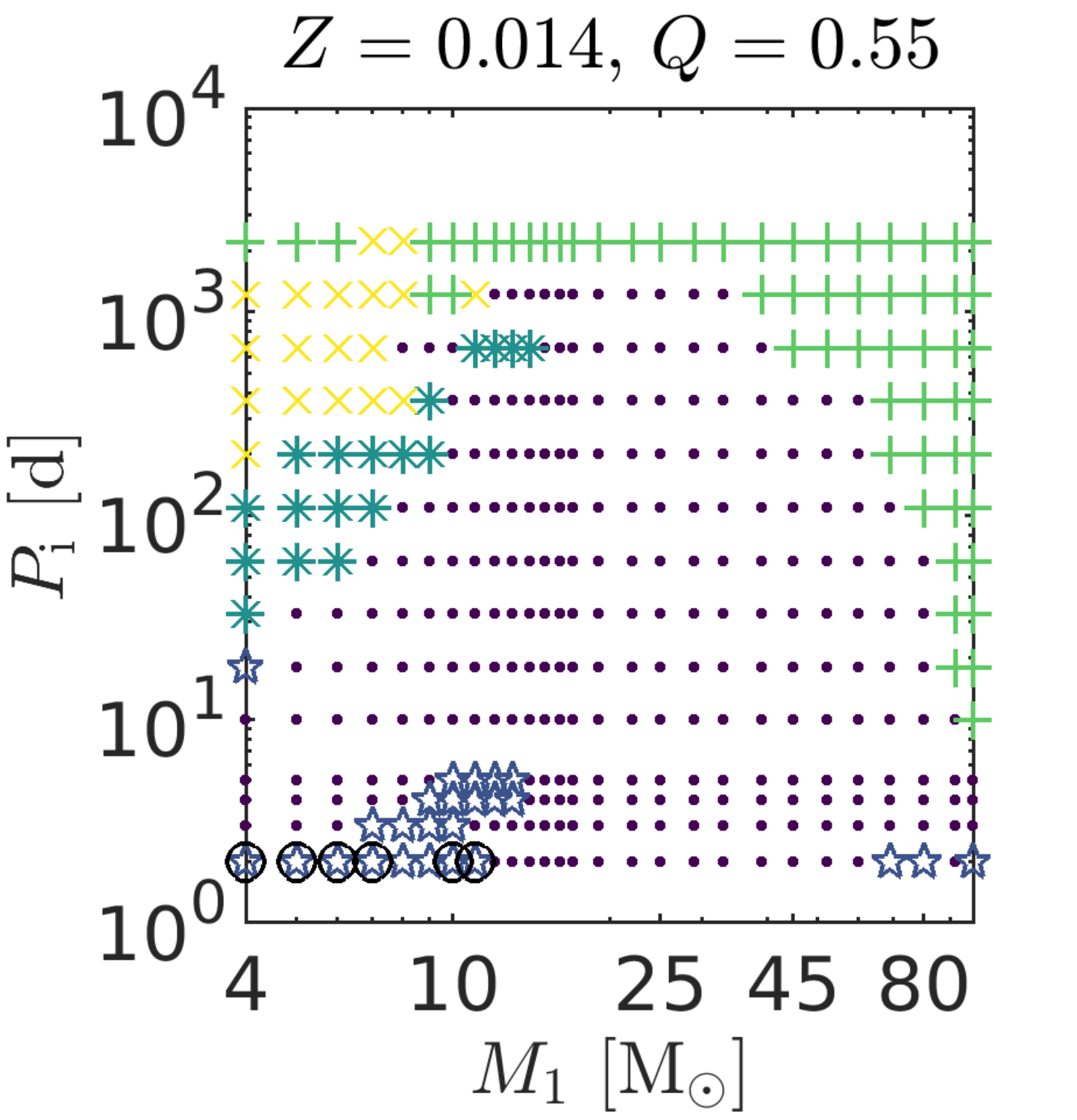}
        \label{fig:stabmapMW55}
    \end{subfigure}
    \begin{subfigure}{0.33\textwidth}   
        \includegraphics[width=1\textwidth]{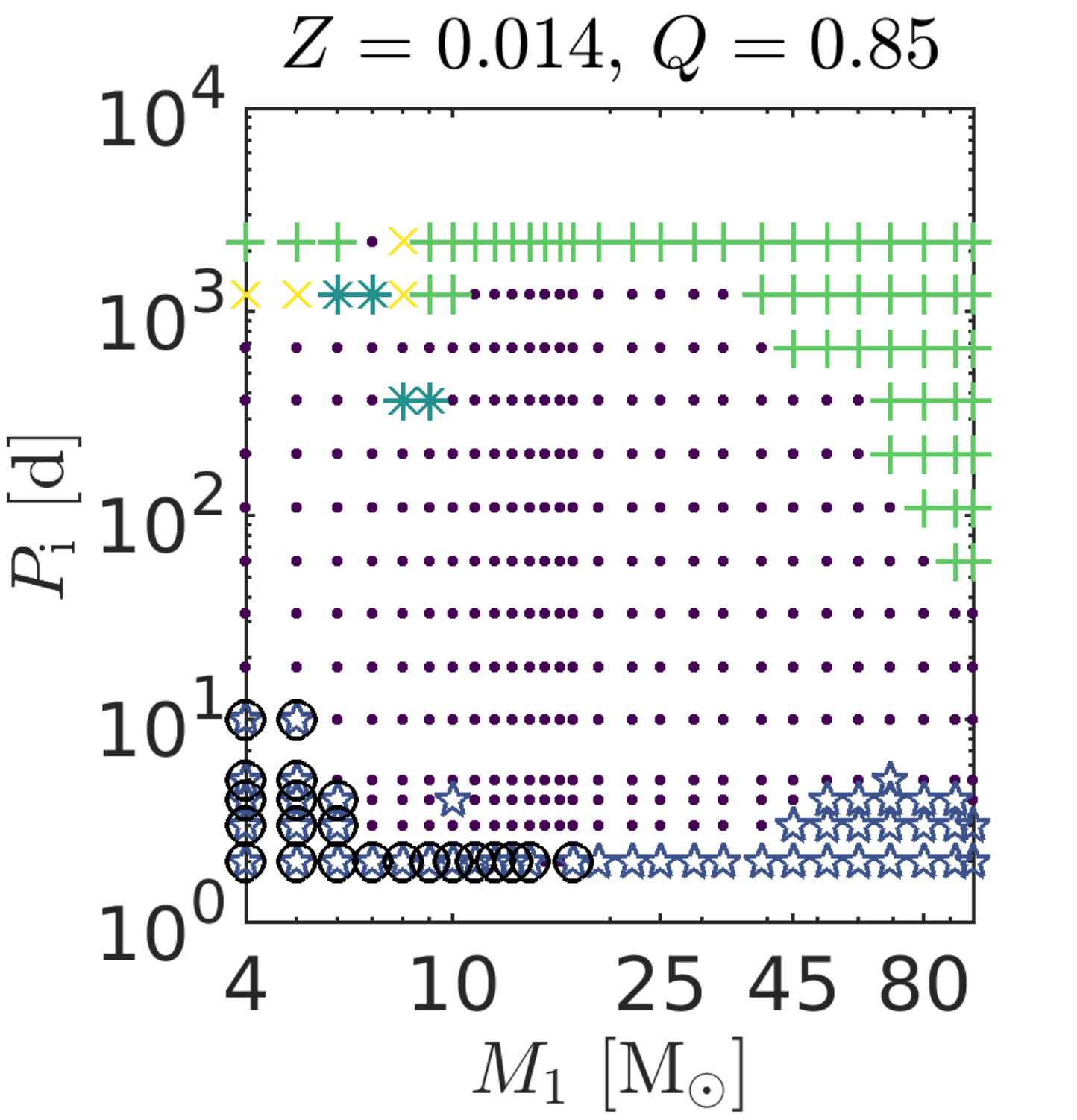}
        \label{fig:stabmapMW85}
    \end{subfigure}\\
\caption{Fate of binaries according to initial primary mass and orbital separation, for metallicity and mass ratio as indicated above each panel. The different points are marked as going through stable mass transfer (`St'), merging after contact or during CEE (`Me'), successfully ejecting the envelope during CEE (`Ej'), experiencing no mass transfer (`No') or being inconclusive (`In') when a simulation ends during the CEE phase because of numerical problems but there is no core overflow which would indicate the stars will merge. Simulations ending during CEE initiated by reverse mass transfer from the initially less massive star onto the primary are marked with black circles.}
\label{fig:stabmap}
\end{figure*}

The stability and outcome of mass transfer in our $3402$ binary simulations are shown in Fig.\,\ref{fig:stabmap}. Each panel contains the outcomes of $378$ simulations, with a specific metallicity and mass ratio,  $27$ different initial primary masses and $14$ different initial orbital periods. In some simulations there is no mass transfer at all, because the initial orbital separation is too wide for any interaction to occur, where in some high mass cases this is a result of strong wind mass loss preventing stellar expansion. For the remaining cases, mass transfer is either stable or unstable, with the latter option leading to CEE. For the most extreme mass ratio, $Q=0.25$, a large part of the parameter space leads to unstable mass transfer, and in most of these simulations the envelope is not ejected during CEE and the stars merge. A narrow strip in the parameter space results in CEE leading to envelope ejection, a result qualitatively similar to the study by \cite{Marchant2021} whose CEE implementation we use, although we consider a wider range of stellar masses. For binary systems where the two components have similar masses ($Q=0.55$ and $Q=0.85$) mass transfer is usually stable, though some short-period high-mass systems are assumed to merge after they achieve contact. For lower-mass systems with short initial periods, the first mass transfer stage is stable, but the evolution of the mass gainer is accelerated so that eventually it expands and initiates a late mass transfer stage that is unstable as the mass ratio between the mass gainer and the stripped star has become extreme. This evolutionary channel involving reverse mass transfer can result in hydrogen-deficient stars with masses lower than those of WR stars with no close companion, as has been suggested for the magnetic quasi-WR star HD~45166 \citep{Shenar2023}. Some simulations initiate CEE after unstable mass transfer but stop because of numerical problems before a conclusive outcome of the CEE phase is reached, and are therefore marked as inconclusive.
%FFFFFFFFFFFFFFFFFFFFFFFFFFFFFFFFFFFFFFFFFFFFFFFFFFFFFFFFFFFFFF
%FFFFFFFFFFFFFFFFFFFFFFFFFFFFFFFFFFFFFFFFFFFFFFFFFFFFFFFFFFFFFF

Mass transfer stability using detailed binary evolution grids has also been investigated with the \textsc{posydon} code \citep{Fragos2023}. Like our models, \textsc{posydon} models are generated with detailed \textsc{mesa} simulations to evaluate the donor’s response to mass loss. Our approach follows the criterion formulated by \cite{CopingWithLoss}, identifying systems as unstable when the donor cannot thermally adjust to avoid entering runaway mass transfer. \textsc{posydon} employs a similar physical basis but includes additional constraints, such as stability limits imposed by the thermal response of the accretor and orbital evolution \citep{Fragos2023}. While the overall trends are in broad agreement (both predicting stable mass transfer for radiative-envelope donors with mass ratios near unity), these extra stability checks can lead to slightly different boundaries between stable and unstable regions. Nonetheless, the physical conclusions are consistent across both approaches: the stability of mass transfer in massive binaries is governed primarily by the internal structure of the donor and the mass ratio at the onset of RLOF.

\section{CO core masses}
\label{sec:appendixb}

\begin{figure}
\centering
\includegraphics[width=0.48\textwidth]{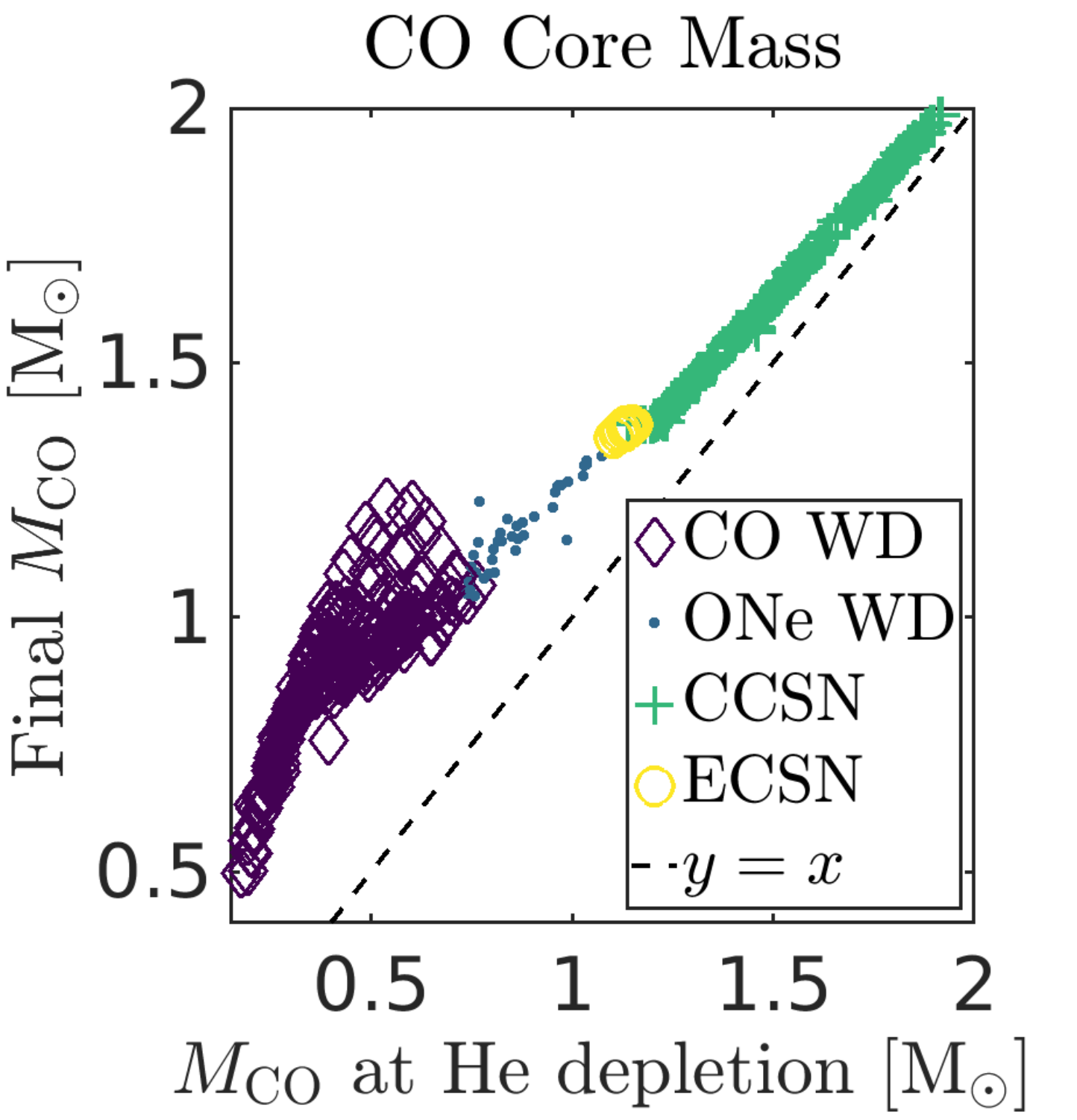} \\
\caption{The final CO core mass as function of the CO core mass at the end of core helium burning.}
\label{fig:MCO}
\end{figure}

In Fig.\,\ref{fig:MCO} we show the final CO core mass as function of the CO core mass at core helium depletion in the `Primary' channel simulations, for $M_\mathrm{CO} < 2\,\mathrm{M}_\odot$ (with increasing masses the CO core mass growth becomes smaller because of the short evolutionary timescale after helium burning). A clear division is found between the formation of CO WDs, ONe WDs, and CCSNe. In a narrow range between the highest CO core mass at helium depletion leading to the formation of ONe WDs and the minimum for iron core collapse we expect the outcome to be an ECSN.
%FFFFFFFFFFFFFFFFFFFFFFFFFFFFFFFFFFFFFFFFFFFFFFFFFFFFFFFFFFFFFF
%FFFFFFFFFFFFFFFFFFFFFFFFFFFFFFFFFFFFFFFFFFFFFFFFFFFFFFFFFFFFFF

\section{Example of evolution with NS companion}
\label{sec:appendixNSexamples}

\begin{figure}
\centering   
    \includegraphics[width=0.48\textwidth]{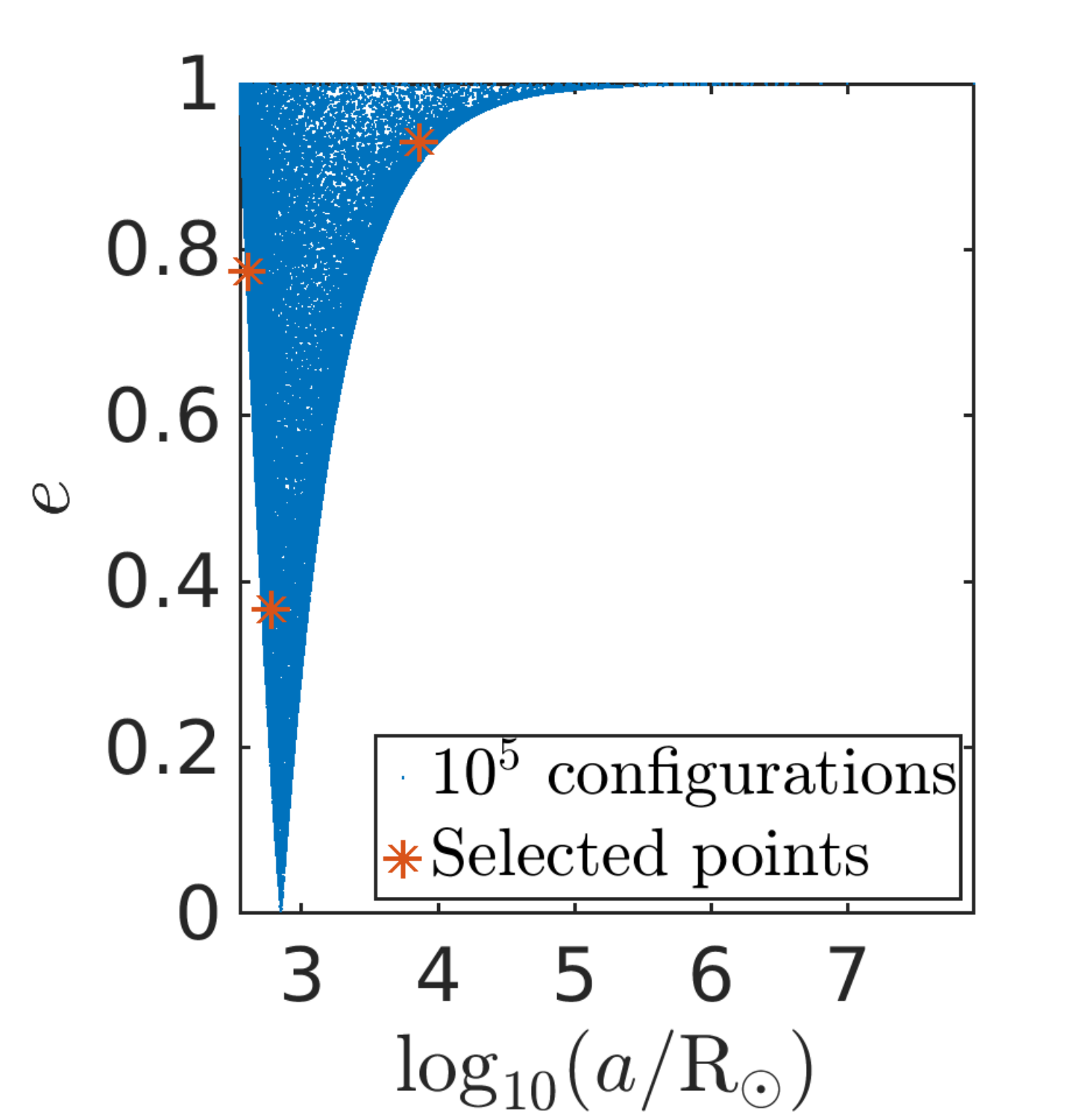}
    \caption{Post-SN configurations of $a$ and $e$ for a binary with metalllicity $Z=0.014$, initial masses $M_1=12\,\mathrm{M}_\odot$ and $M_2=10.2\,\mathrm{M}_\odot$, and initial orbital period $P_\mathrm{i}=60\,\mathrm{d}$.}
    \label{fig:kicks_e_a}
\end{figure}

\begin{figure}
\centering
    \includegraphics[width=0.48\textwidth]{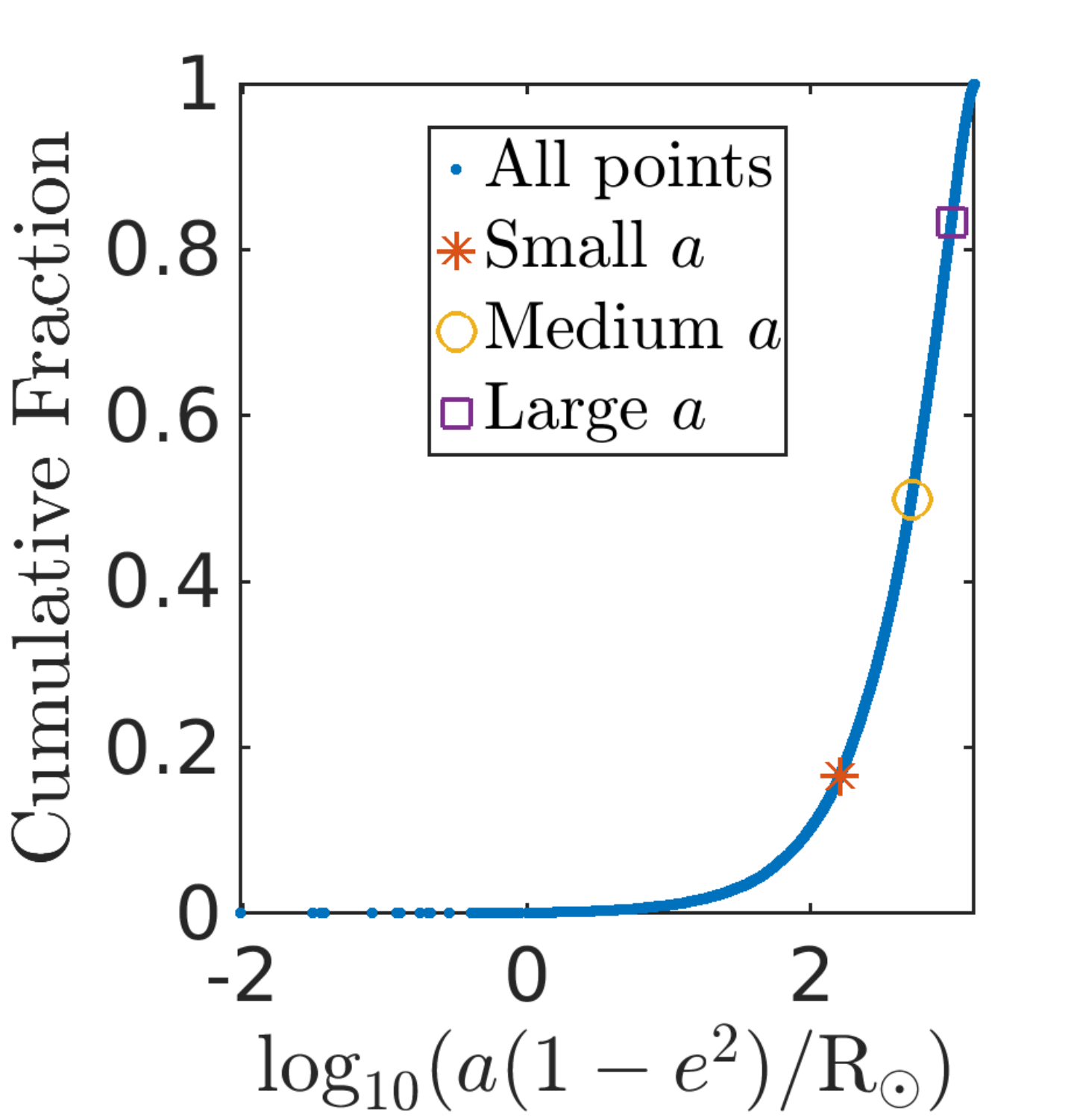}
    \caption{Cumulative distribution of $\log_{10} ( a ( 1 - e^2) / \mathrm{R}_\odot)$. The three selected representative points marked correspond to the $a$ and $e$ pairs marked in Fig.\,\ref{fig:kicks_e_a}.}
    \label{fig:kicks_a}
\end{figure}

Here we demonstrate our method of simulating post-SN binary configurations for a particular combination of initial conditions, $Z=0.014$, $M_1=12\,\mathrm{M}_\odot$, $M_2=10.2\,\mathrm{M}_\odot$, and $P_\mathrm{i}=60\,\mathrm{d}$. As described in Section\,\ref{subsec:channels}, we generate $10^5$ post-SN binary configurations, each with a different combination of the eccentricity $e$ and post-SN separation $a$. All $10^5$ points are shown in Fig.\,\ref{fig:kicks_e_a}. We assume that if there is any binary interaction then tidal circularisation would be fast, and then the orbital separation would reduce to $a(1-e^2)$. The cumulative distribution of $a(1-e^2)$ is presented in Fig.\,\ref{fig:kicks_a}, with three points representing `small', `medium' and `large' post-SN separations marked, each in the middle of a tercile. These values are used in setting up the simulations including a newly-formed NS and the surviving companion. The combinations of $e$ and $a$ corresponding to these selected separations are marked in Fig.\,\ref{fig:kicks_e_a}.
%FFFFFFFFFFFFFFFFFFFFFFFFFFFFFFFFFFFFFFFFFFFFFFFFFFFFFFFFFFFFFF
%FFFFFFFFFFFFFFFFFFFFFFFFFFFFFFFFFFFFFFFFFFFFFFFFFFFFFFFFFFFFFF
%FFFFFFFFFFFFFFFFFFFFFFFFFFFFFFFFFFFFFFFFFFFFFFFFFFFFFFFFFFFFFF
%FFFFFFFFFFFFFFFFFFFFFFFFFFFFFFFFFFFFFFFFFFFFFFFFFFFFFFFFFFFFFF

\begin{figure*}
\centering
\includegraphics[width=\textwidth]{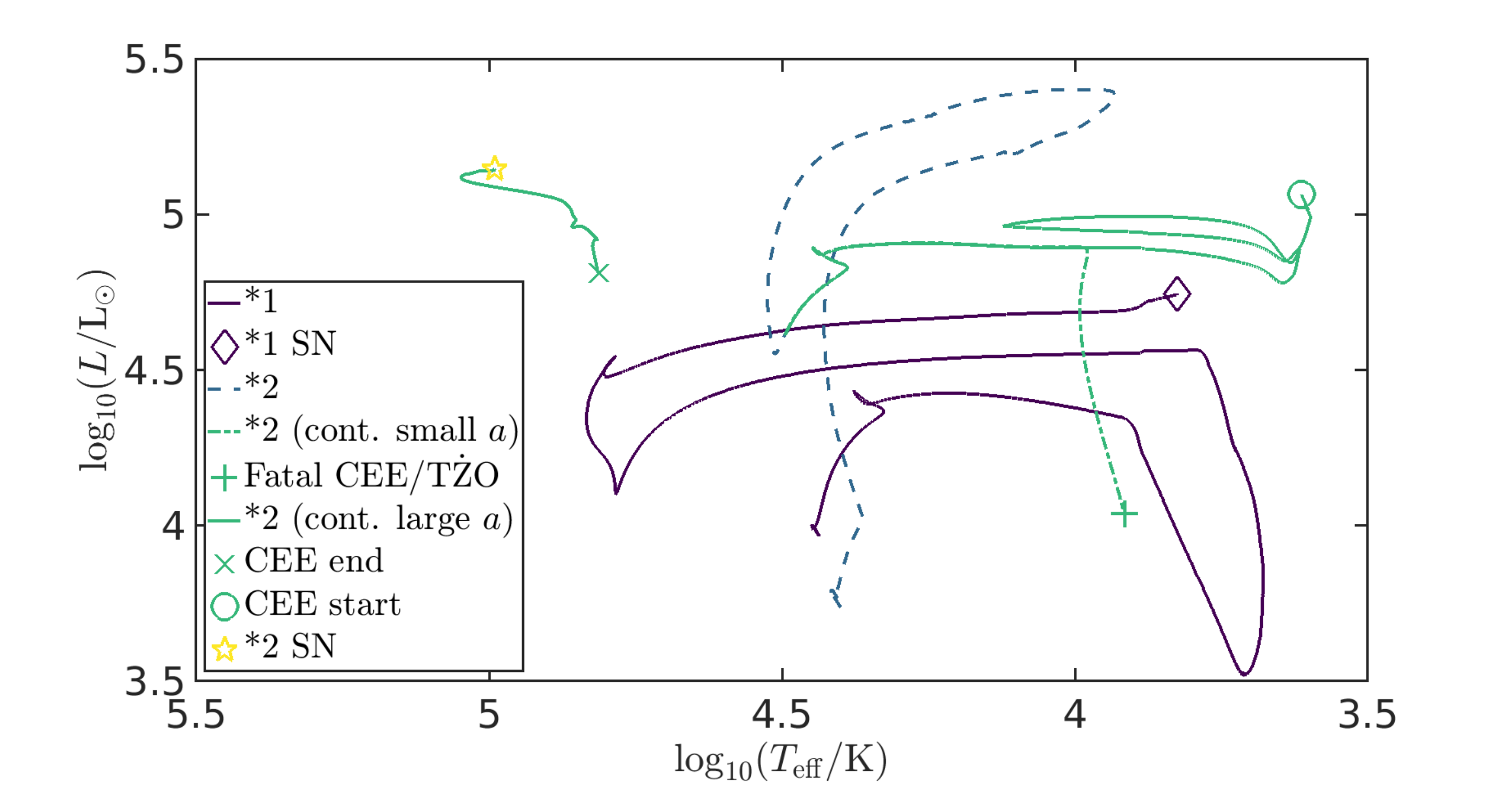} \\
\caption{Effective temperature and luminosity evolution of the two stellar components in a system with a metalllicity $Z=0.014$, initial masses $M_1=12\,\mathrm{M}_\odot$ and $M_2=10.2\,\mathrm{M}_\odot$, and initial orbital period $P_\mathrm{i}=60\,\mathrm{d}$. The configuration with the smallest post-SN separation results in a CEE phase in which the NS merges with the core of a massive star, while the largest post-SN separation results in a CEE phase ending with the ejection of the envelope and a CCSN of a stripped star.}
\label{fig:NSkicksHRD}
\end{figure*}

In Fig.\,\ref{fig:NSkicksHRD} we show two possible evolutionary outcomes, for the `small' and `large' post-SN separations. The first part of the evolution is the same for both cases, including stable mass transfer resulting in the primary going through an intermediate-mass helium star evolutionary phase followed by expansion to a hydrogen-deficient giant that explodes as a stripped-envelope SN. The secondary star gains mass during the first mass transfer phase and becomes more luminous than the initially more massive component of the binary, having $M=16.6\,\mathrm{M}_\odot$ when the primary explodes. For the post-SN configuration with the `small' separation, mass transfer begins when the mass gainer is crossing the Hertzsprung gap. As the mass ratio is high, mass transfer is unstable and a CEE phase ensues. As the envelope of the mass gainer has a high binding energy, it is not ejected, and the stellar core merges with the NS (resulting in a SN-like event or the formation of a T{\.Z}O). For the `large' separation case, mass transfer begins only after the mass gainer becomes a RSG, and its envelope is less tightly bound. The ensuing CEE phase ends with a successful envelope ejection, and the mass gainer explodes $13\,\mathrm{kyr}$ later in a CCSN when it is a WR star with $M=5.6\,\mathrm{M}_\odot$ and $\dot{M}=2\times 10^{-5}\,\mathrm{M}_\odot\,\mathrm{yr}^{-1}$ in a $0.35\,\mathrm{d}$ period with its NS companion.
%FFFFFFFFFFFFFFFFFFFFFFFFFFFFFFFFFFFFFFFFFFFFFFFFFFFFFFFFFFFFFF
%FFFFFFFFFFFFFFFFFFFFFFFFFFFFFFFFFFFFFFFFFFFFFFFFFFFFFFFFFFFFFF

\section{Endpoints by evolutionary channel for different metallicities}
\label{sec:endpointsZ}

\begin{figure*}
   \centering
    \begin{subfigure}{\textwidth}   
        \includegraphics[width=1\textwidth]{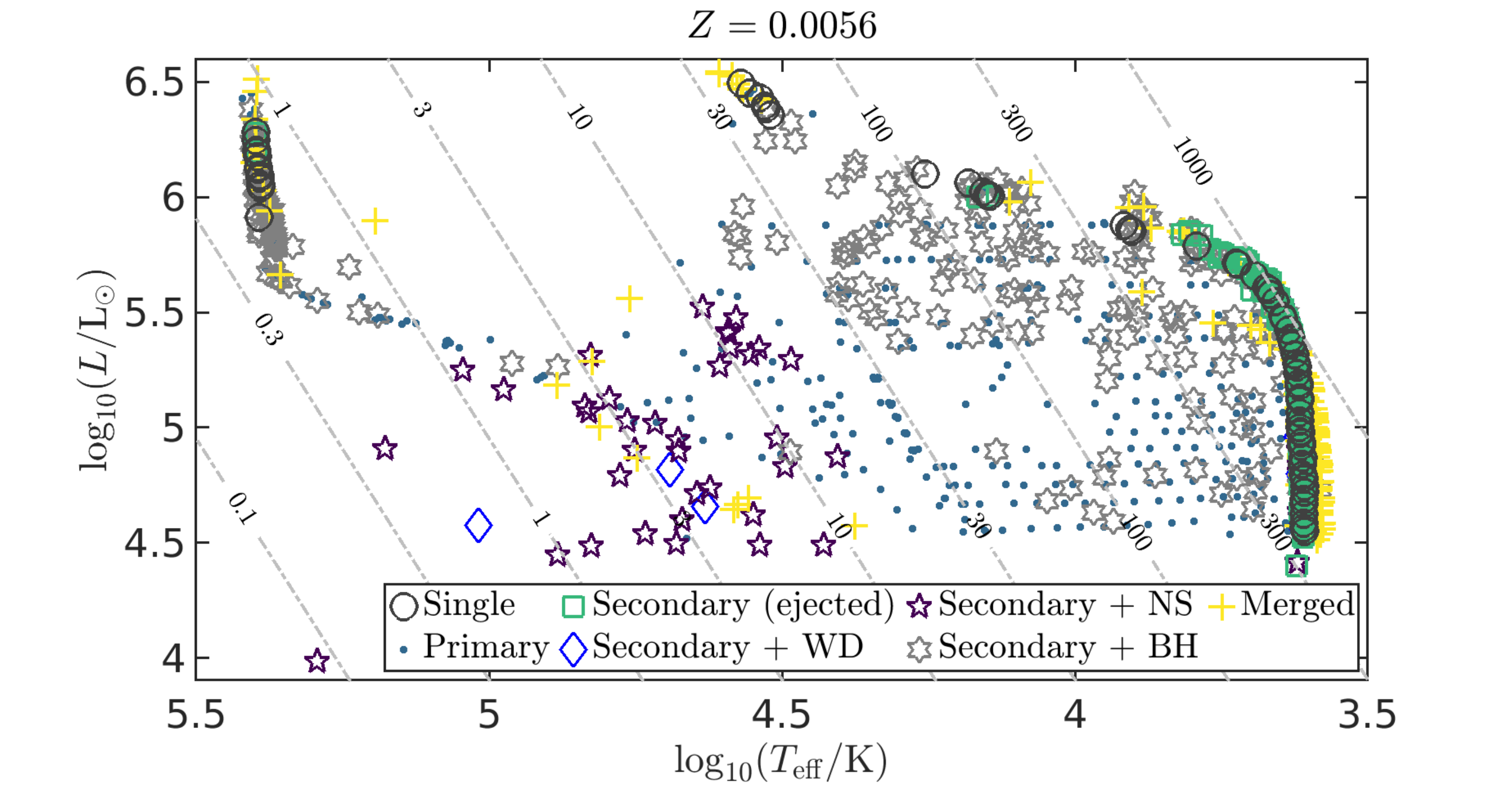}
        \label{fig:HRDchanMW}
    \end{subfigure}
    \begin{subfigure}{\textwidth}   
        \includegraphics[width=1\textwidth]{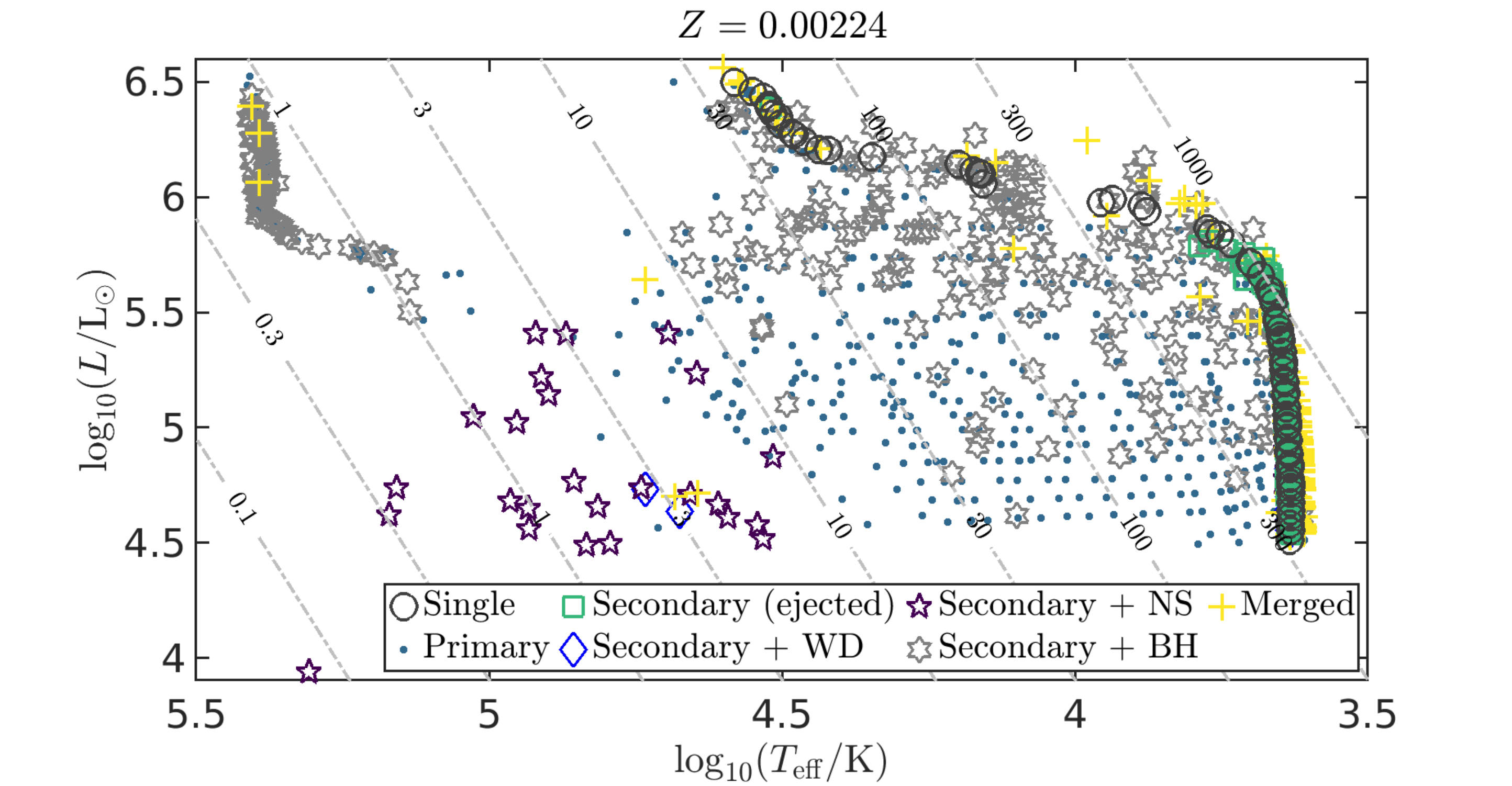}
        \label{fig:HRDchanLMC}
    \end{subfigure}\\
    \caption{Temperature and luminosity of evolutionary endpoints, with metallicity $Z=0.0056$ (top) and $Z=0.00224$ (bottom), marked by evolutionary channel. Lines of constant radius according to the temperature-luminosity relation are plotted, with the numbers indicating the radius in units of $\mathrm{R}_\odot$.}
    \label{fig:HRDchanZ}
\end{figure*}

In Fig.\,\ref{fig:HRDchanZ} we present the effective temperature and luminosity of the models at the evolutionary endpoints of all simulations with metallicities of $Z=0.0056$ and $Z=0.00224$ that reached the end of core carbon burning. Together with Fig.\,\ref{fig:HRDchan} for $Z=0.014$, we see several similarites and differences. The results for different metallicities similarly cover the same overall range of temperatures and luminosities, and contain highly-stripped endpoints from the evolution of the secondary star with a NS or WD companion. Two key differences between metallicities are the shift of the WR sequence to higher luminosities with decreasing metallicity, and the denser population of endpoints in the cooler regimes for lower metallicities. These two aspects correspond to an increasing difficulty in stripping the envelope with decreasing metallicity, and the lower metallicity endpoints retain significantly more hydrogen so that more CSGs are produced rather than helium giants or WR stars.  
%FFFFFFFFFFFFFFFFFFFFFFFFFFFFFFFFFFFFFFFFFFFFFFFFFFFFFFFFFFFFFF
%FFFFFFFFFFFFFFFFFFFFFFFFFFFFFFFFFFFFFFFFFFFFFFFFFFFFFFFFFFFFFF

\section{Formation of an ultra-stripped SN progenitor}
\label{sec:appendixUSSN}

\begin{figure*}
\centering
\includegraphics[width=\textwidth]{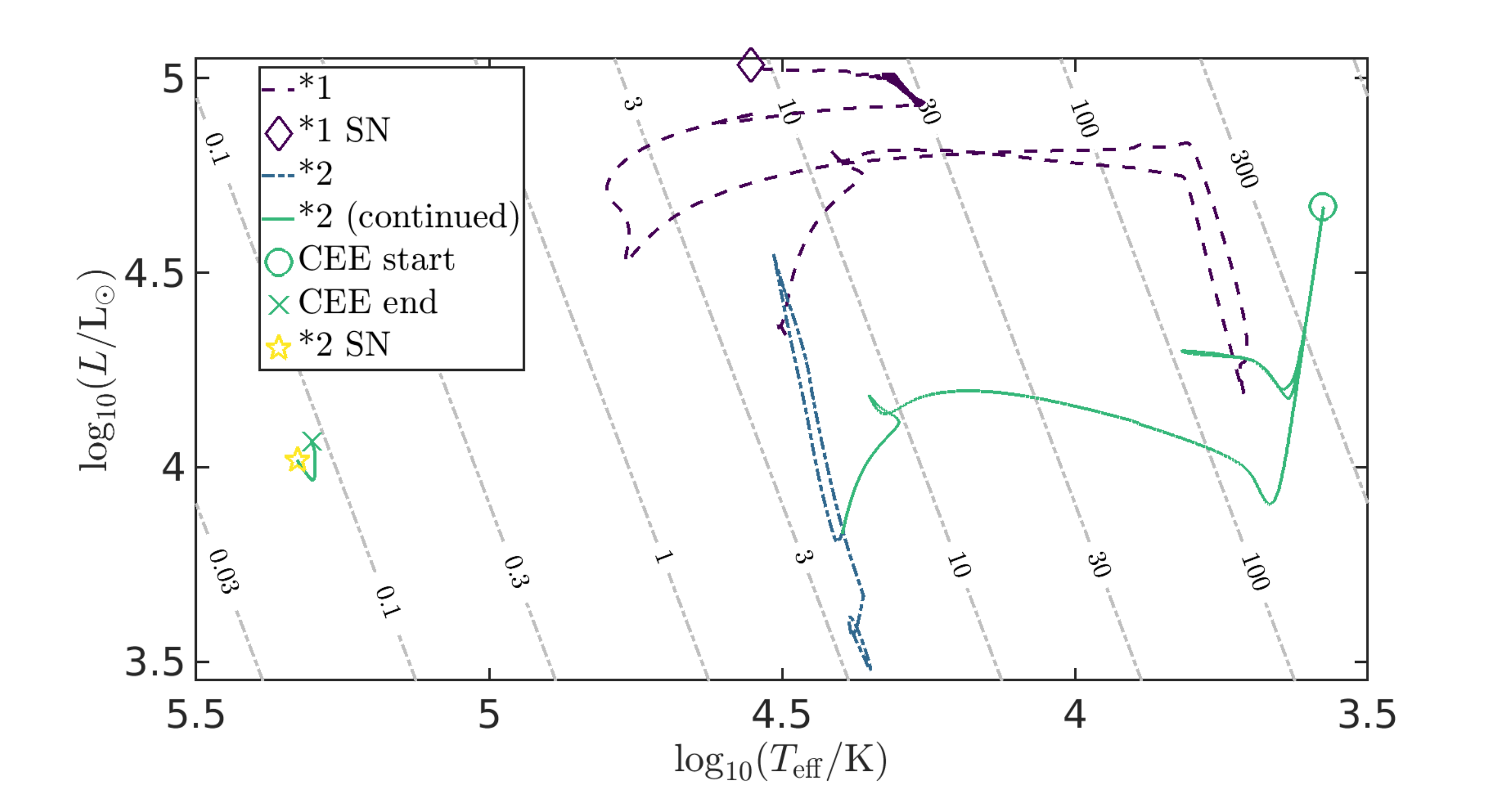} \\
\caption{Effective temperature and luminosity evolution of the two stellar components in a system with a metallicity $Z=0.014$, initial masses $M_1=16\,\mathrm{M}_\odot$ and $M_2=8.8\,\mathrm{M}_\odot$, and initial orbital period $P_\mathrm{i}=201\,\mathrm{d}$. Of the three configurations generated from random natal kicks, the one with the largest post-SN separation results in the ultra-stripped SN progenitor shown here.}
\label{fig:USSN}
\end{figure*}

In Fig.\,\ref{fig:USSN} we depict the evolution towards an USSN progenitor, starting from two stars on the ZAMS. In the first stage of the evolution there is stable mass transfer from the primary onto the secondary, resulting in the primary losing most of its envelope and the secondary gaining mass. After the primary explodes, we generate $10^5$ post-SN binary configurations as described in Section\,\ref{subsec:channels} and Appendix\,\ref{sec:appendixNSexamples}. The `small' and `medium' post-SN separations lead to CEE following unstable mass transfer, with no envelope ejection. For the largest post-SN separation (shown in Fig.\,\ref{fig:USSN}) unstable mass transfer begins when the mass gainer has significantly expanded and its envelope is less tightly bound. The ensuing CEE phase results in a successful envelope ejection and a short orbital period of $\approx 10$ minutes. The originally less massive star in the binary reaches the end of its evolution with $R=0.075\,\mathrm{R}_\odot$, $M=1.75\,\mathrm{M}_\odot$ and $M_\mathrm{CO}=1.64\,\mathrm{M}_\odot$ and $4.9\,\mathrm{kyr}$ after the end of the CEE phase is expected to explode in a CCSN with an ejecta mass of $M_\mathrm{ej}\approx 0.3\,\mathrm{M}_\odot$.
%FFFFFFFFFFFFFFFFFFFFFFFFFFFFFFFFFFFFFFFFFFFFFFFFFFFFFFFFFFFFFF
%FFFFFFFFFFFFFFFFFFFFFFFFFFFFFFFFFFFFFFFFFFFFFFFFFFFFFFFFFFFFFF

\section{Low metallicity CCSN progenitor probability maps}
\label{sec:appendixzmaps}

%FFFFFFFFFFFFFFFFFFFFFFFFFFFFFFFFFFFFFFFFFFFFFFFFFFFFFFFFFFFFFF
\begin{figure*}
\centering
    \begin{subfigure}{0.48\textwidth}   
        \includegraphics[width=1\textwidth]{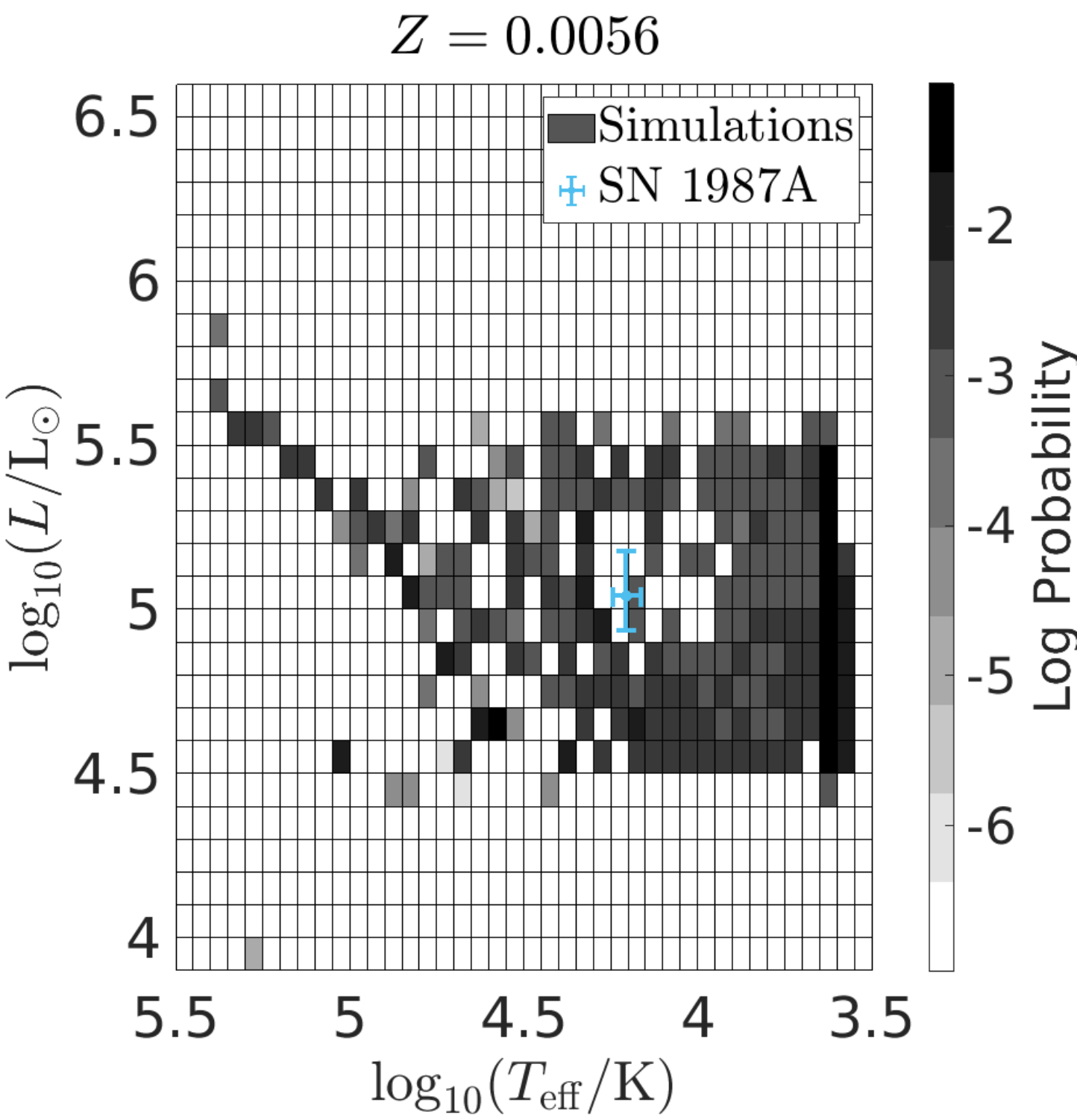}
        \label{fig:progmapLMC}
    \end{subfigure}
    \begin{subfigure}{0.48\textwidth}   
        \includegraphics[width=1\textwidth]{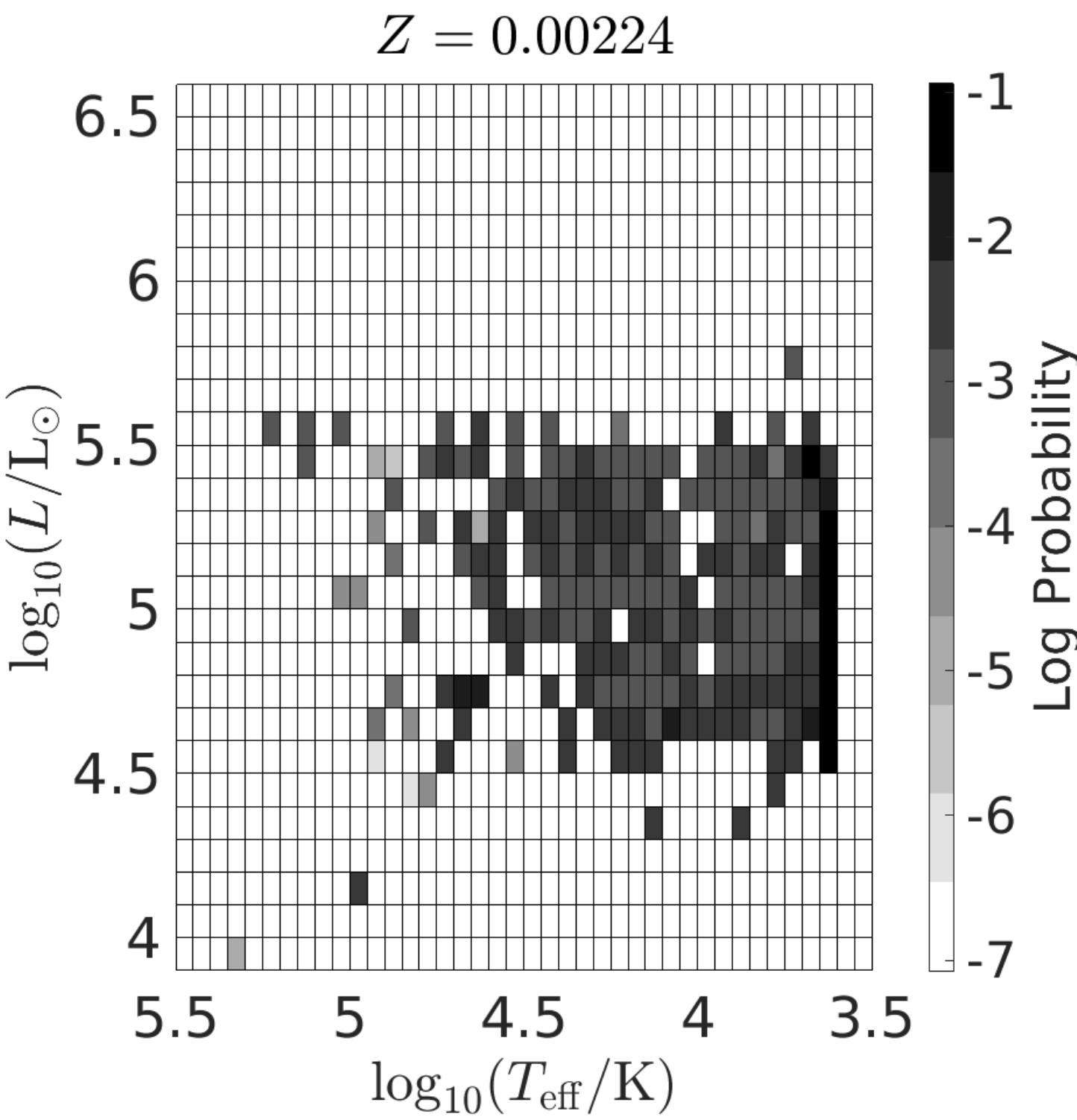}
        \label{fig:progmapSMC}
    \end{subfigure}\\
\caption{Weighted probability map of CCSN progenitors for $Z=0.0056$ (left) and for $Z=0.00224$ (right).}
\label{fig:extraprogmap}
\end{figure*}
%FFFFFFFFFFFFFFFFFFFFFFFFFFFFFFFFFFFFFFFFFFFFFFFFFFFFFFFFFFFFFF
In Fig.\,\ref{fig:extraprogmap} we show the probability of endpoints in bins of luminosity and effective temperature, for $Z=0.0056$ and $Z=0.0056$. The general spread of exploding endpoints is similar for the three metallicities we considered, but there is an apparent trend of numbers of CCSNe originating from the hottest endpoints decreasing with metallicity, with a corresponding increase in the probability for endpoints with intermediate effective temperatures. This is an effect of partial stripping, with lower metallicity stars generally retaining more of their envelope even after binary interactions \citep{Gotberg2017,Laplace2020}.

\label{lastpage}

\end{document}